\documentclass[%
a4paper,       % Seitengroesse
UKenglish,     % Sprache
DIV10,          % evtl. Vergrößerung des Textbereiches
useAMS, referee, doublespace]%
{scrartcl}\usepackage[]{graphicx}\usepackage[]{xcolor}
\makeatletter
\def\maxwidth{ %
  \ifdim\Gin@nat@width>\linewidth
    \linewidth
  \else
    \Gin@nat@width
  \fi
}
\makeatother

\definecolor{fgcolor}{rgb}{0.345, 0.345, 0.345}

\usepackage{framed}
\makeatletter
 {\par\unskip\endMakeFramed%
 \at@end@of@kframe}
\makeatother

\definecolor{shadecolor}{rgb}{.97, .97, .97}
\definecolor{messagecolor}{rgb}{0, 0, 0}
\definecolor{warningcolor}{rgb}{1, 0, 1}
\definecolor{errorcolor}{rgb}{1, 0, 0}
\newenvironment{knitrout}{}{} % an empty environment to be redefined in TeX

\usepackage{alltt}

\usepackage[a4paper, total={6in, 9.25in}]{geometry}
\usepackage{xr-hyper}
\usepackage{authblk}    
\usepackage{hyperref}
\input{header.sty}
\usepackage{color, colortbl}
\usepackage{mathtools}
\usepackage{multirow}
\usepackage{colortbl}
\usepackage{afterpage}
\usepackage{pdflscape}
\usepackage{orcidlink}
\usepackage{txfonts}

\newcommand{\pleft}[1]{\overset{\leftharpoonup}{#1}}
\newcommand{\pright}[1]{\overset{\rightharpoonup}{#1}}

\newcolumntype{C}[1]{>{\centering\arraybackslash}p{#1}}
\IfFileExists{upquote.sty}{\usepackage{upquote}}{}
\begin{document}

\title{A comparison of combined \textit{p}-value functions for meta-analysis}
\ifcase\blinded % 0 (default):
\author{} \or % 1: german cv
\author{
  Leonhard Held\textsuperscript{1}\footnote{Corresponding author, \texttt{leonhard.held@uzh.ch}} ~\orcidlink{0000-0002-8686-5325},
  Felix Hofmann\textsuperscript{1}\orcidlink{0000-0002-3891-6239},
  Samuel Pawel\textsuperscript{1}\orcidlink{0000-0003-2779-320X}}
\affil{
  \large \textsuperscript{1}University of Zurich\\
Epidemiology, Biostatistics and Prevention Institute (EBPI)\\
  and Center for Reproducible Science (CRS) \\
  Hirschengraben 84, 8001 Zurich, Switzerland
} 
\date{\large \today}
\fi

\maketitle

\begin{center}
  \fbox{
    \parbox{0.95\textwidth}{

      \begin{center}
        \Large \textsf{\textbf{Highlights}}
      \end{center}

      \subsubsection*{What is already known}
      \begin{itemize}
        \item $P$-value functions unify 
              hypothesis testing and parameter estimation, and are therefore
              particularly useful for quantitative reporting of statistical
              analyses.
        \item \textit{P}-value combination methods provide a general framework 
              to perform meta-analysis. 
      \end{itemize}

      \subsubsection*{What is new}
      \begin{itemize}
        \item $P$-value functions of different \textit{p}-value combination
          methods are compared.          
            \item Edgington’s method has attractive properties: 
              \begin{itemize}
\item Results do not depend on the orientation of the underlying one-sided $p$-values
\item Confidence intervals are not restricted to be symmetric
% \item The median estimate is essentially unbiased
              \end{itemize}
            \item   A simulation study is performed without and with adjustments for heterogeneity. Comparisons with standard meta-analysis (fixed effect and DerSimonian-Laird random effects) and the Hartung-Knapp method are described.
              \begin{itemize}
           \item The point estimate based on Edgington's method is essentially unbiased for the mean of a normal study effects distribution 
              \item Coverage of Edgington's method is comparable or better than standard meta-analysis, with only slightly wider confidence intervals
\item Confidence intervals based on the Hartung-Knapp method have better coverage, but are also substantially wider
      \end{itemize}
     \end{itemize}
      \subsubsection*{Potential impact} %for \textit{Research Synthesis Methods} readers}
      \begin{itemize}
        \item Edgington's combination method based on the sum of
              \textit{p}-values may complement standard meta-analysis because of its
              ability to reflect data asymmetry, its orientation-invariance, and
              its good operating characteristics.
 \item The usage of $p$-value function methods for meta-analysis is
   faciliated thrugh development of the R package \texttt{confMeta}.
     \end{itemize}
    }
    }
      \end{center}

\newpage

\begin{center}
\begin{minipage}{12cm}
  \textbf{Abstract}:   \\

  $P$-value functions are modern statistical tools that unify effect
  estimation and hypothesis testing and can provide alternative point
  and interval estimates compared to standard meta-analysis methods,
  using any of the many $p$-value combination procedures available
  (Xie et al., 2011, JASA).  %%The resulting confidence intervals are
%%  often not restricted to be symmetric around the point estimate. 
  We provide a systematic comparison of
  different combination procedures, both from a theoretical
  perspective and through simulation. We show that many prominent
  $p$-value combination methods (e.g. Fisher's method) are not
  invariant to the orientation of the underlying one-sided
  $p$-values. Only Edgington's method, a lesser-known combination
  method based on the sum of $p$-values, is orientation-invariant and
  still provides confidence intervals not restricted to be
  symmetric around the point estimate. Adjustments for heterogeneity
  can also be made and results from a simulation study indicate that
  Edgington's method can compete with more standard meta-analytic methods.

\noindent
  \textbf{Key Words}: Confidence curve; Confidence distribution; Heterogeneity; Meta-analysis; $p$-value combination; skewness
\end{minipage}
\end{center}

\section{Introduction}\label{sec:introduction}

%% \todo[inline]{need to discuss \citet{JacksonWhite2018} and \citet{Weber_etal2021}, perhaps also
%%   \citet{vandenNoortgateOnghena2005}}

A pervasive challenge in all areas of research is the assessment of
evidence from multiple studies. Standard meta-analysis aims to
synthesize effect estimates from several studies into an overall
effect estimate, typically a weighted average of the study-specific
effect estimates, combined with an appropriate confidence
interval. Inverse variance weights can be motivated as efficient
choices under homogeneity or heterogeneity between studies
\citep{Rice_etal2018} via either exchangeability or random sampling of
study effects \citep{Higgins_etal2009}. The \citet{DerSimonianLaird1986}
approach to random effects meta-analysis incorporates a measure of
heterogeneity into the weights, but does not incorporate uncertainty
in the variance estimate when making inference on the mean of the
random effects distribution.  This form of weights gives an estimate
that is consistent for the mean of the distribution of study effects
\citep{Normand1999,Jackson_etal2010}, a natural target of inference
where a symmetric (usually normal) distribution can be assumed.

There has been much progress in proposing alternative confidence
intervals for meta-analysis. The approach by \citet{HartungKnapp2001b,HartungKnapp2001} and \citet{Sidik2002} takes into
account the uncertainty in estimating heterogeneity and tends to
produce wider confidence intervals, in particular if the number of studies is small.
The approach by \citet{HenmiCopas2010}
combines the fixed effect point estimate with a standard error from
the random effects model, in order to obtain confidence intervals less
prone to publication bias. However, all these intervals are of a simple
additive form with limits
\begin{equation}\label{eq:additiveCI}
\mbox{ point estimate } \pm \mbox{ additive factor, }
\end{equation}
so symmetric around the point estimate. This may be reasonable if the
number of studies is large or if there is good reason to assume that
the true effect estimates follow a symmetric (normal) distribution
around their mean, but confidence intervals not restricted to be
symmetric around the point estimate may be more suitable if this is
not the case. Non-symmetric confidence intervals can show improved
performance in other applications, for example the ``square-and-add'' Wilson score
interval for the risk difference is non-symmetric and performs better
than symmetric confidence intervals \citep[Section 7.3]{Newcombe1998,
  Newcombe2013}. Also the nonparametric confidence interval for the
median survival time is non-symmetric and performs better than alternativ
parametric and symmetric confidence intervals
\citep{Brookmeyer1982}.  Other prominent examples for non-symmetric
confidence intervals are nonparametric bootstrap confidence intervals
based on the percentile method
\citep{Efron1993,ChiharaHesterberg2022}, deterministic bootstrap
intervals for odds ratios, risk differences and relative risks
\citep[Section 11.4]{Newcombe2013} and confidence intervals for a
population median or the difference of two population medians
\citep{CampbellGardner2000}.  %%\todo[inline,color=blue!10]{discuss KM
%%  square and add? Is it necessary to have square and add + Wilson?
%%  Wilson is for bounded parameter space and already non-symmetric}

In this paper we compare meta-analytic methods based on the combined
$p$-value function \citep{Fraser2019,Infanger2019} or equivalently
confidence curve \citep{Bender_etal2005} and confidence distribution
\citep{Marschner2024,Melilli2024}.  Related meta-analytic approaches
based on $p$-value functions have been proposed by
\citet{Singh2005}. They showed that $p$-value combination based
meta-analysis -- approaches that combine $p$-values of individual
studies, such as Fisher's method -- can be unified with standard
model-based meta-analysis under a common framework using $p$-value
functions. This framework has subsequently been extended
\citep{Xie2011, Liu2014, CunenHjort2021} and different methods have
been recently compared for rare event meta-analysis
\citep{Zabriskie2021}.  \citet[p.~1381]{Yang2016} consider $p$-value
combination methods for rare events based on Fisher's exact test and
note in an application to simulated data that ``\emph{the $p$-value
function based on the exact test preserves the skewness}'' of the
original data. This statement suggests that a desired property of
meta-analytic confidence interval is to reflect data
asymmetry. However, the authors did not investigate this feature any
further.  A related proposal in the meta-analytic literature is
the ``drapery plot'' of \citet{Ruecker2020}. This visualization, an
alternative to the standard forest plot, shows the $p$-value functions
of individual studies and of their pooled effect, providing the reader
with a wealth of information as $p$-values, point estimates, and
confidence intervals (at any level) can be easily read off.

We apply these ideas in our article and provide a systematic
comparison of different types of $p$-value combination procedures
\citep{HedgesOlkin1985,cousins2007annotated} for meta-analysis.
Theory and simulation studies are used to compare the different
confidence intervals and point estimates in terms of coverage, bias, width,
and skewness. The results indicate that four of the five
$p$-value combination methods considered have undesirable properties,
only Edgington's method \citep{Edgington1972,Edgington1972b} based on
the sum of $p$-values can compete with more standard meta-analytic
methods.
%% \todo[inline,color=blue!10]{Maybe also briefly summarize
%%  results for other $p$-value combination methods to shift focus to
%%  comparison of methods rather than only Edgington}

\section{Methodology}\label{sec:methodology}

Suppose results from $k$ studies are to be synthesized and 
let $\hat \theta_i$ denote the effect estimate of the true parameter
$\theta_i$ from the $i$-th study, $i=1,\ldots, k$, and $\sigma_i$ the
corresponding standard error.  As in standard meta-analysis we assume
that the $\hat \theta_i$'s are independent and follow a normal distribution with unknown mean
$\theta_i$ and known variance $\sigma_i^2$ (the squared standard error).  Additional adjustments for
heterogeneity will be discussed in Section \ref{sec:hetero}.

Let
\begin{equation}\label{eq:Zi_mu}
Z_i = \frac{\hat \theta_i - \mu}{\sigma_i}
\end{equation}
denote the $z$-statistic for the null hypothesis $H_{0i}$:
$\theta_i=\mu$, $i=1, \ldots, k$. We can then derive the corresponding one-sided $p$-values
based on the cumulative standard normal distribution function $\Phi(\cdot)$:
\begin{equation}\label{eq:pNormal}
 \pright{p_i} = 1-\Phi(Z_i) \mbox{ and } \pleft{p_i} = \Phi(Z_i) 
 \end{equation}
for the alternatives $H_{1i}$: $\theta_i > \mu$ ({"greater"})
and $H_{1i}$: $\theta_i < \mu$ ({"less"}), respectively. Note
that $\pright{p_i}$ is monotonically increasing and $\pleft{p_i}$ is
monotonically decreasing in $\mu$.

\subsection{\textit{P}-value combination methods}

In what follows we denote with $\pright{p_\bullet}$ a combined
$p$-value based on study-specific one-sided $p$-values $\pright{p_1},
\ldots, \pright{p_k}$ for the alternative {"greater"} and
likewise with $\pleft{p_{\bullet}}$ for the alternative
{"less"}.  The subscript ``$\bullet$'' is a placeholder for a
$p$-value combination method, abbreviated by the first letter of the last
name of the inventor, where we consider the methods listed in Table \ref{tab:pvalcomb}:
\begin{itemize}
%  \item Hartung-Knapp-Sidik-Jonkman \citep{HartungKnapp2001,Sidik2002}
%  \item Random effects meta-analysis \citep{Borenstein2010}
  \item {\textbf E}dgington's method \citep{Edgington1972},
  \item {\textbf F}isher's method \citep{fisher:34},
  \item {\textbf P}earson's method \citep{Pearson1933},
  \item {\textbf W}ilkinson's method \citep{Wilkinson1951}, and
  \item {\textbf T}ippett's method \citep{Tippett1931}.
\end{itemize}
The combined $p$-values $\pright{p_\bullet}$ and
$\pleft{p_\bullet}$ will inherit the monotonicity property from the
$\pright{p_i}$'s and $\pleft{p_i}$'s, respectively: $\pright{p_\bullet}$ is
monotonically increasing and $\pleft{p_\bullet}$ is monotonically decreasing
in $\mu$ for any of the combination methods listed in Table \ref{tab:pvalcomb}.
%% As in \eqref{eq:pNormal}, 
%% if study-specific $p$-values for the alternative ``less'' are used in Table \ref{tab:pvalcomb}, then the combined $p$-value $\pleft{p_\bullet}$ is obtained
%% from  Table \ref{tab:pvalcomb} via
%% $\pleft{p_\bullet}=1-\pright{p_\bullet}$. 

\begingroup
\renewcommand{\arraystretch}{1.4}
\begin{table}[!htb]
  \centering
  \caption{Some methods for combining one-sided $p$-values ${p_{1}}, \dots, {p_{k}}$ from
    $k$ studies into a combined $p$-value ${p_\bullet}$ (in alphabetic order). The floor function ${\lfloor {s} \rfloor}$ denotes
    the greatest integer less than or equal to $s$.     A
    chi-squared random variable with $k$ degrees of freedom is denoted as $\chi^{2}_{k}$.}
  \label{tab:pvalcomb}
  \begin{tabular}{l l}
    \toprule
    \textbf{Method} & \textbf{Combined  \textit{p}-value}  \\
    \midrule
    Edgington % (sum)
           & ${p_E} = \frac{1}{k!} \sum_{j=0}^{\lfloor {s} \rfloor} (-1)^{j} \binom{k}{j} ({s} - j)^{k}$
             with ${s} = \sum_{i=1}^{k}{p_i}$ \\
    Fisher % (product)
           & ${p_F} = \Pr(\chi^{2}_{2k} > {f})$ with ${f} = -2 \sum_{i=1}^{k}\log({p_i})$ \\
    Pearson % (product)
           & ${p_P} = \Pr(\chi^{2}_{2k} \leq {g})$ with ${g} = -2 \sum_{i=1}^{k}\log(1 - {p_i})$ \\
    Tippett % (minimum)
           & ${p_T} = 1 - (1 - \min\{{p}_{1}, \dots, {p}_{k}\})^{k}$ \\
    Wilkinson
%    $n$-trials rule % (maximum)
           & ${p_W} = \max\{{p}_{1}, \dots, {p}_{k}\}^{k}$  \\
    %% Stouffer % (meta-analysis)
    %%        & % $p = 2\{1 - \Phi(|z_{r}|)\}$ with
    %%          $p = 1 - \Phi(z)$ or $p = 2\{1 - \Phi(|z|)\}$ with
    %%          $z = \{\sum_{i=1}^{n}w_i \Phi^{-1}(1 - \pright{p_i}) \}/\sqrt{\sum_{i=1}^{n}w_{i}^2}$\\
    \bottomrule
  \end{tabular}
\end{table}
\endgroup The $p$-value from Edgington's method is based on a
transformation of the sum of the $p$-values ${s}$ with the cumulative
distribution function of the Irwin-Hall distribution
\citep{Irwin1927,Hall1927}. For large $k$ it can be approximated based
on a central limit theorem argument \citep{Edgington1972b}:
\begin{equation}\label{eq:IHNormalApprox}
  {p_E} \approx \Phi(\sqrt{12 \, k}({s}/k-1/2)).
\end{equation}
For $k=12$, this approximation is
considered already ``fairly good'' \citep[Section 3.1]{Ripley1987}, in fact the ``sum
of 12 uniforms'' method was once a popular way to generate samples from
a normal distribution. In order to mitigate
overflow problems of the Irwin-Hall distribution for large $k$, we therefore use
the normal approximation \eqref{eq:IHNormalApprox} if $k \geq 12$. 

Tippett's method is based on the smallest
$p$-value. %% while Wilkinson's method is based on the largest
There is a generalization of Tippett's method based on the $r$-th smallest
$p$-value \citep{Wilkinson1951, HedgesOlkin1985}. For $r=k$ the largest
$p$-value is hence used and we obtain the method denoted here as Wilkinson's
method. Another commonly used $p$-value combination method is Stouffer's method
based on the sum of inverse normal transformed $p$-values \citep{Stouffer1949}.
A weighted version exists \citep{cousins2007annotated}, which is equivalent to
fixed effect and random effects meta-analysis, if the weights are suitably chosen
\citep{Senn2021}. Fixed effect and random effects meta-analysis will be included in our
example (Section \ref{sec:example}) and in the simulation study described in Section \ref{sec:simulation}.

Suppose we combine one-sided $p$-values $\pright{p_1}, \ldots, \pright{p_k}$ for the alternative "greater" into
a combined $p$-value function $\pright{p_\bullet}(\mu)$. 
 The standard point estimate is the median estimate $\hat \mu$ \citep{Fraser2017}, defined as the root of the equation
 \begin{equation}\label{eq:MCE}
\pright{p_\bullet}(\hat \mu) = 0.5.
 \end{equation}
The Irwin-Hall distribution has median $k/2$, therefore the median
estimate $\hat \mu_E$ of Edgington's method is the value of $\mu$ where
the mean $s/k$ of the study-specific one-sided $p$-values is 0.5.
There are even closed-form solutions for the median estimate based on
Tippett's and Wilkinson's method, see Appendix~\ref{app:closedform}.

In order to obtain a two-sided $1-\alpha$ confidence interval $[\mu_l, \mu_u]$ for $\mu$
based on a $p$-value combination method $\pright{p_\bullet}$, we have to find
the roots $\mu_l$ and $\mu_u$ of the two equations
\begin{align}\label{eq:CI}
& &\pright{p_\bullet}(\mu_l) = \alpha/2& &\text{and}& &\pright{p_\bullet}(\mu_u) = 1 - \alpha/2.& &
\end{align}
While in the case of Tippett's and Wilkinson's methods there are closed-form
solutions for the roots in \eqref{eq:MCE} and \eqref{eq:CI} as shown in
Appendix~\ref{app:closedform}, in general they have to be computed using
numerical root-finding algorithms. Equivalently we can find the two roots
$\mu_l$ and $\mu_u$ of the single equation
\begin{align}\label{eq:CI2}
 2 \, \min\left\{\pright{p_\bullet}(\mu), 1- \pright{p_\bullet}(\mu)  \right\} = \alpha
\end{align}
to obtain the $1-\alpha$ confidence interval, while maximization of
the function on the left side of~\eqref{eq:CI2} gives the median
estimate $\hat \mu$. Note that the confidence intervals $[\mu_l,
  \mu_u]$ is not necessarily symmetric around the median estimate. 
The left-hand
side of \eqref{eq:CI2} has been coined the \textit{centrality
  function} by \citet{Xie2011} and is also known as the
\textit{confidence curve}. \citet{Berrar2017} has proposed to use the
area under the confidence curve (AUCC) as a summary measure of
precision. We note that \citet{Schweder.Hjort2016} define the confidence curve as
$1$ minus the left-hand side of \eqref{eq:CI2}, in which case AUCC is
no longer a useful summary measure.

We may also use $\pleft{p_\bullet}(\mu)$ (based on the one-sided $p$-values
$\pleft{p_{1}}, \dots, \pleft{p_{k}}$ for the alternative ``less'') rather than $\pright{p_\bullet}(\mu)$ to
compute a point estimate with two-sided confidence interval. Ideally this should
lead to the same results, but this is only the case for Edgington's
method. 
The other combination methods will generally lead to different results depending
on whether the input $p$-values are oriented "greater" or "less" due to the
following relationships between combined $p$-value and input $p$-value
orientation:
%% We now have the following relationships:
\begin{eqnarray}
  \pright{p_E} & = & 1 - \pleft{p_E}, \label{eq:eq1} \\
  \pright{p_F} & = & 1 - \pleft{p_P}, \label{eq:eq2} \\
  \pright{p_P} & = & 1 - \pleft{p_F}, \label{eq:eq2b} \\
  \pright{p_T} & = & 1 - \pleft{p_W}, \label{eq:eq3} \\
  \pright{p_W} & = & 1 - \pleft{p_T}, \label{eq:eq3b} 
\end{eqnarray}
see Appendix \ref{app:relationships} for a proof. For example, equation
\eqref{eq:eq2} implies that the
combined $p$-value function based on Fisher's method and one-sided
$p$-values for the alternative "greater" is 1 minus the combined
$p$-value function based on Pearson's method and one-sided $p$-values
for the alternative "less". However, in practice
the ultimate goal is to compute a point estimate with two-sided
confidence interval, so the direction of the alternative of the
underlying one-sided $p$-values shouldn't matter. But this is only the
case for Edgington's method due to property \eqref{eq:eq1}.

In principle, we may also use two-sided $p$-values in any of the combination
methods listed in Table \ref{tab:pvalcomb}, to circumvent the lack of
orientation-invariance of Tippett's, Wilkinson's, Fisher's and Pearson's method,
but this comes with new problems. Specifically, the combined $p$-value function
of Fisher's, Pearson's and Edgington's method based on two-sided $p$-values may
then no longer peak at 1, which means that confidence intervals on certain
confidence levels may be empty sets. In contrast, the combined $p$-value
functions of Tippett's and Wilkinson's method will peak at 1, but will have
several modes at the study-specific point estimates. This may lead to confidence
sets consisting of non-overlapping intervals, which are hard to interpret and
not useful in applications. 

%% %In summary, we therefore recommend to use
%% Edgington's method based on
%% one-sided $p$-values is hence the only $p$-value combination method (different from
%% standard meta-analysis) that is orientation-invariant with a unique
%% point estimate and always existing confidence interval. \todo{True? Check with Samuel} A confidence
%% interval based on Edgington's method is not necessarily symmetric
%% around the point estimate, so less restrictive than standard
%% meta-analytic confidence intervals of the form \eqref{eq:additiveCI}.

\subsection{Example: Association between corticosteroids and mortality in COVID-19 hospitalized patients}\label{sec:example}
%% \todo[inline]{add perhaps another example from \citet{JacksonWhite2018}}
\begin{table}[!htb]
  \centering
  \caption{Data from $k=7$ randomized controlled clinical
trials investigating the association between corticosteroids and mortality in hospitalized
patients with COVID-19 \citep{REACT2020}.} \label{tab:covidexample-data}
% latex table generated in R 4.4.2 by xtable 1.8-4 package
% Fri Feb 21 13:22:48 2025
\begin{tabular}{clccrrr}
  \toprule & & \multicolumn{2}{c}{Deaths / Patients}\\ \cmidrule(lr){3-4} Study & Name & Steroids & No Steroids & OR & Lower CI & Upper CI \\ 
  \midrule
  1 & DEXA-COVID 19 & 2 / 7 & 2 / 12 & 2.00 & 0.21 & 18.69 \\ 
    2 & CoDEX & 69 / 128 & 76 / 128 & 0.80 & 0.49 & 1.31 \\ 
    3 & RECOVERY & 95 / 324 & 283 / 683 & 0.59 & 0.44 & 0.78 \\ 
    4 & CAPE COVID & 11 / 75 & 20 / 73 & 0.46 & 0.20 & 1.04 \\ 
    5 & COVID STEROID & 6 / 15 & 2 / 14 & 4.00 & 0.65 & 24.66 \\ 
    6 & REMAP-CAP & 26 / 105 & 29 / 92 & 0.71 & 0.38 & 1.33 \\ 
    7 & Steroids-SARI & 13 / 24 & 13 / 23 & 0.91 & 0.29 & 2.87 \\ 
   \bottomrule
\end{tabular}

\end{table}

We will illustrate the different methods using a meta-analysis
combining information from \mbox{$n = 7$} randomized controlled
clinical trials investigating the association between corticosteroids
and mortality in hospitalized patients with COVID-19
\citep{REACT2020}, see Table \ref{tab:covidexample-data}. We will use one-sided $p$-values for the
alternative "less" as negative log odds ratios indicate treatment
benefit, the results for the alternative "greater" follow from
\eqref{eq:eq1}-\eqref{eq:eq3b}.

\begin{figure}[h!]

\begin{knitrout}
\definecolor{shadecolor}{rgb}{0.969, 0.969, 0.969}\color{fgcolor}
\includegraphics[width=\maxwidth]{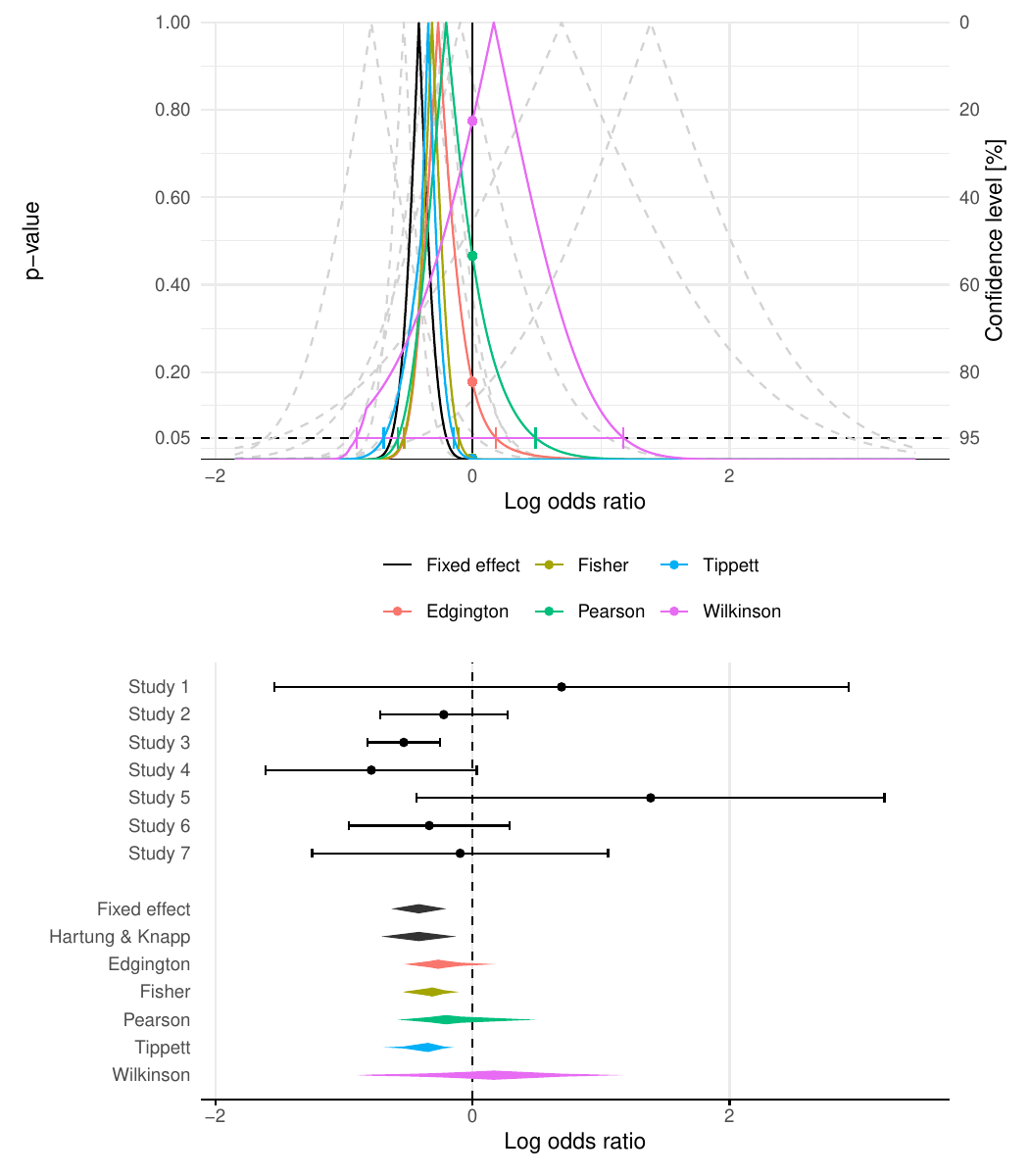} 
\end{knitrout}
\caption{\label{fig:fig1} Drapery plot (top) and forest plot (bottom) from several combination methods with 95\% {confidence} intervals for a meta-analysis of
  $k=7$ randomized controlled clinical trials
investigating the association between corticosteroids and mortality in
hospitalized patients with COVID-19 \citep{REACT2020}.}
\end{figure}

Figure \ref{fig:fig1} shows that the distribution of the study effect estimates is right-skewed. Such skewness can be quantified using Fisher's weighted skewness coefficient \citep[Section 3.3.2]{Ferschl1980} of the meta-analyzed effect estimates, defined as
\begin{align}
  &\gamma = \frac{\left\{\sum_{i=1}^{k} w_{i} (\hat{\theta}_{i} - \bar{\theta})^{3}\right\}\sqrt{\sum_{i=1}^{k} w_{i}}}{\left\{\sum_{i=1}^{k} w_{i}(\hat{\theta}_{i} - \bar{\theta})^{2}\right\}^{3/2}}~\text{ with }~\bar{\theta} = \frac{\sum_{i=1}^{k}\hat{\theta}_{i}w_{i}}{\sum_{i=1}^{k}w_{i}}~\text{ and }~w_{i} = \frac{1}{\sigma^{2}_{i}}. \label{eq:gamma}
\end{align}
In this example we obtain $\gamma=3.72$,
the positive sign reflecting a right-skewed distribution.

%% , see also \citet{Held2021b}.
The results from an inverse variance-weighted fixed effect analysis
are reproduced in Figure \ref{fig:fig1} on the log odds ratio
scale. This was also the prespecified primary analysis in the protocol
registered and made publicly available on the PROSPERO database prior
to data analysis or receipt of outcome data. 
%the data do not indicate conflict with the fixed effect
%assumptions based on Cochran's $Q$-test
%($p=formatPval(meta$pval.Q)$, Higgins'
%$I^2=round(100*I2_hypo)\%$).
Note that the knot point of the
$p$-value function for Wilkinson's method shown in the drapery plot in Figure \ref{fig:fig1} is not an artefact, but caused by its
definition based on the maximum of the different $p$-values, compare
Table \ref{tab:pvalcomb}.

%\todo[inline]{Leo: should we use Paule-Mandel for the Hartung-Knapp method? This is what was done in the NEJM paper}
%\todo[inline,color=blue!10]{Samuel: I don't know how to change the random effects and Hartung Knapp method in confMeta to a different heterogeneity estimator, I think they are hard-coded to REML}

\begin{table}
\centering
\caption{Comparison of different $p$-value combination methods (with alternative ``less'') investigating the association between corticosteroids and mortality in hospitalized patients with COVID-19. Shown are estimates of the log odds ratio and their 95\% CIs and compared with the fixed effect, DerSimonian-Laird (DL) and Hartung-Knapp (HK) random effects approach for meta-analysis of $k=7$ randomized controlled clinical trials. The random effects approach is based on the REML estimate of $\tau^2$, which is so close to zero that its results are indistinguishable from the fixed effect method. 
  The $p$-value shown is two-sided for the standard null hypothesis $\mu=0$ of no effect and calculated based on the left-hand side of \eqref{eq:CI2} (with $\pleft{p_\bullet}$ rather than $\pright{p_\bullet}$) for the different $p$-value combination methods.}
\label{tab:tab2}

\resizebox{\columnwidth}{!}{%
% latex table generated in R 4.4.2 by xtable 1.8-4 package
% Fri Feb 21 13:22:49 2025
\begin{tabular}{lrrrlrrrr}
  \toprule
 & Estimate & Lower CI & Upper CI & $p$-value & CI width & AUCC & CI skewness & AUCC ratio \\ 
  \midrule
Edgington & -0.27 & -0.53 & 0.18 & 0.18 & 0.71 & 0.28 & 0.26 & 0.17 \\ 
  Fisher & -0.31 & -0.54 & -0.10 & 0.003 & 0.43 & 0.17 & -0.03 & -0.02 \\ 
  Pearson & -0.20 & -0.58 & 0.49 & 0.47 & 1.07 & 0.42 & 0.30 & 0.19 \\ 
  Tippett & -0.34 & -0.69 & -0.15 & 0.002 & 0.55 & 0.19 & -0.27 & -0.18 \\ 
  Wilkinson & 0.17 & -0.90 & 1.17 & 0.77 & 2.08 & 0.89 & -0.03 & -0.05 \\ 
   \midrule
Fixed effect & -0.42 & -0.63 & -0.20 & 0.0001 & 0.43 &  & 0.00 &  \\ 
  DL random effects & -0.42 & -0.63 & -0.20 & 0.0001 & 0.43 &  & 0.00 &  \\ 
  HK random effects & -0.42 & -0.71 & -0.13 & 0.013 & 0.58 &  & 0.00 &  \\ 
   \bottomrule
\end{tabular}

}
\end{table}

\clearpage

Results based on the different methods are shown in Table \ref{tab:tab2}.
The point estimates from the different $p$-value combination methods are all closer to zero than the combined effect estimate from the fixed effect, DerSimonian-Laird (DL) and Hartung-Knapp (HK) random effects methods, respectively.
The 95\% confidence intervals also differ substantially.
Wilkinson's method even gives a positive point estimate and has the widest confidence interval.
Wilkinson's method also gives the largest values of
the two-sided $p$-value ${p}_{\bullet}(0)$ for the null hypothesis of no effect, while Tippett's method has the smallest estimate and the smallest $p$-value among all five $p$-value combination methods. 
%\todo[inline]{discuss why this may be the case}
%\todo[inline,color=blue!10]{perhaps give example of the two most extreme $p$-values, i.e., Wilkinson and Tippett?}

To assess
the skewness of the different confidence intervals, we computed the skewness coefficient
\begin{equation}\label{eq:beta}
\beta = \frac{\mbox{upper} + \mbox{lower} - 2 \, \mbox{estimate}}{\mbox{upper} - \mbox{lower}}.
\end{equation}
by \citet{GroeneveldMeeden1984} based on the (median) estimate and the $\mbox{upper}$ and $\mbox{lower}$ interval limits.
Note that $\abs{\beta} \leq 1$ with positive sign for a right-skewed interval and negative sign for a left-skewed one. The coefficient is zero for symmetric confidence intervals, as here for the fixed effect, DL and HK random effects methods.
The penultimate column in Table \ref{tab:tab2} reveals that
three of the different $p$-value combination methods (Fisher, Tippett, Wilkinson) return a left-skewed confidence interval with negative skewness coefficient $\beta$, although the
study effect estimates are right-skewed. Only Edgington's and Pearson's methods preserve the skewness of the data and return
a positive coefficient $\beta$. 

We also calculated the area under the confidence curve (AUCC)
\citep{Berrar2017} for the different $p$-value combination methods,
see the third last column in Table \ref{tab:tab2}.  As the confidence interval width, AUCC is a
measure of precision but has the advantage that it does not depend on
the level of the confidence interval.  In this application AUCC
correlates strongly with the width of the 95\% CI with a correlation
of 0.999.  As
a measure of skewness of the confidence curve we propose to compute
the AUCC below and above the point estimate $\hat \mu$, so that
$\mbox{AUCC}_{\tiny \mbox{below}} + \mbox{AUCC}_{\tiny \mbox{above}} =
\mbox{AUCC}$.  The proposed measure of skewness is the
\[
\mbox{{AUCC ratio}} = \frac{\mbox{AUCC}_{\tiny \mbox{upper}} - \mbox{AUCC}_{\tiny \mbox{lower}}}{\mbox{AUCC}},
\]
which is restricted to the interval $[-1, 1]$, just as the skewness coefficient \eqref{eq:beta}. 
Table \ref{tab:tab2} shows that in this application the AUCC
ratio has the same sign as the skewness coefficient \eqref{eq:beta} for all five $p$-value combination methods,
with correlation 0.997.
%% \todo[inline]{should we also mention ``the proportion of the AUCC that lies above
%% the null value''?}

Two of the studies in this example (study 1 ``DEXA-COVID 19'' and 5 ``COVID STEROID'')
have large confidence intervals due to a small number of events, where
the normality assumption in \eqref{eq:pNormal} may be questionable. To
avoid assuming normality, we can also define $p$-value functions based
on exact one-sided $p$-values from Fisher's exact test, where we
employ the mid-$p$ correction, originally proposed by
\citet{Lancaster1961}.  This ensures that the $p$-values for "greater"
and "less" still sum up to 1, although the distribution of the test
statistic is discrete. Edgington's method is hence still 
orientation-invariant, whereas the other methods are not.

\begin{figure}
\begin{knitrout}
\definecolor{shadecolor}{rgb}{0.969, 0.969, 0.969}\color{fgcolor}
\includegraphics[width=\maxwidth]{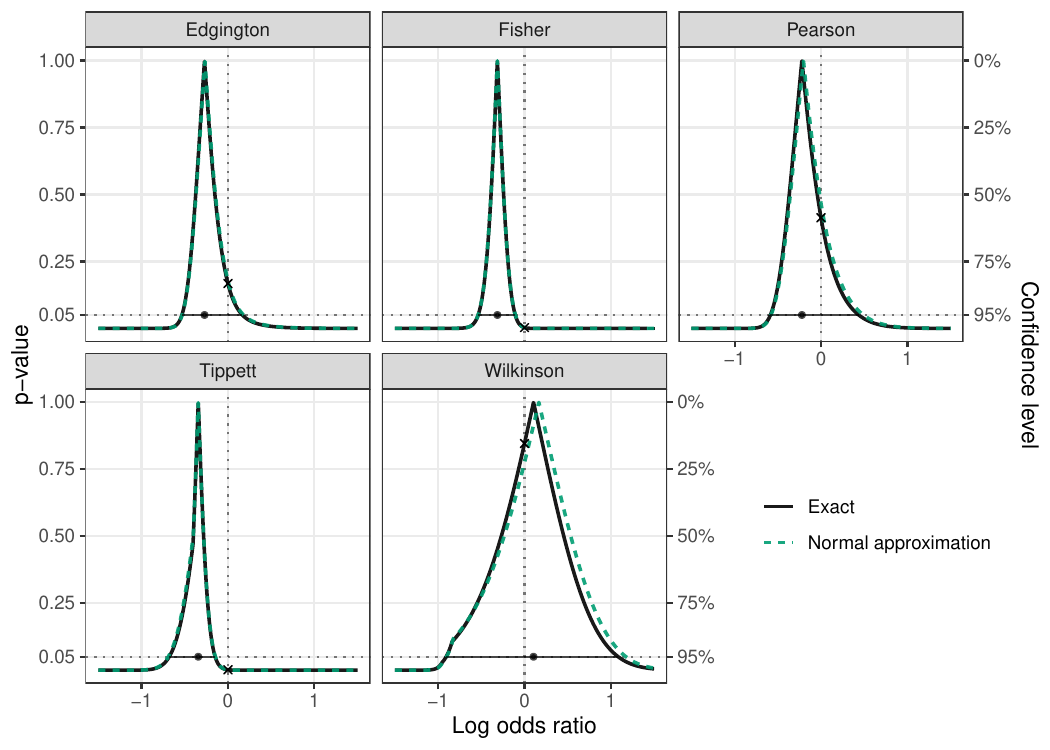} 
\end{knitrout}
\caption{\label{fig:exactP} Combined $p$-value functions with 95\%
  confidence intervals based on exact $p$-values with mid-$p$ correction
for meta-analysis of
  $k=7$ randomized controlled clinical trials
investigating the association between corticosteroids and mortality in
hospitalized patients with COVID-19 \citep{REACT2020}.
Shown are also the
  normal approximation $p$-value functions based on the $z$-statistics
  \eqref{eq:Zi_mu} for comparison.}
\end{figure}

Figure \ref{fig:exactP} shows the corresponding $p$-value functions based on the
exact $p$-values (solid black lines) along with the normal approximation
$p$-value functions based on the $z$-statistics \eqref{eq:Zi_mu} for comparison
(dashed green lines). We see that for most combination methods both curves are
virtually identical, producing almost identical point estimates, $p$-values, and
confidence intervals. Slightly larger differences can only be seen for Wilkinson's method. 
% The point estimate is round(thetahat, 2), and the $p$-value for the
% null hypothesis of no effect is formatPval(p0), very similar to the
% results based on the normal approximation. Also the 95\% confidence interval
% is very similar (round(thetahatL,2) to round(thetahatU,2)).
This suggests that the $p$-value function based on the $z$-statistics provides a
good approximation to the one based exact $p$-values, despite small counts for
some of the studies.

\subsection{Accounting for heterogeneity}\label{sec:hetero}

We consider the case of additive heterogeneity, where heterogeneity is
quantified based on Cochran's $Q$-statistic \citep{Cochran1937}
\begin{equation}\label{eq:Q}
  Q = \sum\limits_{i=1}^k w_i (\hat \theta_i - \hat \theta_w)^2
\end{equation}
with weights $w_i=1/\sigma_i^2$ equal to the inverse squared
standard errors $\sigma_i$ and the standard meta-analytical point estimate $\hat \theta_w$,
the weighted average of the study-specific estimates $\hat \theta_i$
with weights $w_i$.
The $Q$-statistic 
\eqref{eq:Q} depends on $\hat \theta_w$, but can also be written as a weighted sum
of squared paired differences,
\[
  Q = \frac{\sum\limits_{i < j} w_i w_j (\hat \theta_i - \hat \theta_j)^2}{\sum\limits_{i=1}^k w_i},
\]
where $\hat \theta_w$ no longer appears. For example, if there are
only $k=2$ studies we obtain
\[
  Q = \frac{\sigma_1^{-2} \sigma_2^{-2} (\hat \theta_1 - \hat \theta_2)^2}{\sigma_1^{-2} + \sigma_2^{-2}}
= \frac{(\hat \theta_1 - \hat \theta_2)^2}{\sigma_1^{2} + \sigma_2^{2}},
\]
the standard test statistic to assess the evidence for conflict
between two study-specific effect estimates $\hat \theta_1$ and
$\hat \theta_2$. 
A popular measure of
heterogeneity is Higgins'
$$I^2 = \max \left\{Q - (k-1), 0 \right\}/Q,$$ the proportion of the
variance of the study-specific effect estimates that is attributable
to study heterogeneity. Higgins' $I^2$ is used in our simulation study
to specify the amount of heterogeneity.

%% The moment-based estimate of the heterogeneity variance 
%% by \citet{DerSimonianLaird1986} then is 
%% \begin{equation}\label{eq:tau2hat}
%% \hat \tau^2 = \max \left\{Q - (k-1), 0 \right\} / \left\{ \sum w_i - \sum w_i^2 / \sum w_i \right\}
%% \end{equation}
%% and t
The $z$-statistic \eqref{eq:Zi_mu} can be modified to account
for heterogeneity between study effects, represented by the heterogeneity variance $\tau^2$.
The heterogeneity-adjusted $z$-statistic is 
\begin{equation}\label{eq:z.adjusted}
  Z_i = \frac{\hat \theta_i - \mu}{\sqrt{\sigma_i^2 + \hat \tau^2}},
  \end{equation}
  where $\hat \tau^2$ is a suitable estimate of $\tau^2$.  Many
  estimates of the heterogeneity variance $\tau^2$
  exist, we will use the REML estimate in the following due to its
  good performance in simulation studies \citep{Langan2019}.  Note
  that the transformation \eqref{eq:pNormal} is used to compute
  $p$-values based on \eqref{eq:z.adjusted}, which assumes normality of the random effects
  distribution. We will comment on possible relaxations of this
  assumption in the discussion.

%% One may argue that the $Q$-statistic 
%% \eqref{eq:Q} depends on $\hat \theta_w$, so is implicitly based on a common mean model.
%% However, $Q$ can also be written as a weighted sum
%% of squared paired differences,
%% \[
%%   Q = \frac{\sum\limits_{i < j} w_i w_j (\hat \theta_i - \hat \theta_j)^2}{\sum\limits_{i=1}^k w_i},
%% \]
%% where $\hat \theta_w$ no longer appears. For example, if there are
%% only $k=2$ studies we obtain
%% \[
%%   Q = \frac{\sigma_1^{-2} \sigma_2^{-2} (\hat \theta_1 - \hat \theta_2)^2}{\sigma_1^{-2} + \sigma_2^{-2}}
%% = \frac{(\hat \theta_1 - \hat \theta_2)^2}{\sigma_1^{2} + \sigma_2^{2}},
%% \]
%% the standard test statistic to assess the evidence for conflict
%% between two study-specific effect estimates $\hat \theta_1$ and
%% $\hat \theta_2$.

%% \subsection{Visualization}
%% We prefer violin plots over the more standard diamond plot to visualize the  uncertainty regarding the effect estimate \citep{vanderBles_etal2019}.
%% \todo[inline,color=blue!10]{SP: are violion plots used anywhere? I can't see any}

%% \clearpage

\section{Simulation study}\label{sec:simulation}
\subsection{Design}
We first describe the design of our simulation study, following the
structured ``ADEMP'' approach for reporting of simulation studies
\citep{Morris2019}.

\subsubsection{Aims}
The aim of the simulation study was to evaluate the estimation
properties of $p$-value combination methods for meta-analysis and to
compare them with classical meta-analysis methods under different
numbers of studies with potentially different sample sizes and degrees of
heterogeneity.

\subsubsection{Data-generating mechanism}
Our data-generating mechanism follows closely the simulation study of
\citet{IntHout2014}. %% , the main difference being that in our simulation study,
%% true study effects were additionally simulated from skewed distributions.
Specifically, in each simulation repetition, we simulated $k \in \{3,
5, 10, 20, 50\}$ true study effects (on standardized mean difference
scale) and corresponding effect estimates with standard errors. The
mean true study effect was set to $\theta = 0.2$. The true study
effect of study $i$ was then simulated from a normal
distribution $\theta_{i} \sim \Nor(\theta, \tau^2)$ with mean $\theta$
and heterogeneity variance $\tau^2$.

Based on a true effect $\theta_{i}$, the effect estimate $\hat{\theta}_{i}$ of
study $i$ was simulated from a normal distribution
\begin{align*}
  \hat{\theta}_{i} \sim \Nor(\theta_{i}, 2/n_{i}),
\end{align*}
where $n_{i}$ is the sample size per group and set to $n_{i} = 50$
(small studies) or $n_{i} = 500$ (large studies). We considered
scenarios with either $0$, $1$, or $2$ large studies (and the rest as
small studies). As in \citet{IntHout2014}, the variance of the
outcome variable of the studies is assumed to be 1.
The squared standard error of $\hat{\theta}_{i}$ was simulated
from a scaled chi-squared distribution
\begin{align*}
  \sigma_i^{2} \sim \frac{1}{(n_i-1)n_i} \chi^2_{2(n_{i}-1)},
\end{align*}
which were then transformed to standard errors $\sigma_i$ by taking the square root. The
heterogeneity variance $\tau^{2}$ of the study effect distribution was specified
by Higgins' $I^{2}$. Specifically, we first computed the within-study variance
using
\begin{equation}
  \label{eq:eps2}
  \epsilon^2 = \frac{1}{k} \sum\limits_{i=1}^k \frac{2}{n_i},
\end{equation}
from which we then computed the between-study heterogeneity variance 
\begin{equation} \label{eq:eps2}
  \tau^2 = \epsilon^2 \frac{I^2}{1-I^2}
\end{equation}
where $I^{2} = \tau^{2}/(\epsilon^{2} + \tau^{2})$ is Higgins' relative
heterogeneity \citep{Higgins2002}. We considered scenarios with $I^{2} = 0$,
$0.3$, $0.6$ or $0.9$, representing a range from no heterogeneity up to high
relative heterogeneity. All manipulated factors are listed in
Table~\ref{tab:tab1} and were varied in a fully factorial manner, resulting in 5
(number of studies) $\times$ 3 (number of large studies) %% $\times$ 3 (effect distribution)
$\times$ 4 (relative heterogeneity) $= 60$ simulation scenarios.
Additional simulation results based on a skew normal study effect distribution are described in the supplementary material.

\begin{table}[!htb]
  \caption{Factors considered in simulation study (varied in fully-factorial
    way).}
  \label{tab:tab1}
  \begin{center}
    \begin{tabular}{ll}
      \toprule
      Factor & Levels \\
      \midrule
      Number of studies $k$ & 3, 5, 10, 20, 50 \\
      Number of large studies & 0, 1, 2 \\
%%      Distribution of study-specific effects & normal, left-skew normal, right-skew normal \\
      Higgins' $I^2$ & 0.0, 0.3, 0.6, 0.9 \\
      \bottomrule
    \end{tabular}
  \end{center}
\end{table}

%% \todo[inline]{Leo: not sure about the next section, perhaps we should only
%%   report operating charactreristics for the mean and describe the skew normal simulation as studies of model misspecification}
\subsubsection{Estimands}
In meta-analysis the estimand of interest is typically the mean true
study effect, which coincides with the median for a normal study
effects distribution.  The mean $\theta = 0.2$ is therefore used to
evaluate coverage and bias of the point estimates of the different
methods.  In the supplementary material we also use the median as estimand
if the study effect distribution was assumed to be skew-normal.

\subsubsection{Methods}
Each set of simulated effect estimates and standard errors were analyzed using
different methods, each producing a point estimate and a 95\% confidence
interval for the true effect, namely
\begin{enumerate}
  \item Standard fixed and DL random effects meta-analysis \citep{DerSimonianLaird1986,Borenstein2010}
  \item HK random effects meta-analysis \citep{HartungKnapp2001,Sidik2002}
  %% \item Edgington's method \citep{Edgington1972}
  %% \item Fisher's method \citep{fisher:34}
  %% \item Pearson's method \citep{Pearson1933}
  %% \item Wilkinson's method \citep{Wilkinson1951}
  %% \item Tippett's method \citep{Tippett1931}
\end{enumerate}
and the five $p$-value combination methods listed in Table \ref{tab:pvalcomb}.
All input $p$-values were one-sided and oriented in positive effect direction
(alternative "greater"). This setup exhausts all possible method/orientation
combinations as Edgington's method is orientation-invariant whereas
Fisher/Pearson and Wilkinson/Tippett are orientation mirrored as described in
Section~\ref{sec:methodology}.

Section \ref{sec:resultsWithout} gives results without adjustments for
heterogeneity (using $p$-values derived from \eqref{eq:Zi_mu}) and
compared with fixed effect (FE) meta-analysis.  In Section
\ref{sec:resultsWith} adjustments have been made for potential
between-study heterogeneity as described in Section \ref{sec:hetero}
and compared with random effects meta-analysis (DerSimonian-Laird (DL)
and Hartung-Knapp (HK)).  The restricted maximum likelihood (REML)
estimate of the heterogeneity variance $\tau^{2}$ was used, which is
usually recommended as a default choice \citep{Langan2019}. The fixed
effect and random effects meta-analysis methods were computed with the
\texttt{metagen} function from the R package \texttt{meta}
\citep{Balduzzi2019}, while the remaining $p$-value combination
methods were computed with the \texttt{confMeta} R package
\citep{Hofmann2024}.

\subsubsection{Performance measures}
Our primary performance measure was coverage of the 95\% confidence interval
which we estimated by
\begin{align*}
  \widehat{\mathrm{Cov}} = \frac{\text{\# 95\% CI includes estimand}}{n_{\mathrm{sim}}}
\end{align*}
with Monte Carlo standard error (MCSE)
\begin{align*}
  \mathrm{MCSE}_{\widehat{\mathrm{Cov}}} = \sqrt{\frac{\widehat{\mathrm{Cov}} (1 - \widehat{\mathrm{Cov}})}{n_{\mathrm{sim}}}}.
\end{align*}

We conducted $n_{\mathrm{sim}} = 20'000$ simulation repetitions. This
ensures a maximum MCSE of $0.35\%$
(attained when the estimated coverage is 50\%), which we consider as
sufficiently small to detect relevant differences. Our secondary
performance measures were bias %%(with respect to the mean and
%%\todo{Leo: do we need the median?} median effect)
and 95\% confidence interval width \citep[see Table~3 in][for
  definitions and MCSE formulas]{Siepe_etal2024}. To assess the
skewness properties of the different methods, we computed the skewness
coefficient \eqref{eq:beta} for each 95\% confidence interval. We then
evaluated the distribution (mean, median, minimum, maximum) of the
skewness coefficients for a given method and simulation scenario. To
assess the relationship between confidence interval skewness and data
skewness, we also computed the Pearson correlation between the 95\%
confidence interval skewness $\beta$ and Fisher's weighted skewness
coefficient \eqref{eq:gamma} of the meta-analyzed effect estimates
with weights $w_i = 1/\sigma^{2}_{i}$ and $w_{i} =
  ({\sigma^{2}_{i} + \hat{\tau}^{2}})^{-1}$ without and with
  heterogeneity adjustment, respectively.
  Finally, to assess agreement between
confidence interval skewness and data skewness, we also computed
Cohen's $\kappa$ of the sign of $\gamma$ and the sign of $\beta$,
using the function \texttt{cohen.kappa} from the \texttt{psych} R
package \citep{Revelle2024}.

\subsection{Computational aspects}
The simulation study was performed using R version 4.4.1 (2024-06-14) on a
server running Debian GNU/Linux. More information on the computational
environment and code to reproduce the simulation study are available at
\url{https://github.com/felix-hof/confMeta_simulation}.
%% \todo[inline]{Leo:  Update needed?}
%% \todo[inline,color=blue!10]{SP: Done}

\subsection{Results}
We will now describe the results of the simulation study, first without
adjustments for heterogeneity (Section \ref{sec:resultsWithout}) and then with
(Section \ref{sec:resultsWith}). Without heterogeneity adjustments there were a
few cases where Pearson and Wilkinson CIs did not converge (lowest convergence
rate $99.37\%$ for Pearson when $I^2 = 0.9$ and $k=50$ studies with 2 large
studies, see Table~\ref{tab:convergence} in the supplement for details), and
method performance was then estimated based on the convergent
repetitions only (case-wise deletion). With heterogeneity adjustments, no
non-convergent confidence intervals or point estimates occured.

\subsubsection{Without adjustments for heterogeneity}\label{sec:resultsWithout}

\afterpage{
\begin{landscape}
\begin{figure}[!htb]
\begin{knitrout}
\definecolor{shadecolor}{rgb}{0.969, 0.969, 0.969}\color{fgcolor}
\includegraphics[width=\maxwidth]{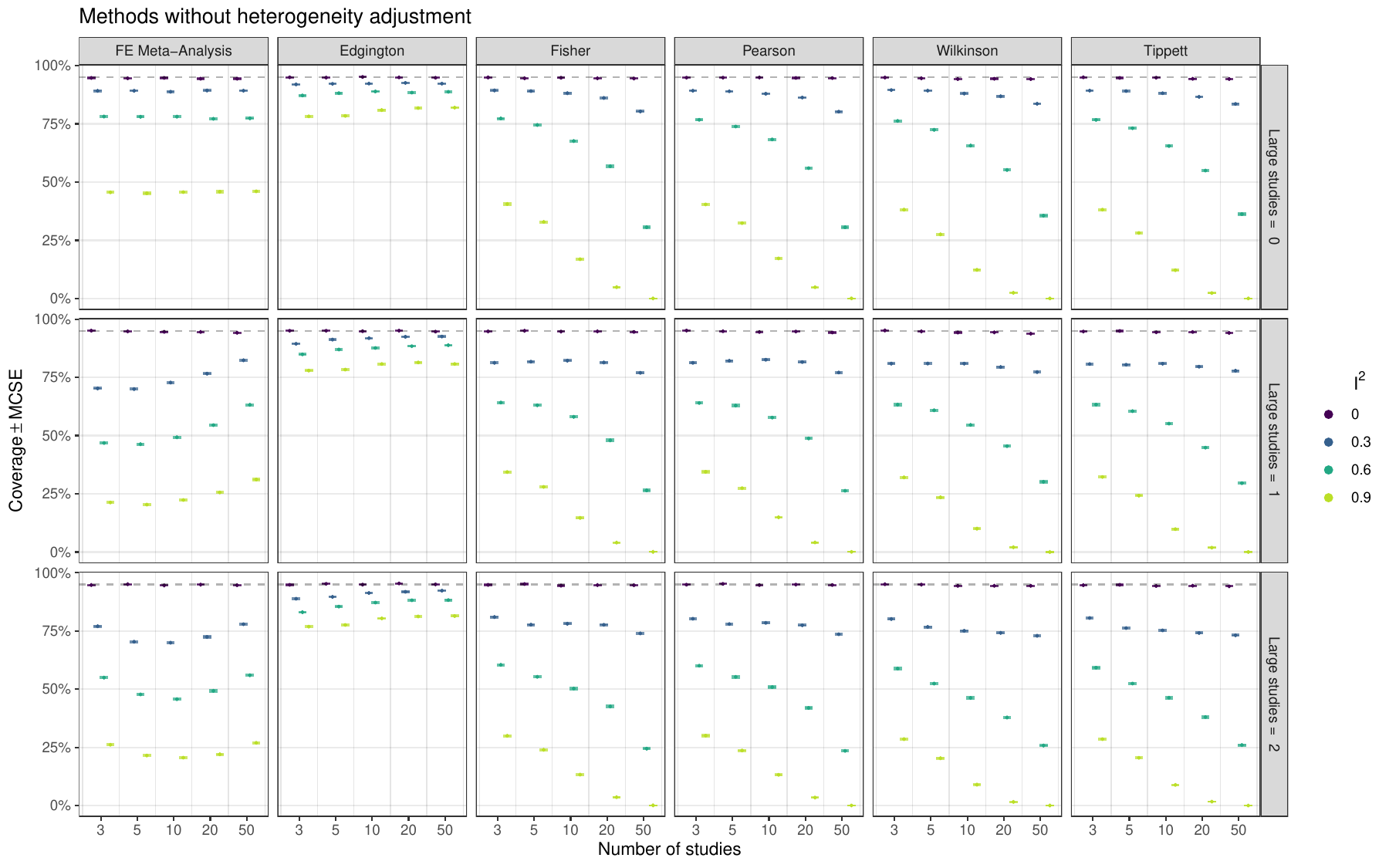} 
\end{knitrout}
\caption{Empirical coverage of the 95\% confidence intervals based on
  20'000 simulation repetitions.}
\label{fig:coverage-unadjusted}
\end{figure}
\begin{figure}[!htb]
\begin{knitrout}
\definecolor{shadecolor}{rgb}{0.969, 0.969, 0.969}\color{fgcolor}
\includegraphics[width=\maxwidth]{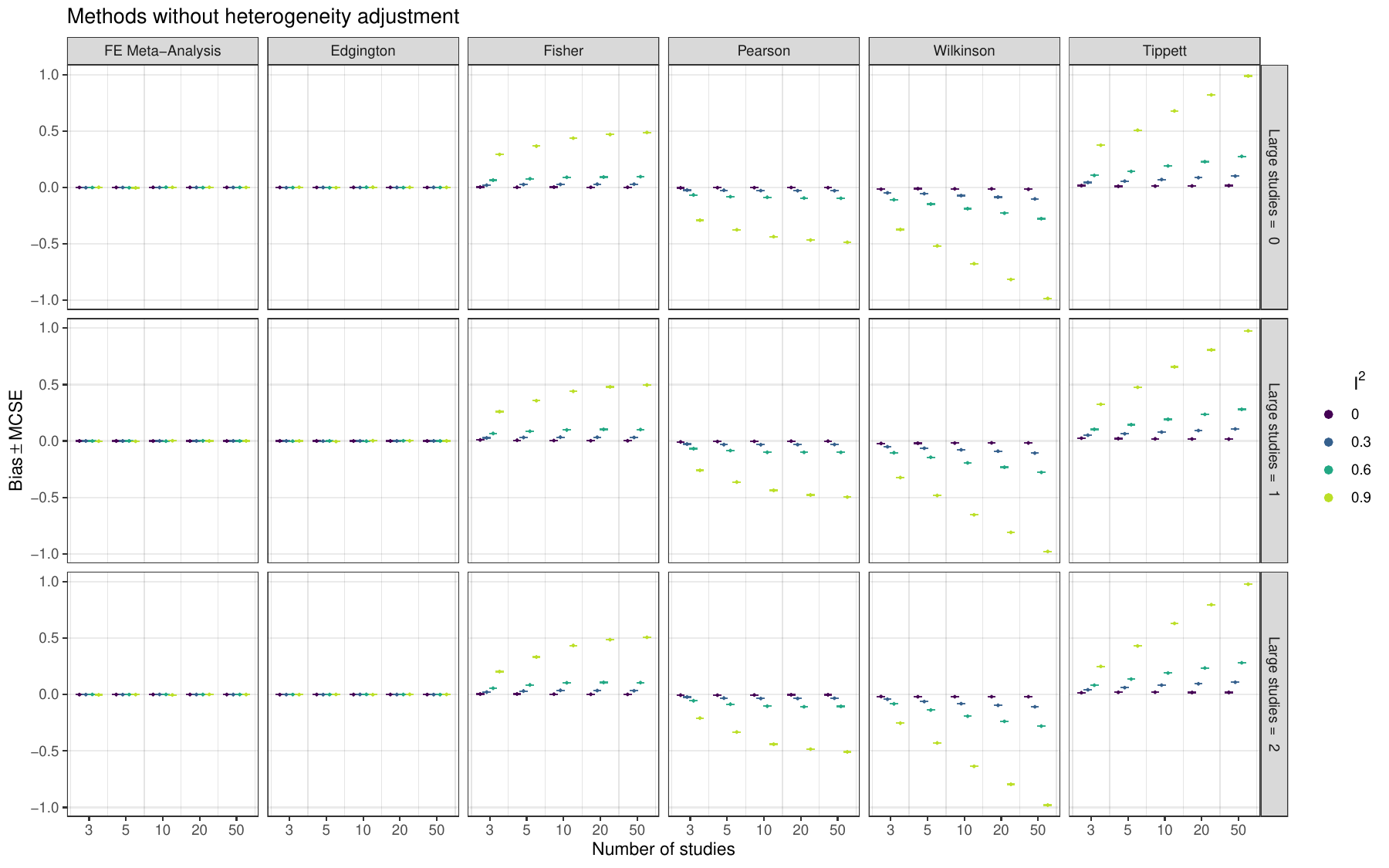} 
\end{knitrout}
\caption{Empirical bias of the point estimates for the true effect based on
  20'000 simulation repetitions.
}
\label{fig:bias-unadjusted}
\end{figure}
\begin{figure}[!htb]
\begin{knitrout}
\definecolor{shadecolor}{rgb}{0.969, 0.969, 0.969}\color{fgcolor}
\includegraphics[width=\maxwidth]{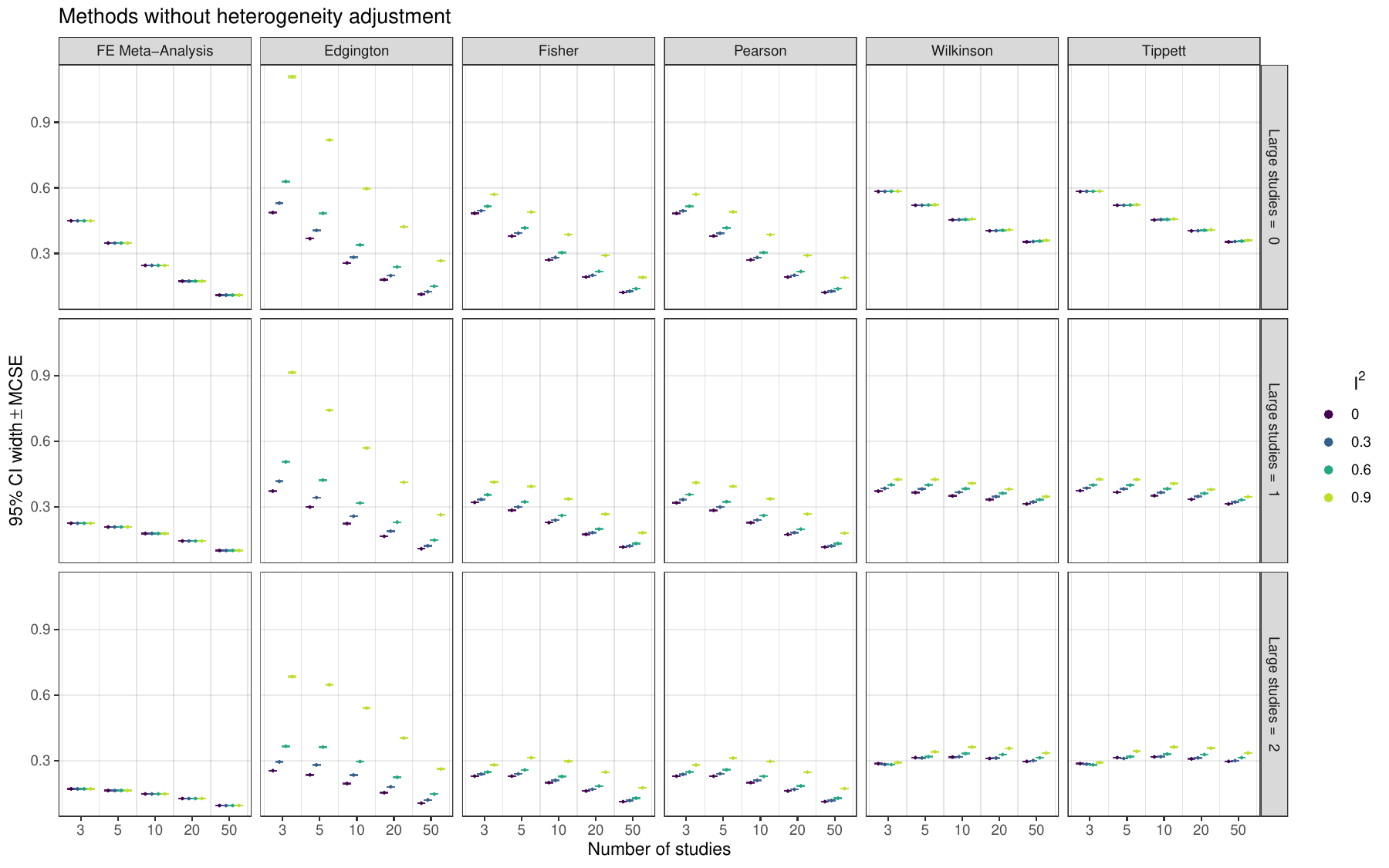} 
\end{knitrout}
\caption{Mean width of 95\% confidence intervals based on 20'000 simulation repetitions.}
\label{fig:width-unadjusted}
\end{figure}

\begin{figure}[!htb]
\begin{knitrout}
\definecolor{shadecolor}{rgb}{0.969, 0.969, 0.969}\color{fgcolor}
\includegraphics[width=\maxwidth]{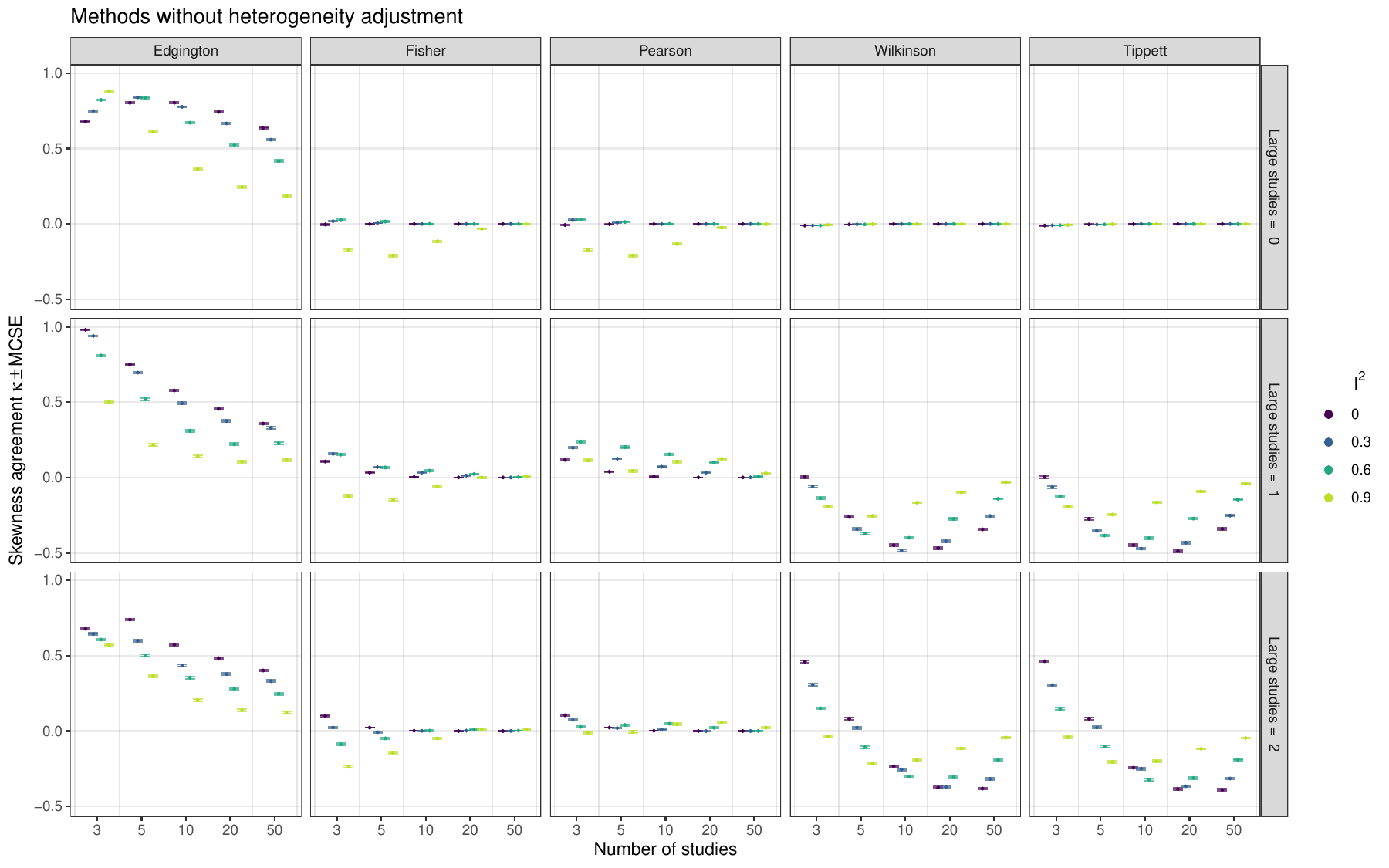} 
\end{knitrout}
\caption{Cohen's $\kappa$ sign agreement between 95\% confidence interval
  skewness and data skewness based on 20'000
  simulation repetitions.}
\label{fig:kappa-unadjusted}
\end{figure}

\end{landscape}
}

\paragraph{Coverage}
Figure \ref{fig:coverage-unadjusted} shows the empirical coverage of the
different methods without adjustments for heterogeneity. The coverage
of all methods is at the nominal 95\% level when data are generated
under effect homogeneity ($I^2=0$), as expected from theory. The
coverage drops below the nominal level for all methods under
simulation with heterogeneity ($I^2>0$). However, the drop is the
smallest for Edgington's method, which always remains above 75\%, even
for conditions with high heterogeneity ($I^2=0.9$). In contrast, the
coverage of all other methods (including fixed effect meta-analysis)
can drop to values below 25\%.

\paragraph{Bias}
Figure \ref{fig:bias-unadjusted} shows the performance of the compared methods in
terms of bias. We see that fixed effect meta-analysis and all $p$-value
combination methods are essentially unbiased under effect homogeneity ($I^2=0$).
However, when there is heterogeneity ($I^2>0$), only fixed effect meta-analysis
and Edgington's method remain unbiased, while the remaining $p$-value
combination methods show increasing bias with increasing relative heterogeneity
$I^2$. The bias patterns of Fisher/Pearson and Wilkinson/Tippett are mirrored
around zero, because these methods' are mirrored with respect to
$p$-value orientation.

\paragraph{Confidence interval width}
Figure \ref{fig:width-unadjusted} shows the average width of the
method's 95\% confidence intervals. We see that the width of fixed
effect meta-analysis, Wilkinson's, and Tippett's methods remains
constant for different levels of relative heterogeneity $I^2$. In contrast,
Edgington's, Fisher's, and Pearson's confidence interval widths
increase as $I^2$ increases, adapting to greater heterogeneity. This
adaptation is most pronounced for Edgington's method. A very similar
pattern can be observed for the AUCC, shown in Figure
\ref{fig:AUCC-unadjusted} in the supplement.
Figure~\ref{fig:relwidth-unadjusted} in the supplement shows the
confidence interval width relative to the fixed effect meta-analysis
method. Fisher's method tends to have smaller width
than all the other methods including fixed effect meta-analysis, most
pronounced for large $I^2$.

\paragraph{Confidence interval skewness}
Figure \ref{fig:ciskew-unadjusted} in the supplement displays the median
and the range (min -- max) of the CI skewness of the different
methods. Both fixed effect meta-analyis and Edgington have median
skewness of zero, which is desirable, as we simulate from a non-skewed
normal distribution. Whereas the skewness of fixed meta-analysis is always zero,
Edgington method shows considerable symmetric variation of the skewness coefficient around zero.
This variation increases with $I^2$ and decreases with the number of
studies. The other methods often have a median skewness different from
zero and also show non-symmetry of the range around the median.

To investigate how well the skewness of the confidence interval
captures the skewness of the data, Figure
\ref{fig:correlation-unadjusted} in the supplement displays the
correlation of the skewness of the data and the skewness of the
confidence interval.  There is always positive correlation for
Edgington's method, while this is not the case for the other methods.
Fisher's and Pearson's methods exhibit only sometimes a negative correlation
(for simulations with $I^2=0.9$) whereas Wilkinson's and Tippett's methods
have negative correlations most of the time. 

Figure \ref{fig:kappa-unadjusted}
shows the Cohen's $\kappa$ agreement between the sign of the skewness
of the confidence intervals and the sign of the skewness of the
data. Fixed effect meta-analysis is not shown because the confidence
intervals are always symmetric and thus always produce a skewness
coefficient of zero. We can see that Edgington's method shows
consistently high agreement with a decreasing trend with increasing
$I^2$. Agreement also decreases with increasing number of
studies. This is to be expected as we simulate from a normal study
effect distribution with zero skewness. The distribution of the
skewness of the data will therefore more and more concentrate around
zero with increasing number of studies. The other methods have
surprisingly poor performace, sometimes not better than what would be
expected by chance ($\kappa \approx 0$, Wilkinson and Tippett for no
large study) and sometimes even worse ($\kappa < 0$, Wilkinson and
Tippett with one or two large studies). This illustrates that only
Edgington's confidence intervals are capable to reflect the skewness
of the data. A very similar picture can be observed for the agreement
of the AUCC ratio with data skewness, as shown in Figure
\ref{fig:kappa-auccratio-unadjusted} in the supplement, which suggests
that the AUCC ratio as a measure of the skewness of a confidence
curve is a useful generalization of the confidence interval skewness
at level 95\%.

\subsubsection{With adjustments for heterogeneity}\label{sec:resultsWith}

\afterpage{
\begin{landscape}
\begin{figure}[!htb]
\begin{knitrout}
\definecolor{shadecolor}{rgb}{0.969, 0.969, 0.969}\color{fgcolor}
\includegraphics[width=\maxwidth]{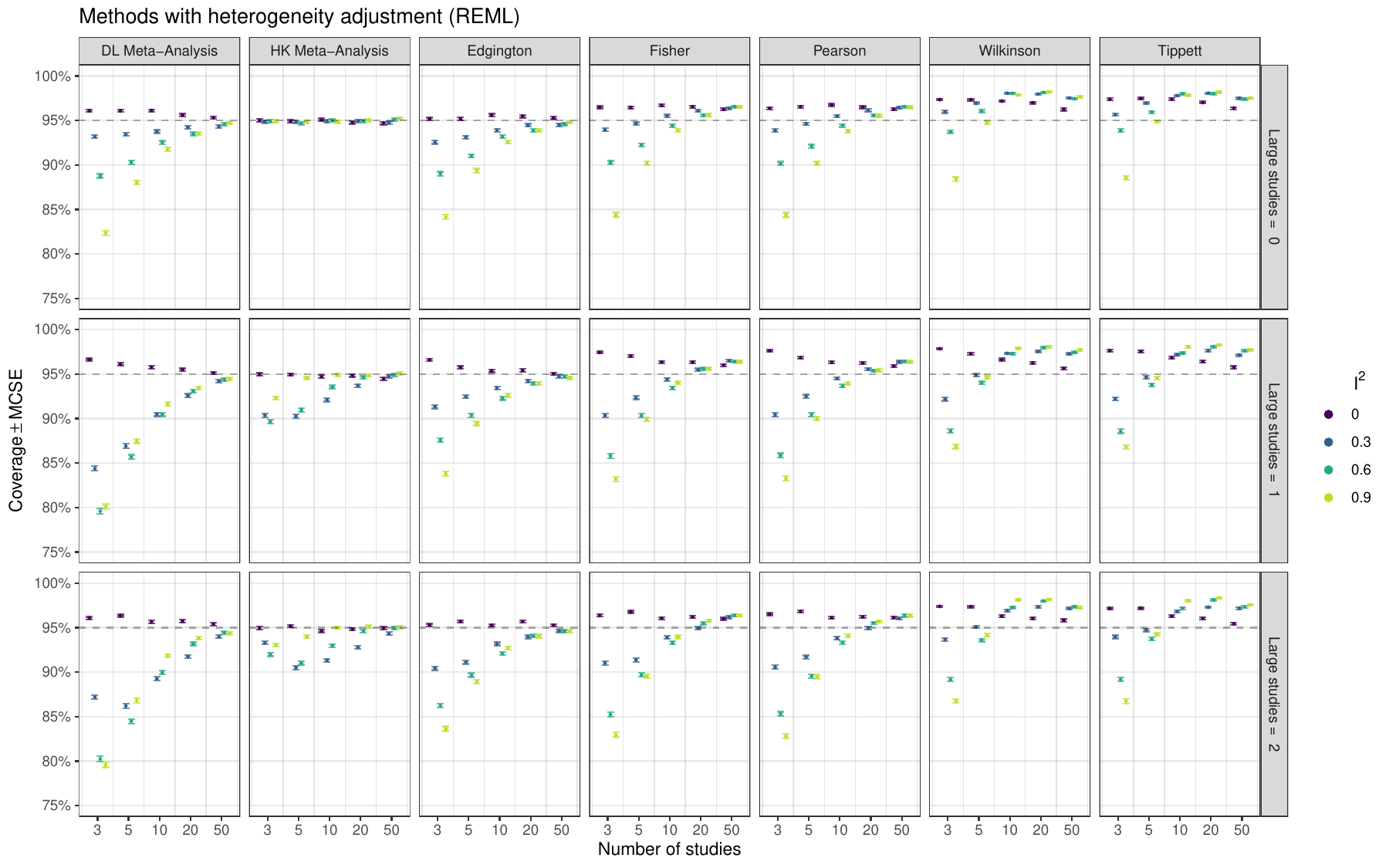} 
\end{knitrout}
\caption{Empirical coverage of the 95\% confidence intervals based on
  20'000 simulation repetitions.}
\label{fig:coverage}
\end{figure}
\begin{figure}[!htb]
\begin{knitrout}
\definecolor{shadecolor}{rgb}{0.969, 0.969, 0.969}\color{fgcolor}
\includegraphics[width=\maxwidth]{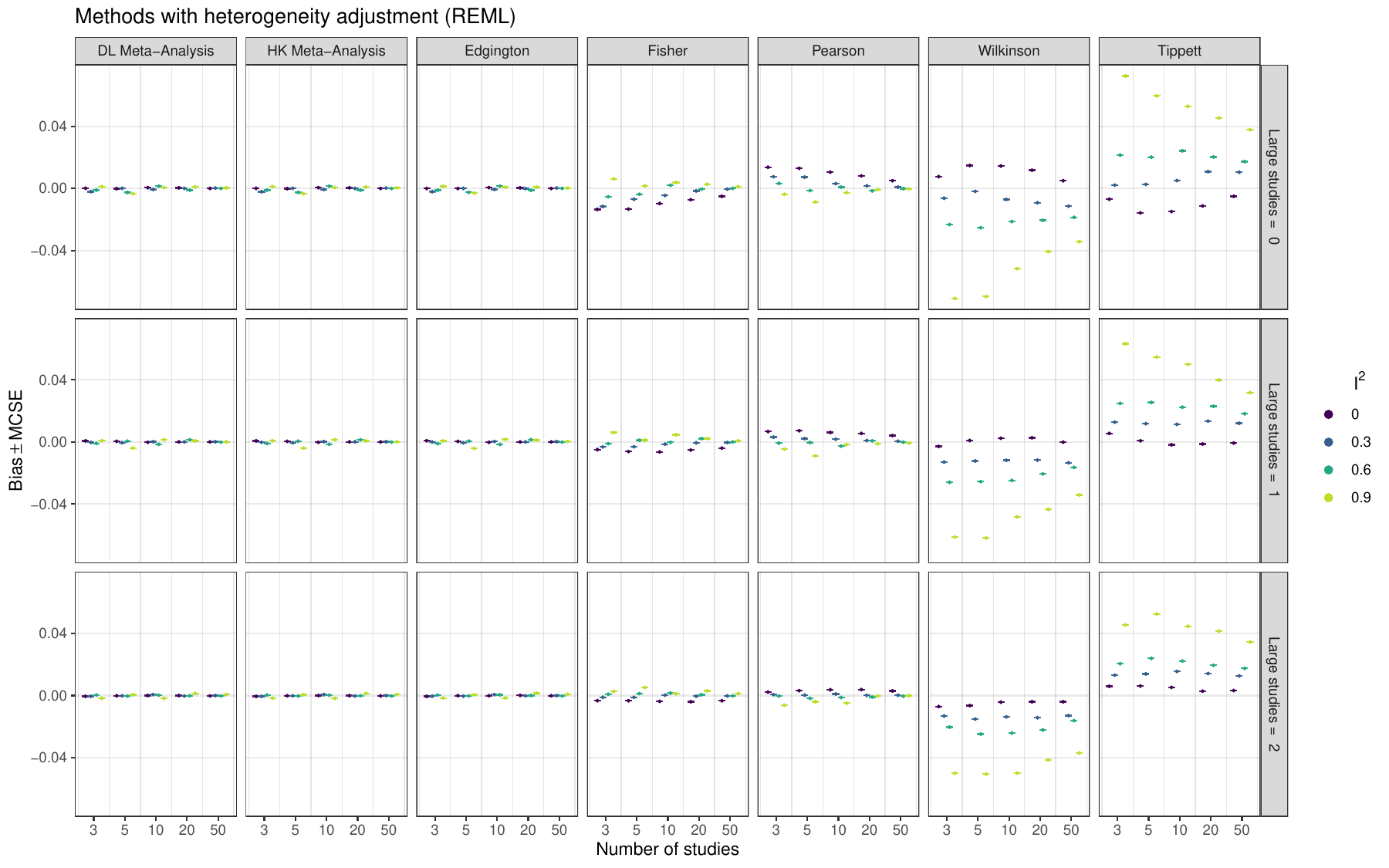} 
\end{knitrout}
\caption{Empirical bias of the point estimates for the true effect based on
  20'000 simulation repetitions.}
\label{fig:bias}
\end{figure}
\begin{figure}[!htb]
\begin{knitrout}
\definecolor{shadecolor}{rgb}{0.969, 0.969, 0.969}\color{fgcolor}
\includegraphics[width=\maxwidth]{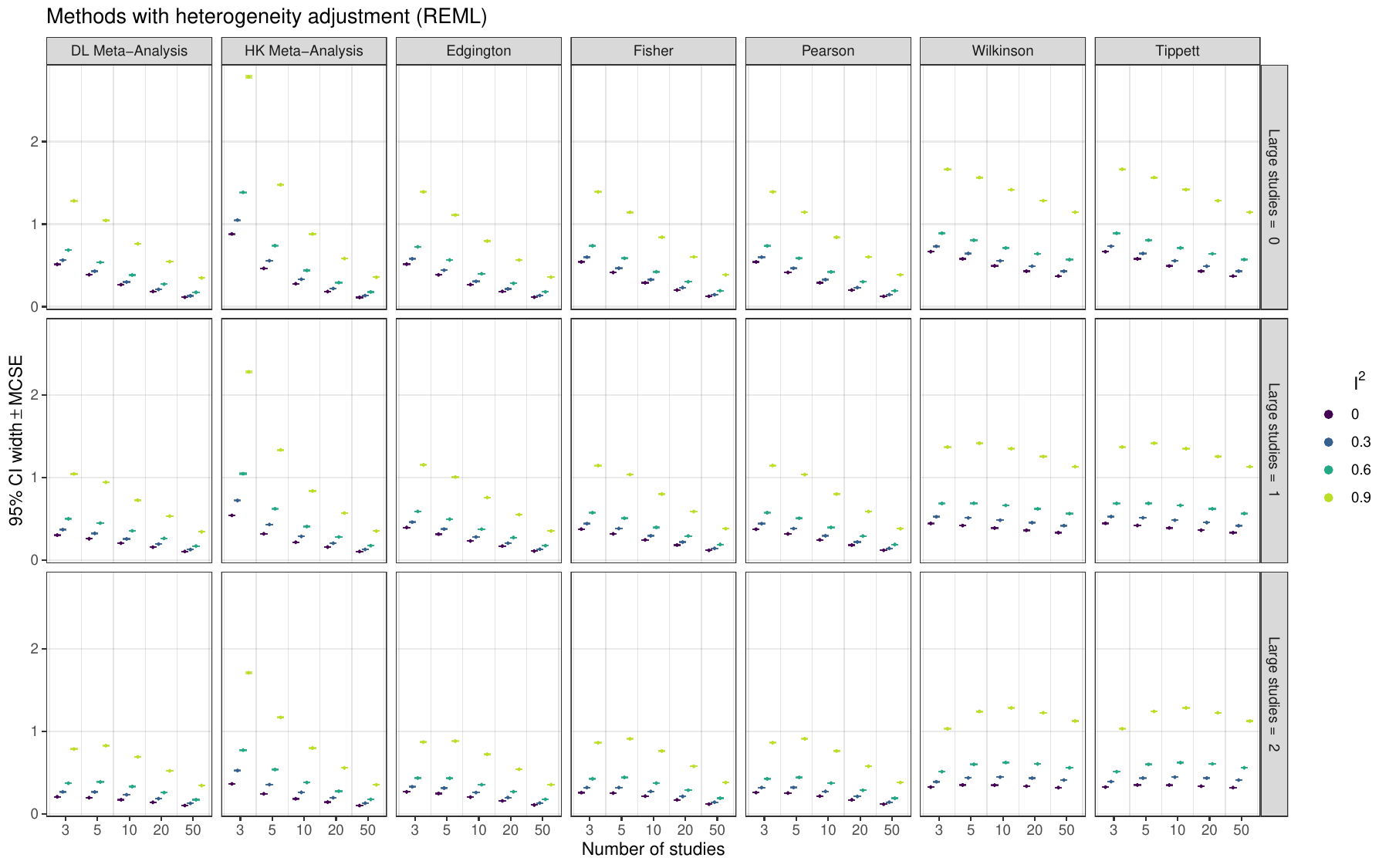} 
\end{knitrout}
\caption{Mean width of 95\% confidence intervals based on 20'000 simulation repetitions.}
\label{fig:width}
\end{figure}

\begin{figure}[!htb]
\begin{knitrout}
\definecolor{shadecolor}{rgb}{0.969, 0.969, 0.969}\color{fgcolor}
\includegraphics[width=\maxwidth]{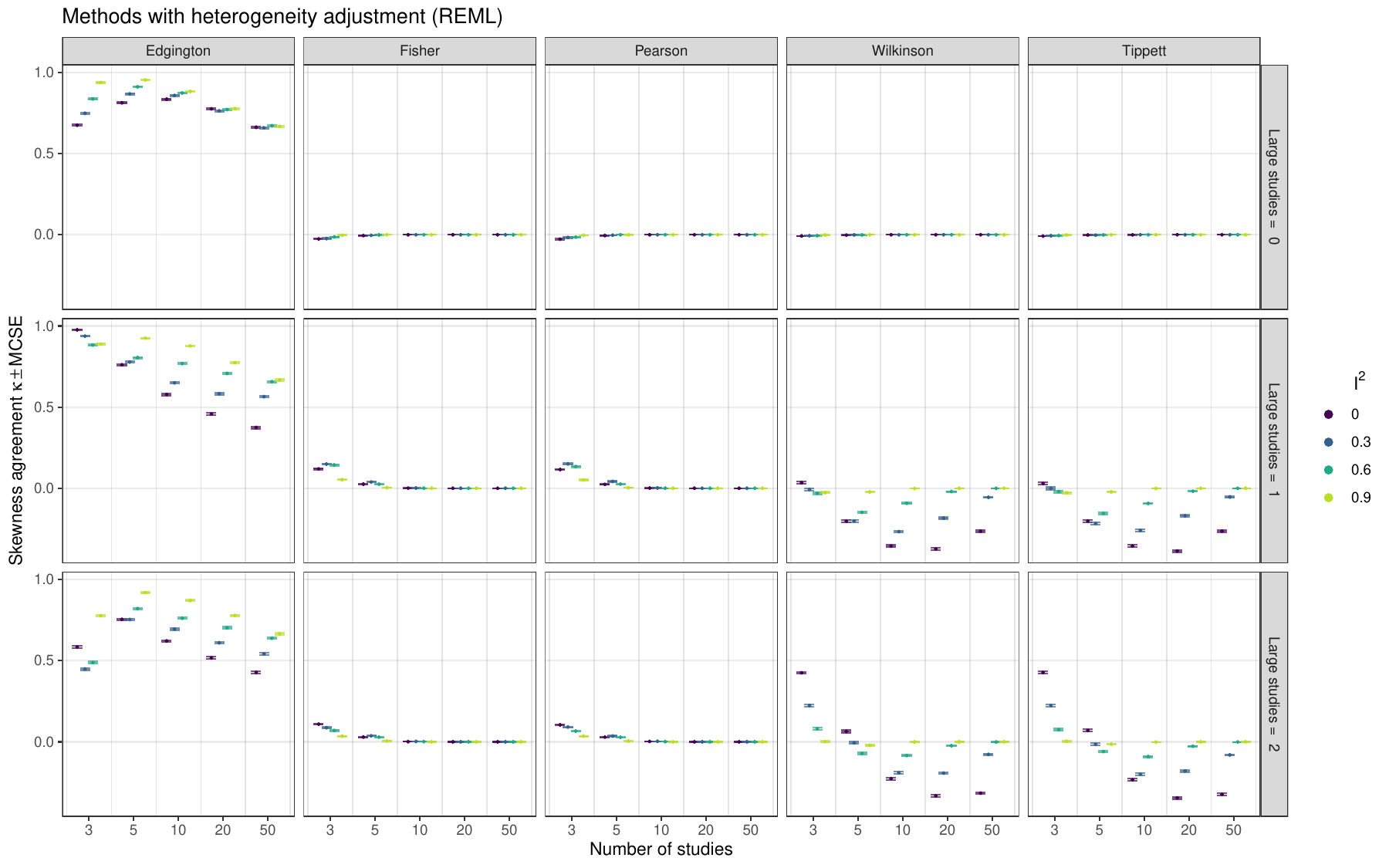} 
\end{knitrout}
\caption{Cohen's $\kappa$ sign agreement between 95\% confidence interval
  skewness and data skewness based on 20'000
  simulation repetitions.}
\label{fig:kappa}
\end{figure}

\end{landscape}
}

% coverage
\paragraph{Coverage}
Figure~\ref{fig:coverage} shows the empirical coverage of the
different methods with heterogeneity adjustment, which is, as
expected, considerably better than without adjustments for
heterogeneity (Figure~\ref{fig:coverage-unadjusted}).  We see that the
DL random effects meta-analysis (leftmost panels) has either too high
(for $I^{2} = 0$) or too low (for $I^{2} > 0$) coverage for small
numbers of studies, but seems to stabilize at the nominal 95\% level
as the number of studies increases. In contrast, for scenarios with no
large studies (top panels), the HK method shows almost perfect nominal
coverage over all numbers of studies, while for scenarios with one or
two large studies (middle and bottom panels), the coverage is slightly
too low for small numbers of studies, consistent with the results of
\citet{IntHout2014}. Focusing now on the $p$-value combination
methods, we can see that Edgington's method shows a qualitatively
similar behavior to DL random effects meta-analysis, but with somewhat
better coverage in most conditions. The coverage is in general not as
good as with Hartung-Knapp, but close to the nominal level for a large
number of studies.  In contrast, the coverage of Fisher, Pearson, Wilkinson, and
Tippett methods does not stabilize at the nominal 95\% but increases
above as the number of studies increases.

% bias
\paragraph{Bias}
Figure~\ref{fig:bias} shows the performance of the methods in terms of bias. We
see that the DL and HK random-effects meta-analysis as well as Edgington's methods are
essentially unbiased, while Fisher's, Pearson's, Wilkinson's, and Tippett's
method have substantial bias in most conditions. The bias patterns of
Fisher/Pearson and Wilkinson/Tippett are mirrored around zero, because
these methods' are mirrored with respect to $p$-value orientation.

% width
\paragraph{Confidence interval width}
Figure~\ref{fig:width} shows the average width of the method's 95\%
confidence intervals, which is now considerably larger than without
adjustments for heterogeneity (Figure~\ref{fig:width-unadjusted}). 
Edgington's, Fisher's, and Pearson's methods all have somewhat
wider confidence intervals than DL random effects meta-analysis, while
Hartung-Knapp has substantially wider intervals in conditions with a small
number of studies, as also noted by \citet{Weber_etal2021}. 
All of these widths shrink and become narrower as the
number of studies increases. However, Wilkinson's and Tippett's methods seem
to shrink much more slowly and remain relatively wide even with larger numbers
of studies. This is even better seen in Figure~\ref{fig:relwidth} reported in
the supplement, which shows confidence interval width relative to the DL random
effects meta-analysis method. We can see that the relative width of Wilkinson's
and Tippett's methods increases, while the Hartung-Knapp and to a lesser extent
Edgington's, Fisher's, and Pearson's method remain constant or decrease with
increasing number of studies. Interestingly, the figure also shows that the
Hartung-Knapp method 
can have narrower
confidence intervals than the DL random effects meta-analysis, which has been
described as a potential shortcoming %% of the Hartung-Knapp method
in the literature \citep{Jackson2017}.
In rare cases this may also happen with Edgington's method, but only
if there are no large studies and small amounts of relative heterogeneity. 

\paragraph{Confidence interval skewness}
%% The correlation of the skewness of the confidence interval and the
%% skewness of the data, shown in Figure \ref{fig:correlation}
%% in the supplement, is always positive and generally high for
%% Edgington's, Fisher's and Pearson's method.
{Figure~\ref{fig:kappa} shows the Cohen's $\kappa$ agreement between the sign of
the skewness of the confidence intervals and the skewness of the data. The
DL random effects meta-analysis and Hartung-Knapp methods are not shown because
their confidence intervals are always symmetric and thus always produce a
skewness coefficient of zero. We can see that Edgington's method shows
consistently high agreement with a decreasing trend as the number of studies
increases, while the agreement of the other methods is not better than what
would be expected by chance ($\kappa \approx 0$, Fisher and Pearson) and
sometimes even worse ($\kappa < 0$, Wilkinson and Tippett).
Figure~\ref{fig:ciskew} reported in the supplement shows the median skewness of
the confidence intervals and the corresponding min-max range, illustrating why
all but Edgington's method often show exactly zero agreement: As the number of
studies increases, their confidence intervals tend to be skewed in only one
direction. For ten or more studies, Pearson's method produced only confidence
intervals with negative skewness, while confidence intervals based on Fisher's
method were all positively skewed. Thus, the confidence interval cannot
represent the skewness type of the data, even though the confidence interval
skewness tends to be correlated with the data skewness (see
Figure~\ref{fig:correlation} in the supplement).

\subsection{Summary of simulation results}

Edgington's method produced
unbiased point estimates, its confidence intervals had comparable or
better coverage, and were only slightly wider than the confidence
intervals from a fixed effect and DL random effects meta-analysis, respectively. In addition, it was the
only method that could accurately represent data
skewness. %% Surprisingly, however, the performance of the method was
%% worse when study effects were simulated from a skewed distribution,
%% possibly because the method targets neither the mean nor the median of
%% the distribution, but something in between. 
The remaining $p$-value combination methods, Fisher/Pearson and, to a
greater extent, Wilkinson/Tippett, could not achieve satisfactory
performance. Their point estimates were more biased, their coverage
for a large number of studies was too high after adjustments for
heterogeneity, and their confidence intervals could not reliably
represent the skewness of the data.

The Hartung-Knapp method leads to known improvements in coverage
compared to DL random effects meta-analysis, although nominal coverage is
still not guaranteed when a meta-analysis includes a few studies that
are much larger than the remaining ones \citep{IntHout2014}. The
improved coverage of the Hartung-Knapp method also comes at the cost
of substantially wider confidence intervals on average, in particular
if the number of studies is small (see Figure \ref{fig:relwidth}).  %% In addition, our study showed that
%% when the true study effects are simulated from a skewed distribution
%% and the estimand is the mean study effect, both methods appear to be
%% unbiased and show similar coverage patterns as for symmetric study
%% effect distributions, but when the estimand of interest is the median
%% study effect, both methods become biased and their coverage becomes
%% worse.

\section{Discussion and extensions}
\label{sec:discussion}

% \todo[inline,color=blue!5]{SP: the discussion is mostly technical and the paper
%   ends a bit ``abruptly'', perhaps the main take away of the paper could be
%   summarized along with some practical recommendations? (perhaps at the last
%   paragraph to have a nicer ending?)}

We have
compared different $p$-value combination methods for meta-analysis theoretically
and through simulation. 
The $p$-value function approach based on Edgington
combination method constitutes a promising avenue for further research
and applications. Its ability to reflect the skewness of the data will
be attractive to applied meta-analysts. The good simulation
performance (compared to standard fixed effect and DL random effects
meta-analysis, respectively) for a small number of studies
suggests possible applicability in health technology assessment,
where a Bayesian approach has recently been proposed as an alternative
to the ``overly conservative'' Hartung-Knapp method
\citep{Lilienthal2024}.
A possible extension of the method provided is cumulative meta-analysis, where
studies are added one at a time in a specific order, usually time of publication
\citep{Egger_etal2022}. It would be interesting to compare the width of the
confidence interval with the length of the standard random effects confidence
interval as studies accumulate. %% It may also be worthwhile to conduct a
%% simulation study with focus on error rates, as standard random effects
%% cumulative meta-analysis is prone to Type-I error rate inflation
%% \citep{terSchureGrunwald2019,KulinskayaMah2022}, which may result in too narrow
%% confidence intervals.

Adjustments for heterogeneity have been made based on
the standard additive approach. Alter\-natively multiplicative heterogeneity can
be incorporated, where the squared standard errors $\sigma_i^2$ are multiplied
with a factor $\phi > 0$. Then we would use the adjusted $z$-statistic
\[
  Z_i = \frac{\hat \theta_i - \mu}{\sqrt{\hat \phi} \, \sigma_i}
\]
where $\hat \phi = \max\{Q/(k-1), 1\}$ is the appropriate overdispersion estimate \citep{StanleyDoucouliagos2015,Mawdsley2017}.
However, both additive and multiplicative adjustments are based on a plug-in
approach, which ignores the uncertainty of the heterogeneity estimate
$\hat \tau^2$ and $\hat \phi$, respectively. An interesting alternative
would be to profile-out the heterogeneity parameter, % as in
%\citet{Nagashima_etal2019},
%see \citet{Viechtbauer2007} and
see \citet{CunenHjort2021} and \citet{Zabriskie2021}.

%% \todo[inline,color=blue!10]{Discuss implicit normality distribution and how we want to move away from it in future work}

%% There is only little literature on meta-analysis if the true effect sizes do
%% not follow a normal distribution.
%% \citet{Higgins2008} focus on possible transformations of skewed
%% outcome data, where results are reported on a log-transformed scale
%% for some studies, but on the raw scale for other studies.
\citet{JacksonWhite2018} raise concerns about the usual between-study
normality assumption in meta-analysis, but also note that this issue
has not received sufficient attention.  \citet{BakerJackson2008,
  BakerJackson2016} propose long-tailed, but still symmetric random
effects distributions while \citet{Beath2014}
considers a mixture of two normals model to accomodate for
outliers. The two components have differenct variances but the same
means, so the resulting mixture distribution is still symmetric.
Finally, \citet{Kontopantelis_etal2012} and \citet{Weber_etal2021} compare existing (symmetric)
meta-analytic interval estimates in a simulation study with normal and
skew normal random effects distributions.
A possible area for future research is therefore to assume a non-normal
distribution of the random effects, for example a skew normal. The
heterogeneity variance could then be estimated based on a moment-based
estimate (not assuming normality) together with the skewness
parameter. The resulting distribution of the $z$-value
\eqref{eq:z.adjusted} is then no longer normal and needs to be
calculated through numerical integration and can then be used to
convert $z$-values to $p$-values.

It would also be interesting to
extend the approach to compute prediction intervals for future study
effects
\citep{Higgins_etal2009,Hamaguchi_etal2021,Nagashima_etal2019}.  This
would involve numerical integration of a $\Nor(\mu, \tau^2)$ (or
skew normal) distribution with respect to the confidence density for
$\mu$, which can be obtained from any (monotonically increasing)
one-sided $p$-value function (for alternative ``greater'') through differentiation.  For example,
differentiation of the underlying exact one-sided $p$-value function
from Edgington's method in Figure \ref{fig:exactP}
gives the confidence density shown in
Figure \ref{fig:exactP2}. The confidence density is clearly skewed,
which would then also be the case for the corresponding prediction
interval. We plan to consider this in future work.

\begin{figure}
\begin{knitrout}
\definecolor{shadecolor}{rgb}{0.969, 0.969, 0.969}\color{fgcolor}
\includegraphics[width=\maxwidth]{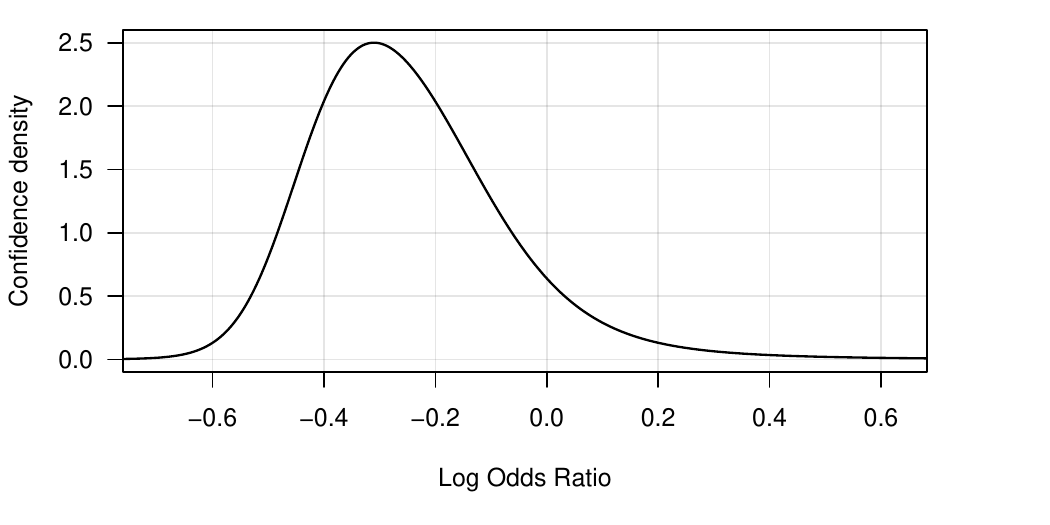} 
\end{knitrout}
\caption{\label{fig:exactP2} Confidence density based on Edgington's combined $p$-value function and exact $p$-values with mid-$p$ correction for meta-analysis of
  $k=7$ randomized controlled clinical trials
investigating the association between corticosteroids and mortality in
hospitalized patients with COVID-19 \citep{REACT2020}.}
\end{figure}

\paragraph*{Competing Interests} 
The authors have declared no conflict of interest.
\paragraph*{Data and Software Availability} 
Meta-analysis with $p$-value combination methods is implemented in the R-package
\texttt{confMeta} available on GitHub
(\url{https://github.com/felix-hof/confMeta}). The package will be soon
submitted to CRAN. R code to reproduce our results is available
at \url{https://doi.org/10.17605/OSF.IO/JE8XB}. R code to reproduce the simulation study is
available at \url{https://github.com/felix-hof/confMeta_simulation}.

\paragraph*{Funding}
Partial support by the Swiss National Science Foundation (Project \# 189295) in the initial phase of this project is
gratefully acknowledged.

\paragraph*{Acknowledgments}
We appreciate helpful comments by two anomynous referees and are grateful for additional suggestions by Ralf Bender and Frank Weber on an earlier version of this manuscript. We also appreciate inputs from Florian Gerber and Lisa Hofer in earlier stages of the project. 

\paragraph*{Conflict of Interest Statement}
The authors declare that no conflict of interest exists for all authors.

\bibliographystyle{apalike}
\bibliography{antritt}

\begin{appendix}

\section{Closed-form solution of Wilkinson's and Tippett's methods}
\label{app:closedform}
Assuming $p$-values under normality of the form~\eqref{eq:pNormal},
fixing the combined $p$-values $\pright{p_W}$ and $\pleft{p_W}$, respectively, of Wilkinson's method from
Table~\ref{tab:pvalcomb} to $\alpha$ and solving for $\mu$, we obain
the following closed-form solutions
\begin{align*}
  \hat{\mu}_W(\alpha) =
  \begin{cases}
    \min\limits_{i=1, \ldots, k}\left\{ \hat{\theta}_i + \sigma_i \, z_{\alpha^{1/n}} \right\} & \text{for~alternative = "greater"} \\
    \max\limits_{i=1, \ldots, k}\left\{ \hat{\theta}_i - \sigma_i \, z_{\alpha^{1/n}} \right\} & \text{for~alternative = "less"} \\
  \end{cases}
\end{align*}
%% \begin{align*}
%%   \hat{\mu}_W(\alpha) =
%%   \begin{cases}
%%     \min\left\{\hat{\theta}_1 + \sigma_1 z_{\alpha^{1/n}}, \dots, \hat{\theta}_n + \sigma_n z_{\alpha^{1/n}} \right\} & \text{for~alternative = "greater"} \\
%%     \max\left\{\hat{\theta}_1 - \sigma_1 z_{\alpha^{1/n}}, \dots, \hat{\theta}_n - \sigma_n z_{\alpha^{1/n}}\right\} & \text{for~alternative = "less"} \\
%%   \end{cases}
%% \end{align*}
with $z_q$ the $q \times 100\%$ quantile of the standard normal distribution.
Plugging in $\alpha = 0.5$ produces the median estimate, while $\alpha = 0.025$
and $\alpha = 0.975$ give the limits of a 95\% confidence interval. A similar
approach applied to the combined $p$-value from Tippett's method leads to
\begin{align*}
  \hat{\mu}_T(\alpha) =
  \begin{cases}
    \max\limits_{i=1, \ldots, k}\left\{\hat{\theta}_i - \sigma_i \, z_{(1 - \alpha)^{1/n}} \right\} & \text{for~alternative = "greater"} \\
    \min\limits_{i=1, \ldots, k}\left\{\hat{\theta}_i + \sigma_i \, z_{(1 - \alpha)^{1/n}} \right\} & \text{for~alternative = "less"} \\
  \end{cases}
\end{align*}
%% \begin{align*}
%%   \hat{\mu}_T(\alpha) =
%%   \begin{cases}
%%     \max\left\{\hat{\theta}_1 - \sigma_1 z_{(1 - \alpha)^{1/n}}, \dots, \hat{\theta}_n - \sigma_n z_{(1 - \alpha)^{1/n}}\right\} & \text{for~alternative = "greater"} \\
%%     \min\left\{\hat{\theta}_1 + \sigma_1 z_{(1 - \alpha)^{1/n}}, \dots, \hat{\theta}_n + \sigma_n z_{(1 - \alpha)^{1/n}}\right\} & \text{for~alternative = "less"} \\
%%   \end{cases}
%% \end{align*}

\section{Relationships of \textit{p}-value combination methods}
\label{app:relationships}

\subsection{Edgington's method}
Define
\[
  \pleft{s} = \sum_{i=1}^{k} \pleft{p_{i}}
  = \sum_{i=1}^{k} (1 - \pright{p_{i}})= k - \sum_{i=1}^{k} \pright{p_{i}}
  = k - \pright{s}, 
\]
so $\pright{p_{E}} = \Pr(\mathrm{I}_{k} \leq \pright{s}) = 1 - \Pr(\mathrm{I}_{k} > \pright{s})$
where $\mathrm{I}_{k} \in [0, k]$ is an Irwin-Hall random variable with
parameter $k$. Since the Irwin-Hall distribution is symmetric around its mean
$k/2$, it holds that
$\Pr(\mathrm{I}_{k} \leq {s}) = \Pr(\mathrm{I}_{k} > k - {s})$ for any $s \in [0,k]$.
%% = 1 - \Pr(\mathrm{I}_{k} \leq k - \pright{s})$.
Hence, it follows that
\begin{eqnarray*}
  \pright{p}_{E} & = &  \Pr(\mathrm{I}_{k} \leq \pright{s}) \\
  & = &  \Pr(\mathrm{I}_{k} > k - \pright{s}) \\
  & = & \Pr(\mathrm{I}_{k} > \pleft{s}) \\
%%  & = & \Pr(\mathrm{I}_{k} > \pleft{s}) \\
  & = & 1- \Pr(\mathrm{I}_{k} \leq \pleft{s}) \\
  & = & 1 - \pleft{p}_{E}.
\end{eqnarray*}

\subsection{Fisher's and Pearson's methods}
Define
%% We have %% that
%% \[
%%   \pleft{f} = -2 \sum_{i=1}^{n} \log(\pleft{p_{i}})
%%   = -2 \sum_{i=1}^{n} \log(1 - \pright{p_{i}})
%%   = \pright{k}
%% \]
%% and
\[
  \pright{f} = -2 \sum_{i=1}^{k} \log(\pright{p_{i}})
  = -2 \sum_{i=1}^{k} \log(1 - \pleft{p_{i}})
  = \pleft{g}.
\]
%% Since \(p_{F} = \Pr(\chi_{2k}^{2} \geq f)\) and
%% \(p_{P} = \Pr(\chi_{2n}^{2} < k)\),
It follows that
\begin{eqnarray*}
  \pright{p}_{F} & = & \Pr(\chi_{2k}^{2} > \pright{f}) \\
& = & \Pr(\chi_{2k}^{2} > \pleft{g}) \\
& = & 1 - \Pr(\chi_{2k}^{2} \leq \pleft{g}) \\
& = & 1 - \pleft{p}_{P}.
\end{eqnarray*}
Similarly we obtain
\mbox{\(\pright{p}_{P} = 1 - \pleft{p}_{F}\)}.

\subsection{Tippett's and Wilkinson's methods}
We have 
%% \[
%%   \min\{\pright{p}_{1}, \dots, \pright{p}_{n}\}
%%   = \min\{1 - \pleft{p}_{1}, \dots, 1 - \pleft{p}_{n}\}
%%   = 1 - \max\{\pleft{p}_{1}, \dots, \pleft{p}_{n}\}
%% \]
%% and
%% \begin{eqnarray*}
%%   \pright{p_{T}} &=& 1 - (1 - \min\{\pright{p_{1}}, \dots, \pright{p_{n}}\})^{n}, \\
%%   \pleft{p_{T}} &=& 1 - \min\{\pleft{p_{1}}, \dots, \pleft{p_{n}}\})^{n}, \\
%%   \pright{p_{W}} &=& \max\{\pright{p_{1}}, \dots, \pright{p_{n}}\}^{n}, \\
%%   \pleft{p_{W}} &=& 1-\max\{\pleft{p_{1}}, \dots, \pleft{p_{n}}\}^{n}. \\
%%   \end{eqnarray*}
%% It follows that
\begin{eqnarray*}
  \pright{p}_{T} & = & 1 - (1 - \min\{\pright{p_{1}}, \dots, \pright{p_{k}}\})^{k} \\
                & = & 1 - (1 - \min\{1-\pleft{p_{1}}, \dots, 1-\pleft{p_{k}}\})^{k} \\
  & = & 1 - \max\{\pleft{p_{1}}, \dots, \pleft{p_{k}}\}^{k} \\
  & = & 1 - \pleft{p}_{W} %%\\
%%  & = & \pleft{p}_{W}
\end{eqnarray*}
and similarly
\mbox{\(\pright{p}_{W} = 1 - \pleft{p}_{T}\)}.

\newpage

% reset counter of pages, tables and figures, and add preceeding S
\setcounter{figure}{0}
\renewcommand{\thefigure}{S\arabic{figure}}
\setcounter{table}{0}
\renewcommand{\thetable}{S\arabic{table}}
\setcounter{page}{1}

\begin{center}
  {\huge \textsf{Supplementary material for}}

  {\huge \textsf{\textbf{A comparison of combined \textit{p}-value functions for meta-analysis}}}

  {\Large
  \begin{tabular}{c c c}
  Leonhard Held\orcidlink{0000-0002-8686-5325} &
  Felix Hofmann\orcidlink{0000-0002-3891-6239} &
  Samuel Pawel\orcidlink{0000-0003-2779-320X}
  \end{tabular}
  }

  {\Large \today}

\end{center}

\section*{Additional simulation results}
\label{app:additionalsim}
In the following, we report figures and tables with additional simulation
results comparing properties of the different $p$-value combination methods, see
Table~\ref{tab:summaryaddresults} for an overview.

\begin{table}[!h]
  \centering
      \caption{Location of additional simulation results.}
      \label{tab:summaryaddresults}
      \small
  \begin{tabular}{l c c}
    \toprule
    \textbf{Result} & \textbf{Without heterogeneity adjustment} & \textbf{With heterogeneity adjustment} \\
    \midrule
    Non-convergence & Table~\ref{tab:convergence} & All methods always converged \\
    AUCC & Figure~\ref{fig:AUCC-unadjusted} & Figure~\ref{fig:AUCC} \\
    AUCC ratio & Figure~\ref{fig:AUCCratio-unadjusted} & Figure~\ref{fig:AUCCratio} \\
    %% AUCC ratio for Edgington & Figure~\ref{fig:CIauccrEdgington-unadjusted} & Figure~\ref{fig:CIauccrEdgington} \\
    AUCC ratio skewness agreement & Figure~\ref{fig:kappa-auccratio-unadjusted} & Figure~\ref{fig:kappa-auccratio} \\
    Relative CI width & Figure~\ref{fig:relwidth-unadjusted} & Figure~\ref{fig:relwidth} \\
    Median CI skewness & Figure~\ref{fig:ciskew-unadjusted} & Figure~\ref{fig:ciskew} \\
    %% Mean CI skewness for Edgington & Figure~\ref{fig:CIskewEdgington-unadjusted} & Figure~\ref{fig:CIskewEdgington} \\
    CI skewness correlation & Figure~\ref{fig:correlation-unadjusted} & Figure~\ref{fig:correlation} \\
    \bottomrule
  \end{tabular}
\end{table}

\begin{table}[!htb]
  \centering
  \caption{Simulation conditions for which convergence rate was not 100\% (i.e.,
    some methods did not produce a confidence interval in some repetitions).}
  \label{tab:convergence}

% latex table generated in R 4.4.2 by xtable 1.8-4 package
% Fri Feb 21 13:23:13 2025
\begin{tabular}{ccccc}
  \toprule
Convergence rate & Method & $I^2$ & $k$ & Large studies \\ 
  \midrule
99.372\% & Pearson & 0.90 &  50 &   2 \\ 
  99.696\% & Pearson & 0.90 &  20 &   1 \\ 
  99.785\% & Pearson & 0.90 &   5 &   1 \\ 
  99.790\% & Pearson & 0.90 &  20 &   2 \\ 
  99.800\% & Pearson & 0.90 &  50 &   1 \\ 
  99.840\% & Pearson & 0.90 &   5 &   2 \\ 
  99.945\% & Pearson & 0.90 &  10 &   2 \\ 
  99.945\% & Wilkinson & 0.90 &  10 &   2 \\ 
  99.960\% & Pearson & 0.90 &  10 &   1 \\ 
  99.975\% & Wilkinson & 0.90 &  10 &   1 \\ 
  99.985\% & Wilkinson & 0.90 &  20 &   1 \\ 
  99.985\% & Pearson & 0.90 &  50 &   0 \\ 
  99.990\% & Wilkinson & 0.90 &  20 &   2 \\ 
  99.995\% & Wilkinson & 0.90 &  50 &   2 \\ 
   \bottomrule
\end{tabular}

\end{table}

\begin{landscape}

\begin{figure}[!htb]
\begin{knitrout}
\definecolor{shadecolor}{rgb}{0.969, 0.969, 0.969}\color{fgcolor}
\includegraphics[width=\maxwidth]{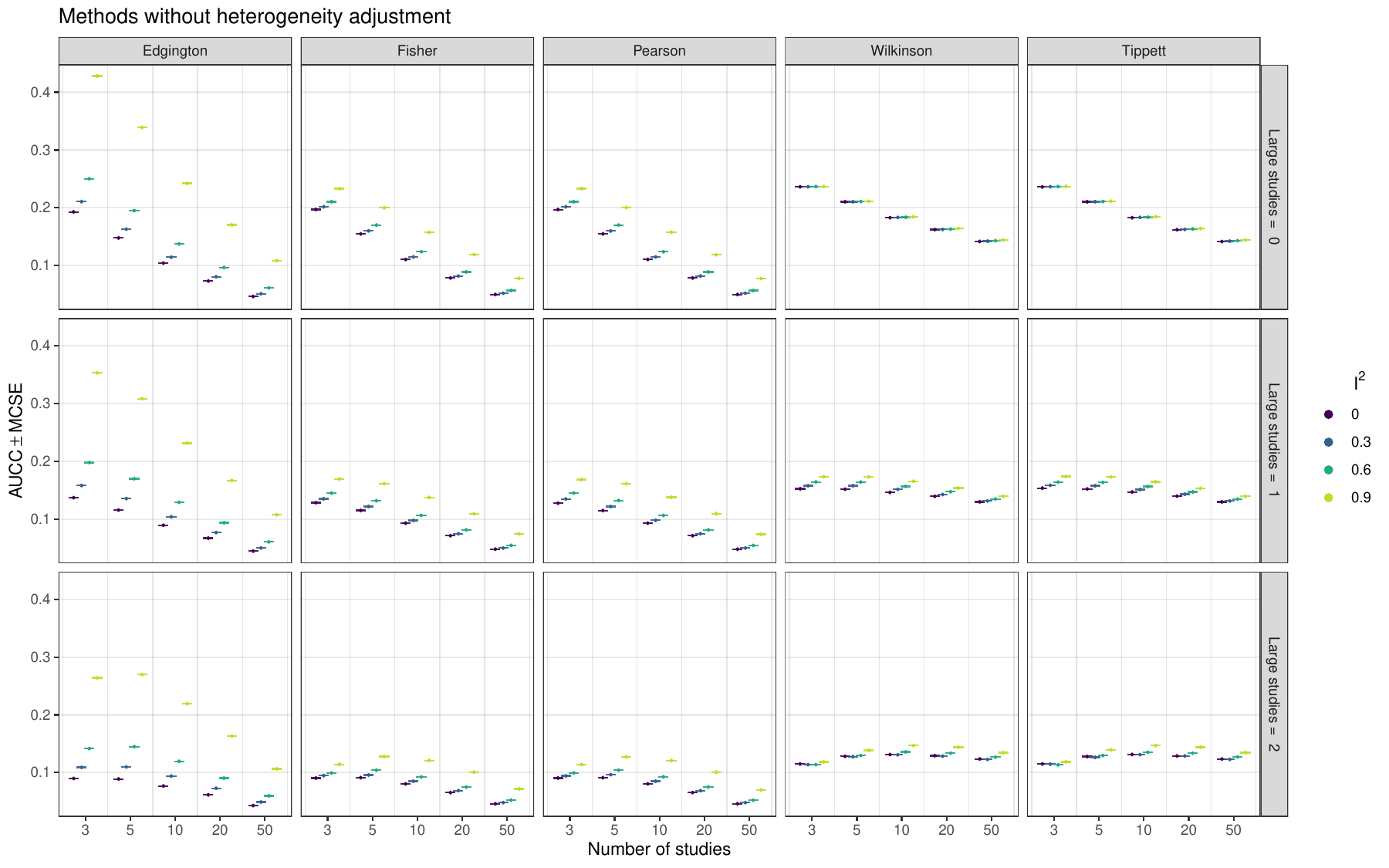} 
\end{knitrout}
\caption{Mean area under the confidence curve (AUCC) based on 20'000 simulation repetitions.}
\label{fig:AUCC-unadjusted}
\end{figure}

\begin{figure}[!htb]
\begin{knitrout}
\definecolor{shadecolor}{rgb}{0.969, 0.969, 0.969}\color{fgcolor}
\includegraphics[width=\maxwidth]{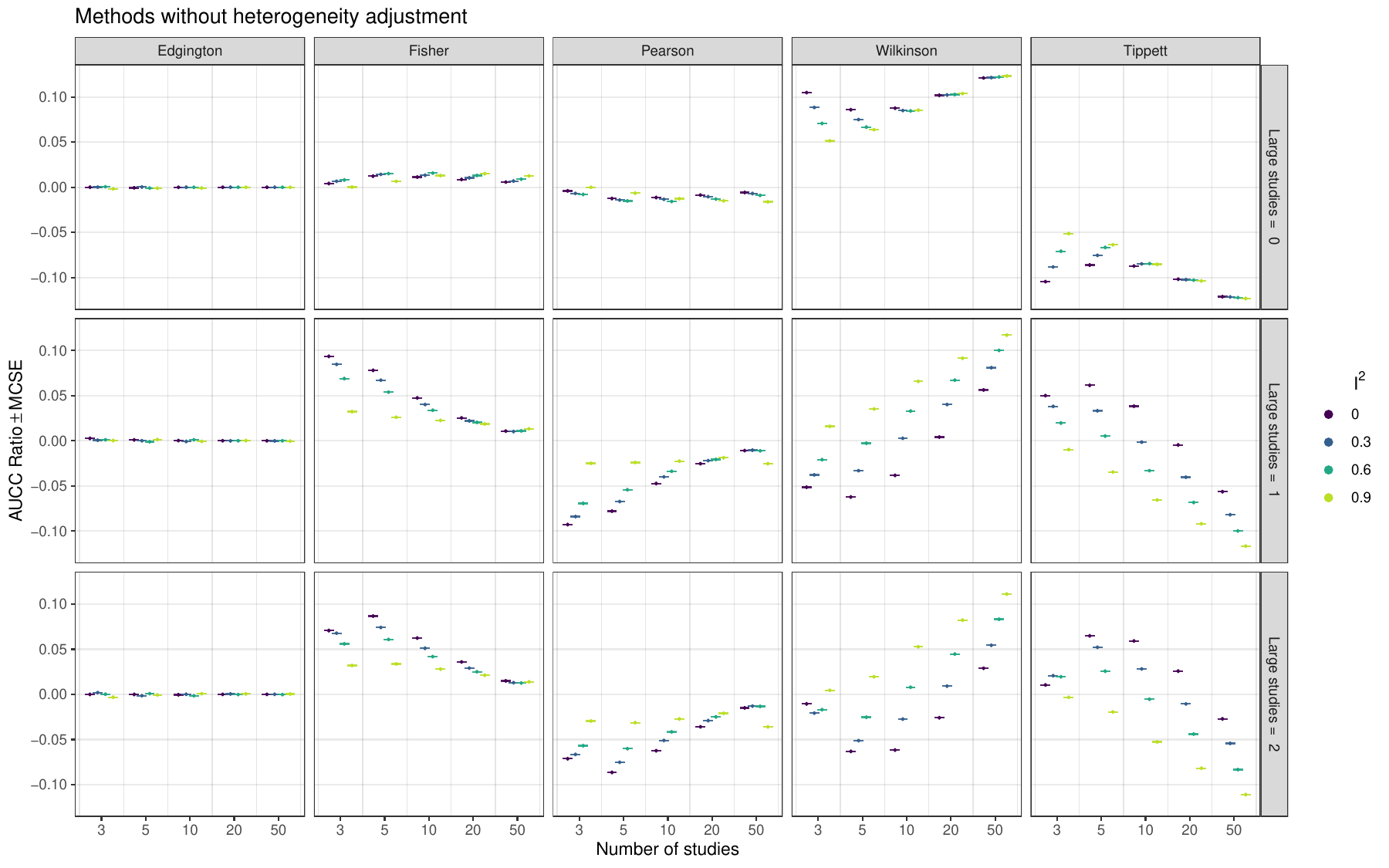} 
\end{knitrout}
\caption{Mean area under the confidence curve (AUCC) ratio based on 20'000 simulation repetitions.}
\label{fig:AUCCratio-unadjusted}
\end{figure}

\begin{figure}[!htb]
\begin{knitrout}
\definecolor{shadecolor}{rgb}{0.969, 0.969, 0.969}\color{fgcolor}
\includegraphics[width=\maxwidth]{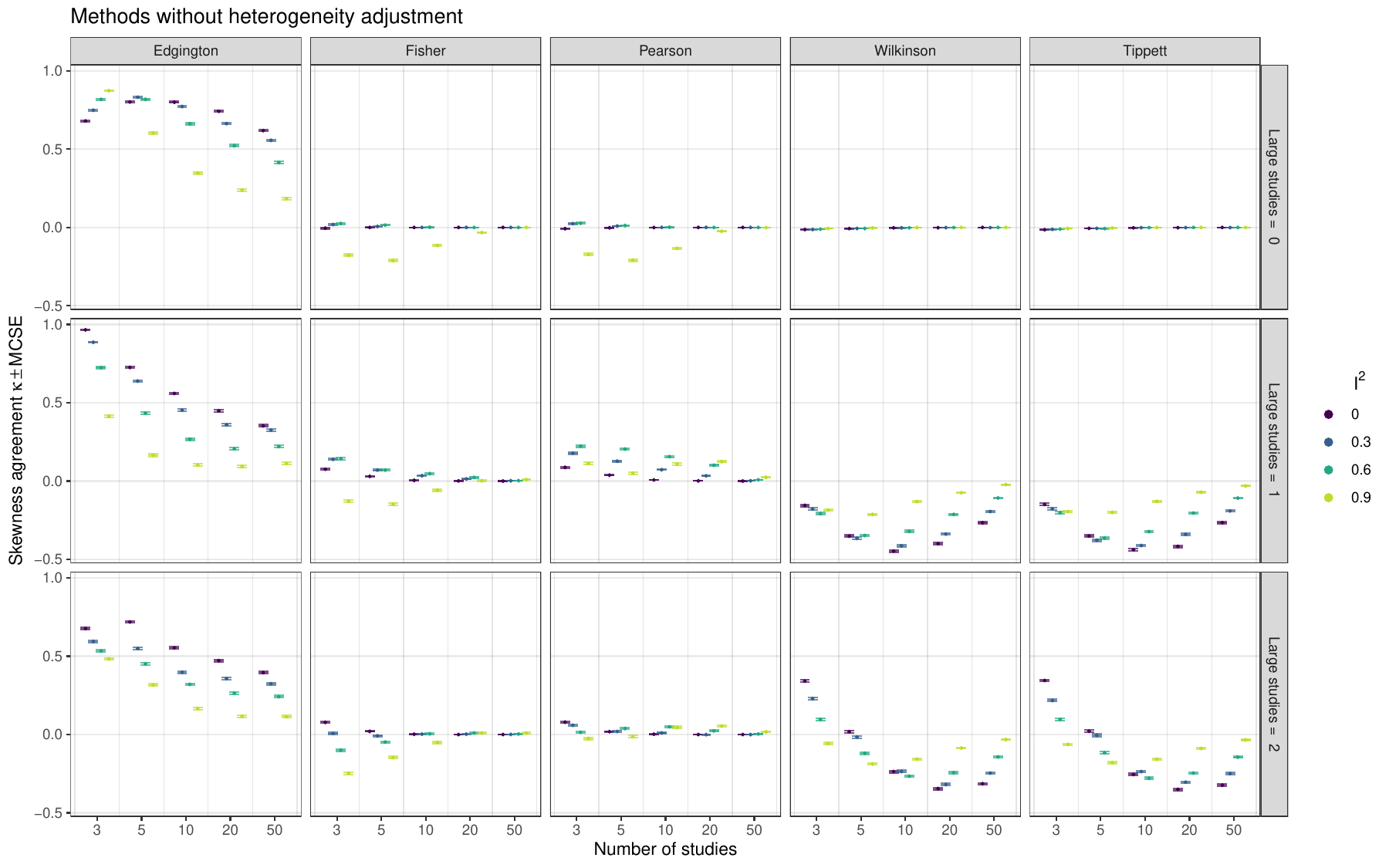} 
\end{knitrout}
\caption{Cohen's $\kappa$ sign agreement between AUCC ratio skewness and data
  skewness based on 20'000 simulation
  repetitions.}
\label{fig:kappa-auccratio-unadjusted}
\end{figure}

\begin{figure}[!htb]
\begin{knitrout}
\definecolor{shadecolor}{rgb}{0.969, 0.969, 0.969}\color{fgcolor}
\includegraphics[width=\maxwidth]{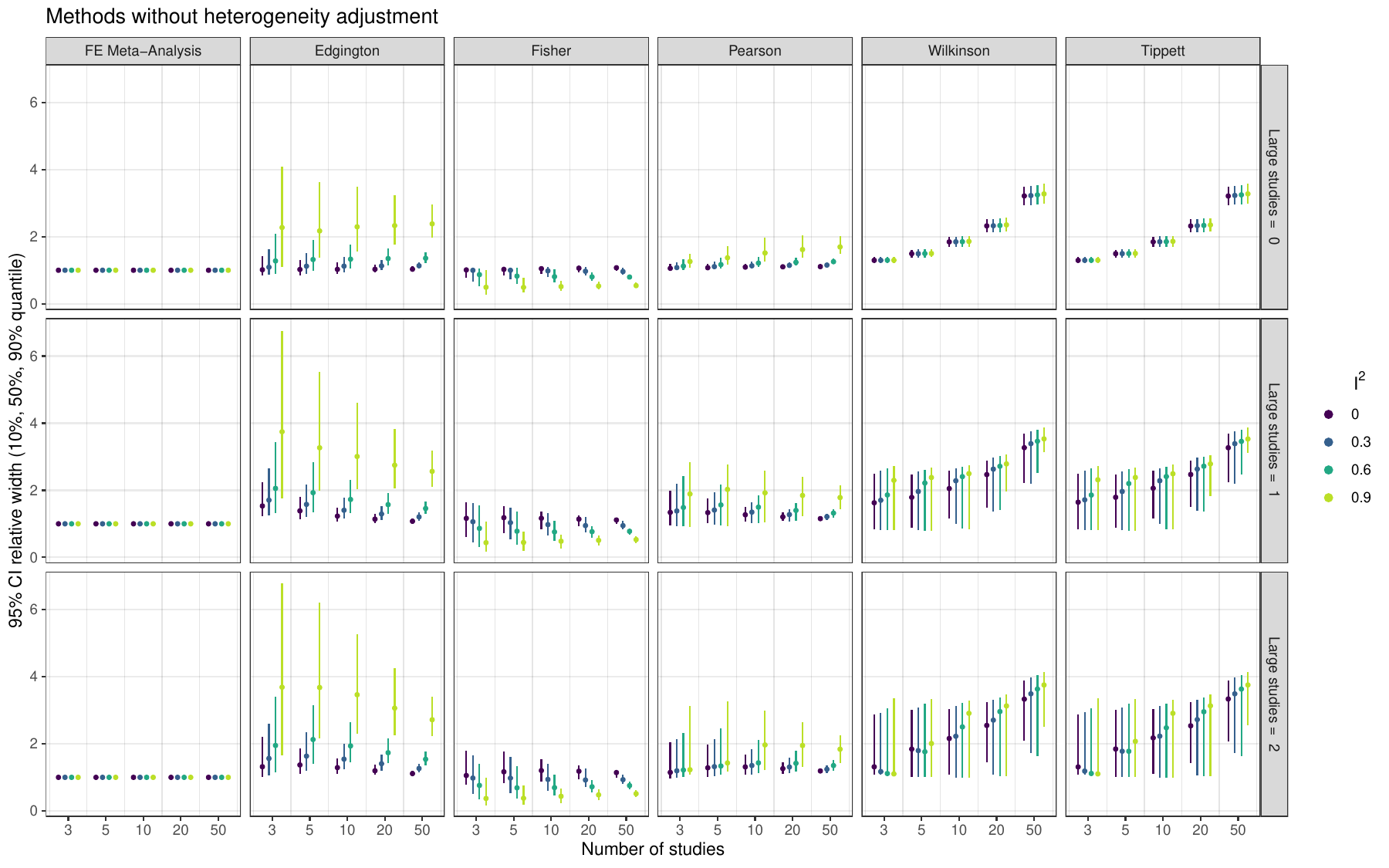} 
\end{knitrout}
\caption{Relative width of 95\% confidence intervals (relative to fixed effect
  meta-analysis) based on 20'000 simulation
  repetitions.}
\label{fig:relwidth-unadjusted}
\end{figure}

\begin{figure}[!htb]
\begin{knitrout}
\definecolor{shadecolor}{rgb}{0.969, 0.969, 0.969}\color{fgcolor}
\includegraphics[width=\maxwidth]{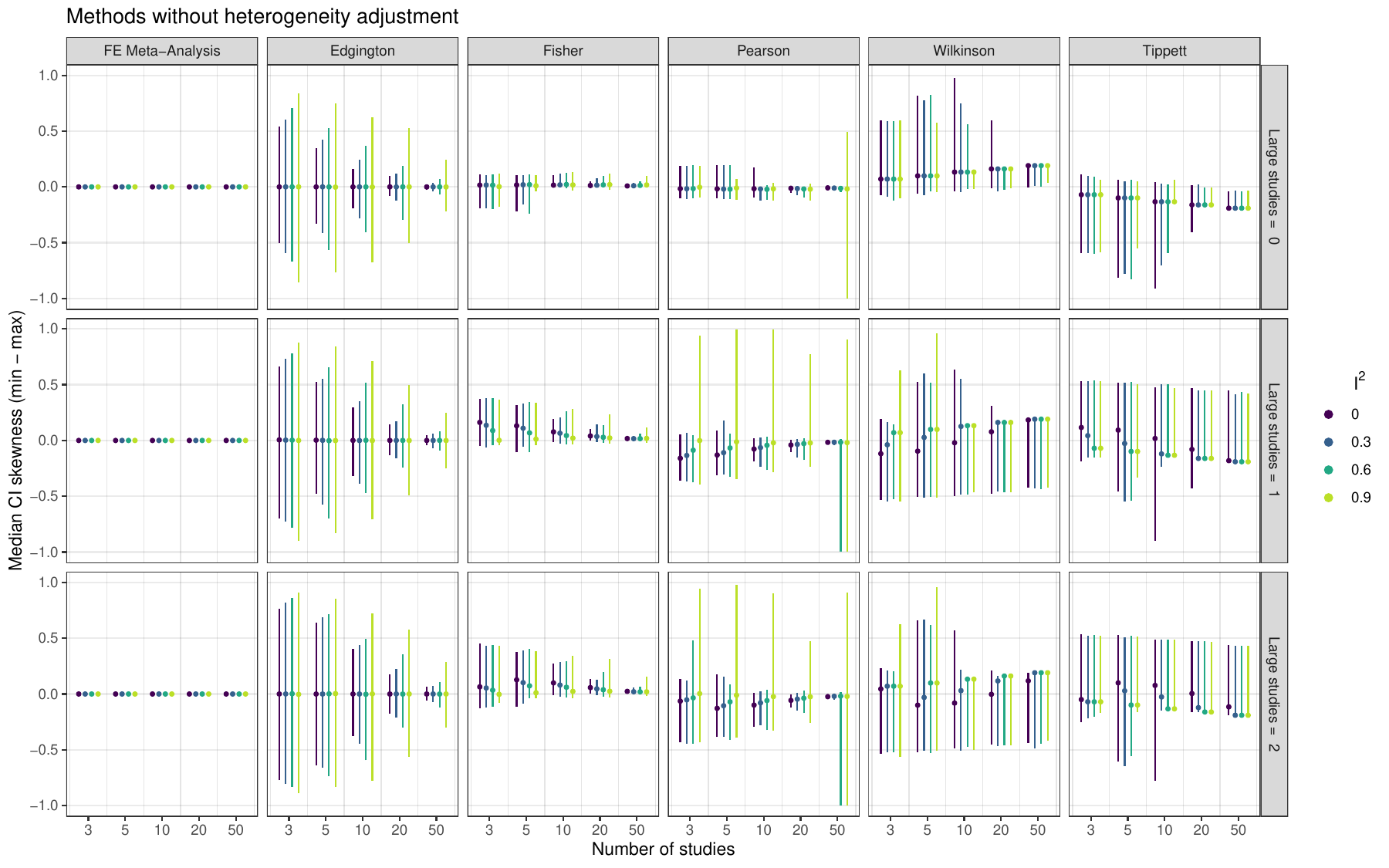} 
\end{knitrout}
\caption{Skewness of 95\% confidence interval based on 20'000 simulation repetitions.}
\label{fig:ciskew-unadjusted}
\end{figure}

\begin{figure}[!htb]
\begin{knitrout}
\definecolor{shadecolor}{rgb}{0.969, 0.969, 0.969}\color{fgcolor}
\includegraphics[width=\maxwidth]{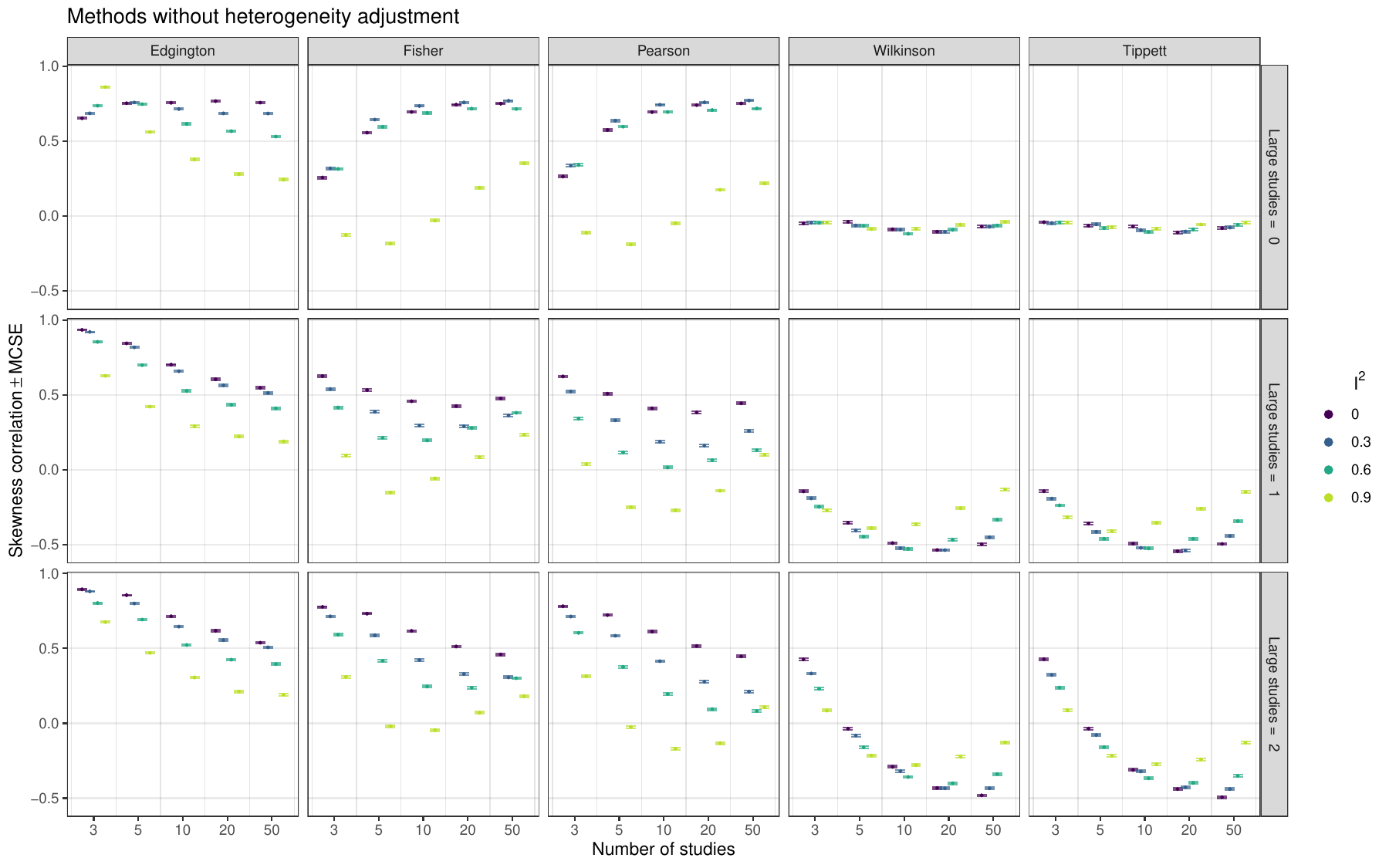} 
\end{knitrout}
\caption{Correlation between 95\% confidence interval skewness and data skewness
  based on 20'000 simulation repetitions.}
\label{fig:correlation-unadjusted}
\end{figure}

\begin{figure}[!htb]
\begin{knitrout}
\definecolor{shadecolor}{rgb}{0.969, 0.969, 0.969}\color{fgcolor}
\includegraphics[width=\maxwidth]{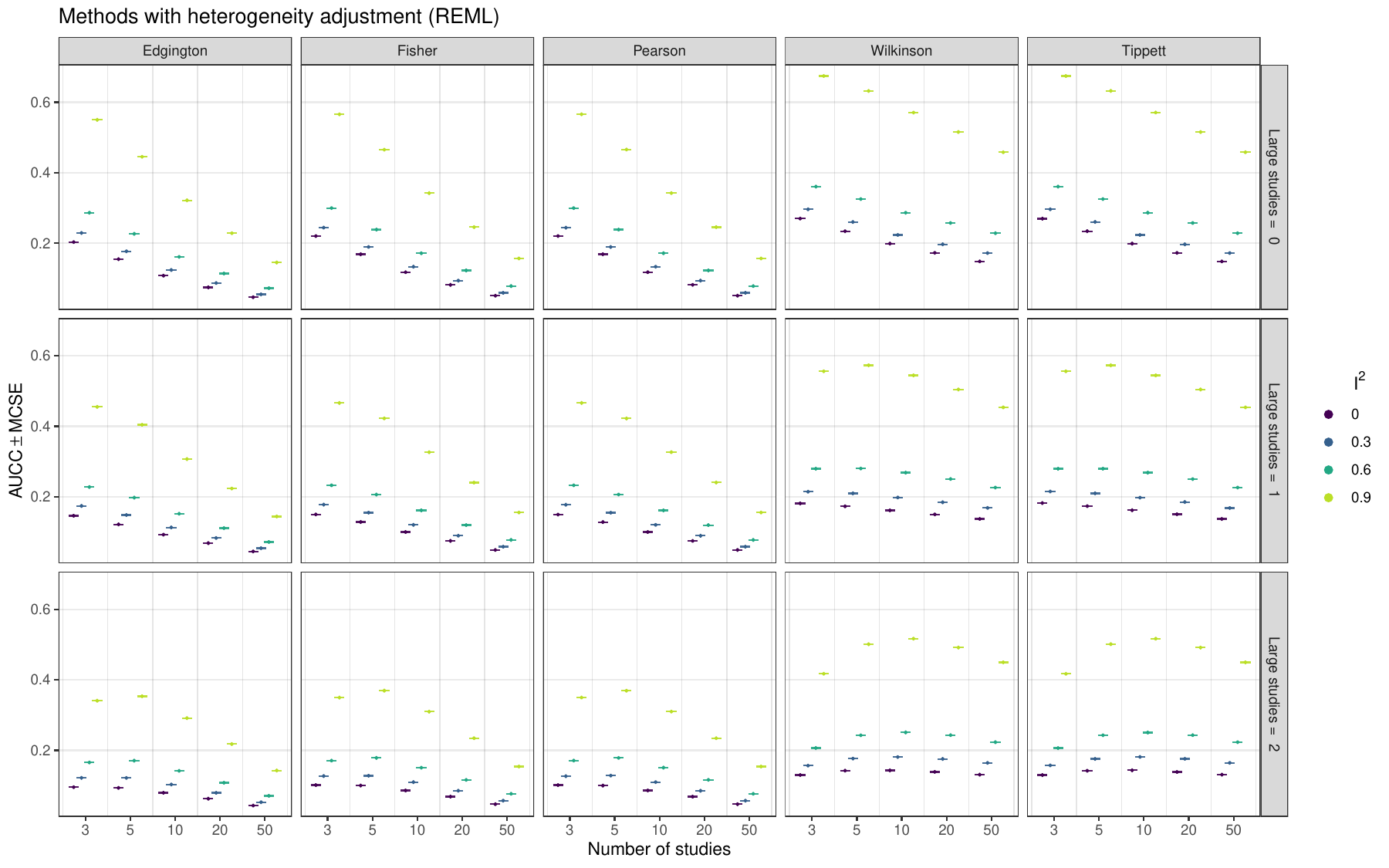} 
\end{knitrout}
\caption{Mean area under the confidence curve (AUCC) based on 20'000 simulation repetitions.}
\label{fig:AUCC}
\end{figure}

\begin{figure}[!htb]
\begin{knitrout}
\definecolor{shadecolor}{rgb}{0.969, 0.969, 0.969}\color{fgcolor}
\includegraphics[width=\maxwidth]{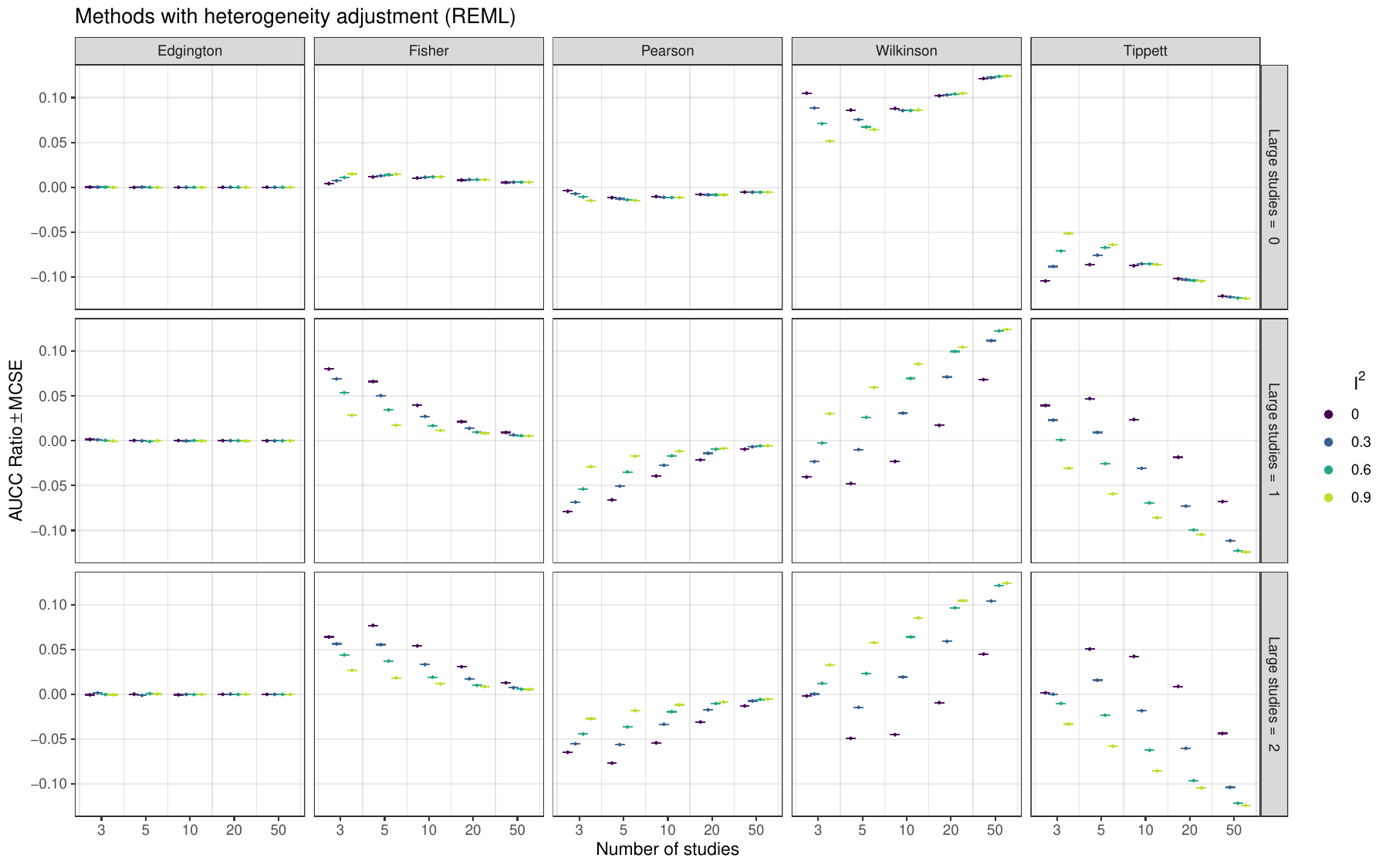} 
\end{knitrout}
\caption{Mean area under the confidence curve (AUCC) ratio based on 20'000 simulation repetitions.}
\label{fig:AUCCratio}
\end{figure}

\begin{figure}[!htb]
\begin{knitrout}
\definecolor{shadecolor}{rgb}{0.969, 0.969, 0.969}\color{fgcolor}
\includegraphics[width=\maxwidth]{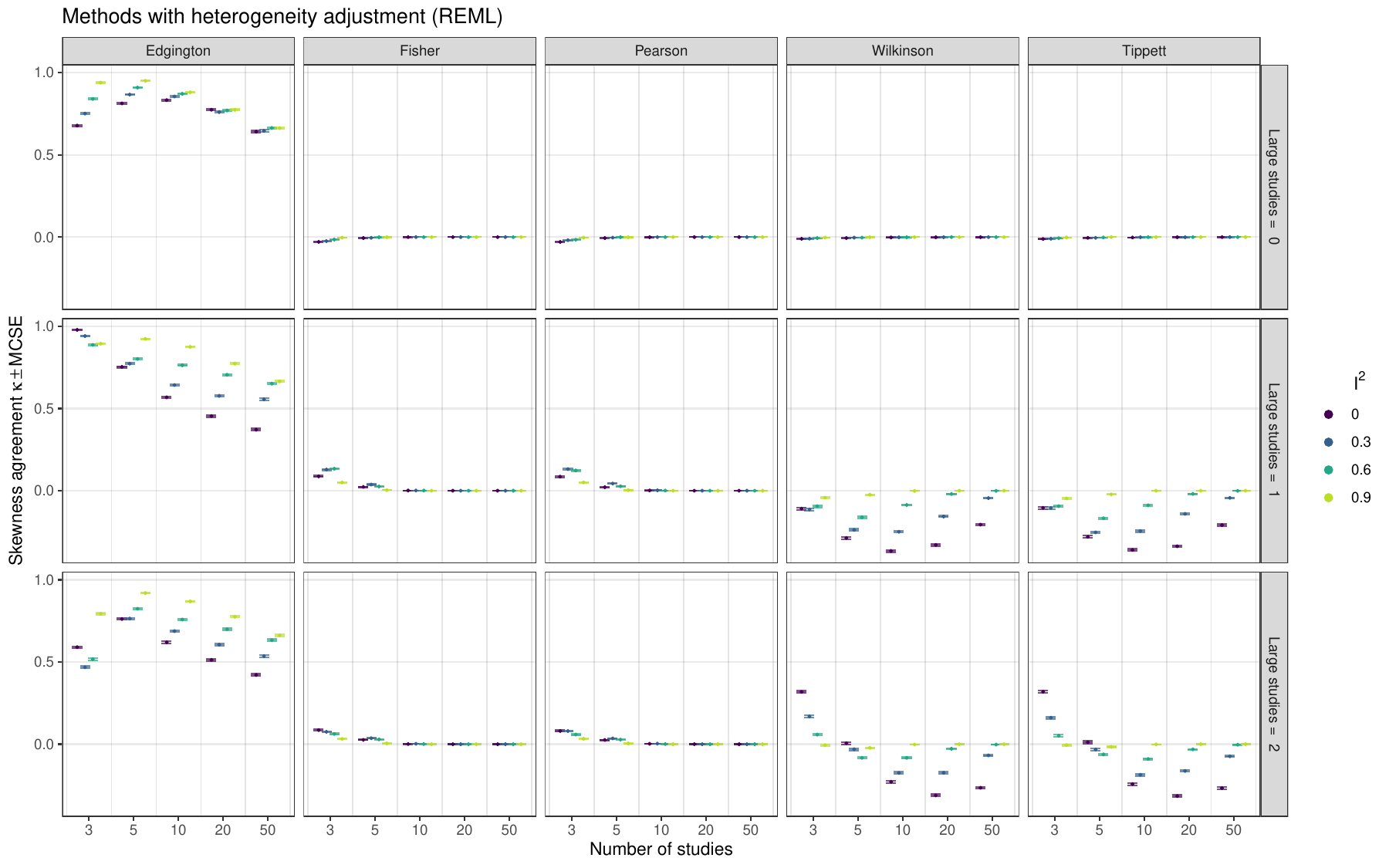} 
\end{knitrout}
\caption{Cohen's $\kappa$ sign agreement between AUCC ratio skewness and data
  skewness based on 20'000 simulation
  repetitions.}
\label{fig:kappa-auccratio}
\end{figure}

\begin{figure}[!htb]
\begin{knitrout}
\definecolor{shadecolor}{rgb}{0.969, 0.969, 0.969}\color{fgcolor}
\includegraphics[width=\maxwidth]{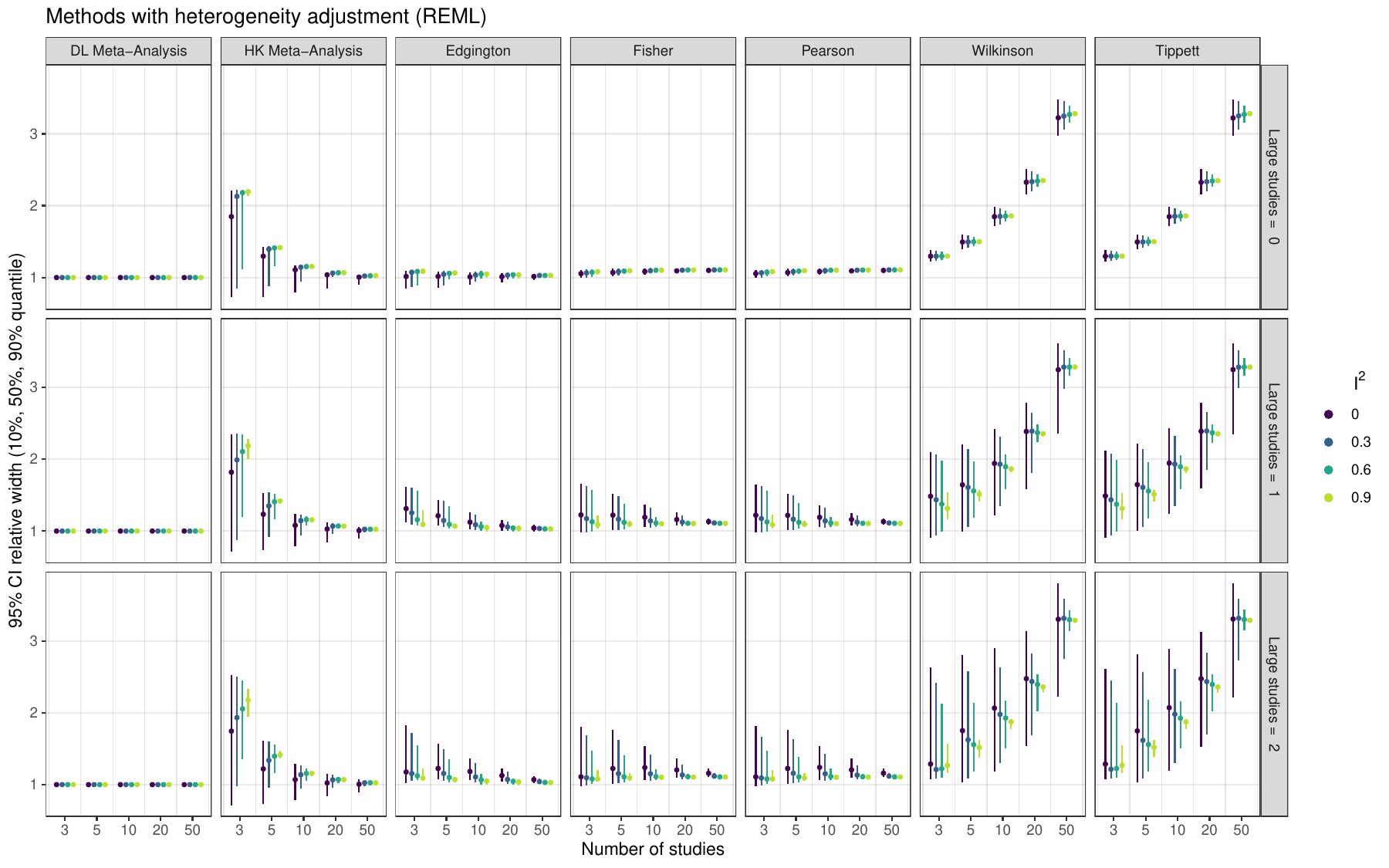} 
\end{knitrout}
\caption{Relative width of 95\% confidence intervals (relative to random effects
  meta-analysis) based on 20'000 simulation
  repetitions.}
\label{fig:relwidth}
\end{figure}

\begin{figure}[!htb]
\begin{knitrout}
\definecolor{shadecolor}{rgb}{0.969, 0.969, 0.969}\color{fgcolor}
\includegraphics[width=\maxwidth]{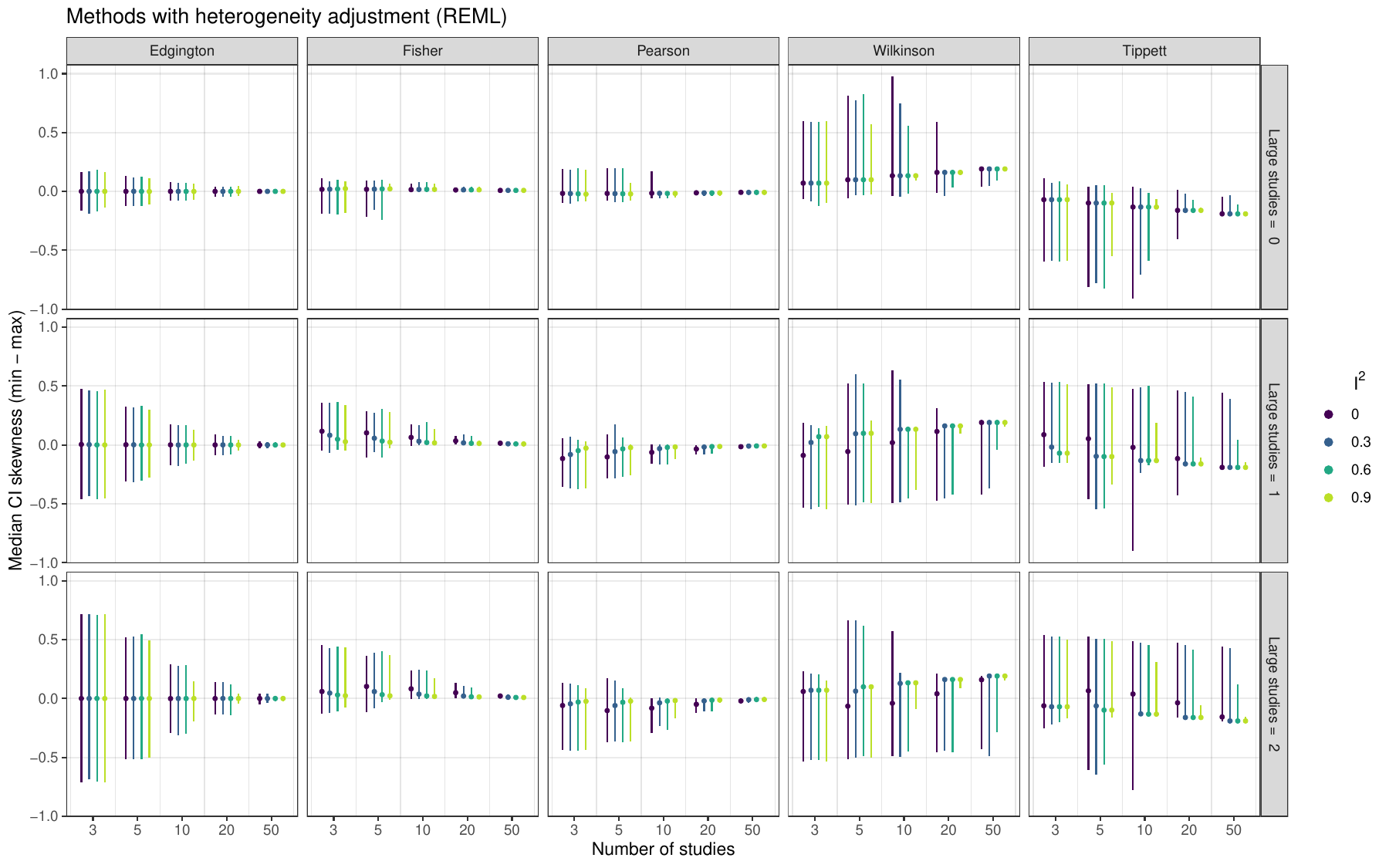} 
\end{knitrout}
\caption{Skewness of 95\% confidence interval based on 20'000 simulation repetitions.}
\label{fig:ciskew}
\end{figure}

\begin{figure}[!htb]
\begin{knitrout}
\definecolor{shadecolor}{rgb}{0.969, 0.969, 0.969}\color{fgcolor}
\includegraphics[width=\maxwidth]{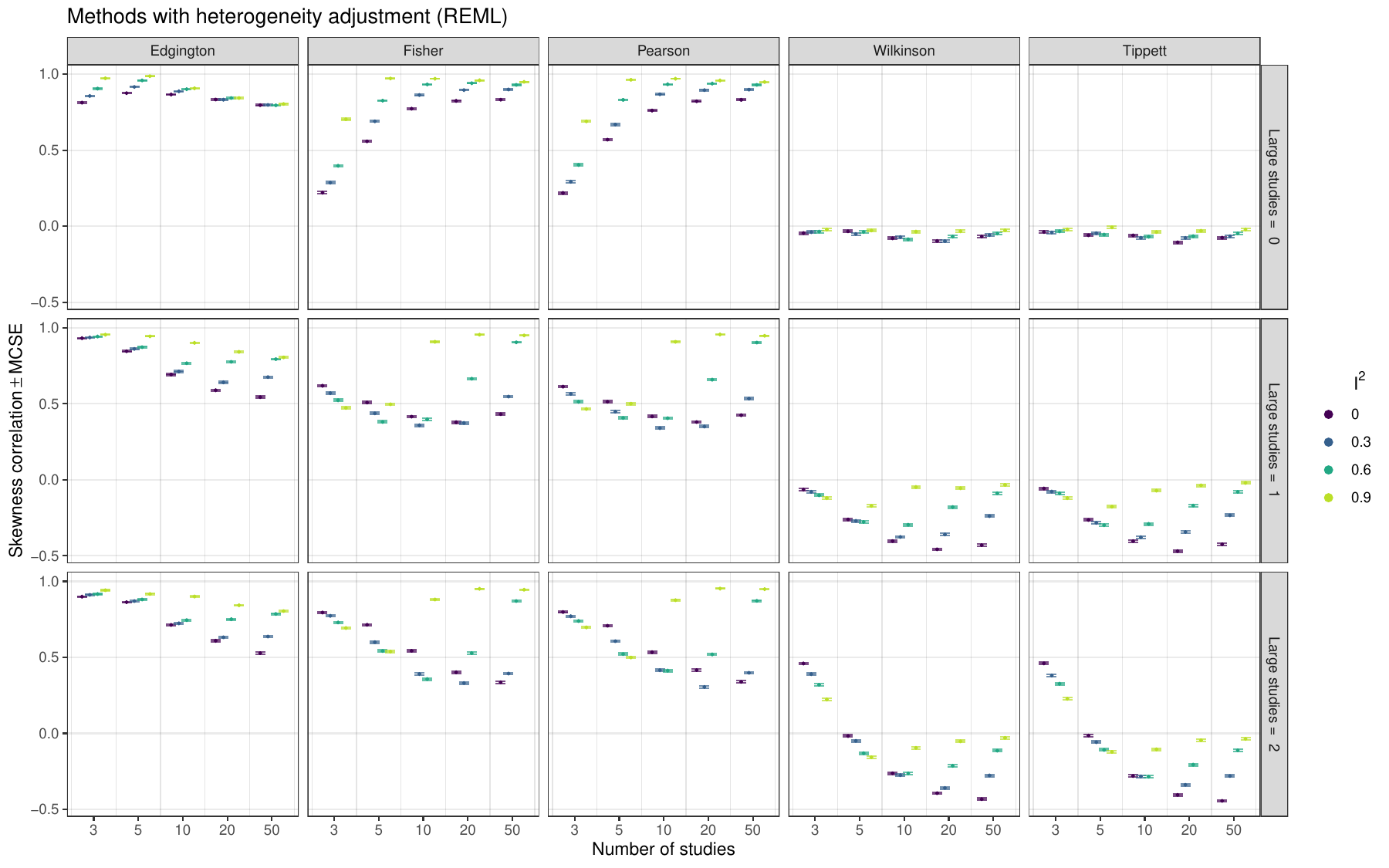} 
\end{knitrout}
\caption{Correlation between 95\% confidence interval skewness and data skewness
  based on 20'000 simulation repetitions.}
\label{fig:correlation}
\end{figure}

\end{landscape}

\section*{Simulations with skew normal study effect distribution}
\label{app:additionalsimSkewnormal}

We also considered the performance of the different methods (adjusted for
heterogeneity) under model misspecification, where the study effects
distribution is non-normal. Specifically, we considered a skew normal
distribution \citep{Azzalini2013} $\theta_{i} \sim \mathrm{SN}(\xi, \omega,
\alpha)$ with location $\xi = \theta - \omega \delta \sqrt{2/\pi}$, scale
$\omega = \tau/\sqrt{1 - 2 \delta^{2}/\pi}$, and shape parameter $\alpha$,
alternatively parameterized as $\delta = \alpha/\sqrt{1 + \alpha^{2}}$. The
location and scale were chosen so that the expectation and variance of the
distribution are $\theta$ and $\tau^{2}$, respectively. The skew normal
distribution reduces to the normal distribution when $\alpha = 0$, we considered
scenarios with $\alpha = 8$ (right skewed) or $\alpha = -8$ (left skewed), see
Figure~\ref{fig:skewnormal} for an illustration. As in
\citet{Kontopantelis_etal2012} and \citet{Weber_etal2021}, simulation from the
skew normal distribution aims to study the effects of model misspecification on
the different methods. We note that \citet{Weber_etal2021} use a different
parametrisation of the skew normal distribution.

\begin{figure}[!htb]
\begin{knitrout}
\definecolor{shadecolor}{rgb}{0.969, 0.969, 0.969}\color{fgcolor}
\includegraphics[width=\maxwidth]{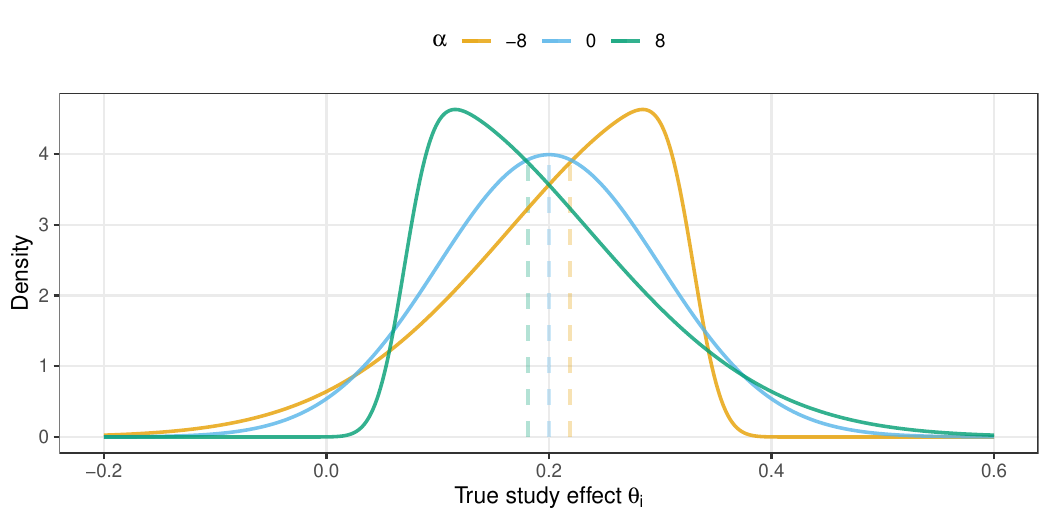} 
\end{knitrout}
\caption{Illustration of skew normal distribution of true study effects
  $\theta_{i}$ with mean \mbox{$\theta = 0.2$} and variance
  \mbox{$\tau^{2} = 0.1^{2}$} for different shape parameters $\alpha$.
  A left-skewed distribution is obtained for $\alpha=-8$, a right-skew distribution for
  $\alpha=8$.
  The dashed lines depict the median of the corresponding distribution.}
\label{fig:skewnormal}
\end{figure}

Under the assumption that the true study effects are generated from a skewed
distribution, it is no longer clear whether the mean or another measure of
central tendency, such as the median, should be of primary interest as the mean
becomes more ``atypical'' with increasing skewness. The median may even be
preferable because it is more robust to outliers than the mean. Therefore, we
considered both the mean and the median of the true study effect distribution as
estimands of interest. For a skew normal study effect distribution, the mean is
$\theta = 0.2$ while the median depends on the heterogeneity and skewness, and
was computed numerically using the function \texttt{qsn(p = 0.5, ....)} from the
\texttt{sn} R package \citep{sn2023}.

\subsection*{Results}

Table~\ref{tab:summaryaddresults-skewed} shows the location of the simulation
results based on the scenarios where the true study effects were simulated from
left or right skew normal distributions. The results for confidence interval
width and skewness were generally similar to those from the symmetric normal
scenarios, while the results for bias and coverage showed some differences which
we will now discuss.

\begin{table}[!h]
  \centering
      \caption{Additional simulation results with skew normal study effect
        distribution (all methods with heterogeneity adjustment using REML).}
      \label{tab:summaryaddresults-skewed}
  \small
  \begin{tabular}{l c c}
    \toprule
    \textbf{Result} & \textbf{Left-skewed} & \textbf{Right-skewed} \\
    \midrule
    Non-convergence & Table~\ref{tab:convergence-skewed} & Table~\ref{tab:convergence-skewed} \\
    CI Coverage (mean) & Figure~\ref{fig:covmeanleft} & Figure~\ref{fig:covmeanright} \\
    CI Coverage (median) & Figure~\ref{fig:covmedleft} & Figure~\ref{fig:covmedright} \\
    Bias (mean) & Figure~\ref{fig:biasmeanleft} & Figure~\ref{fig:biasmeanright} \\
    Bias (median) & Figure~\ref{fig:biasmedleft} & Figure~\ref{fig:biasmedright} \\
    CI width & Figure~\ref{fig:widthleft} & Figure~\ref{fig:widthright} \\
    AUCC & Figure~\ref{fig:AUCCleft} & Figure~\ref{fig:AUCCright} \\
    Relative CI width & Figure~\ref{fig:relwidthleft} & Figure~\ref{fig:relwidthright} \\
    Median CI skewness & Figure~\ref{fig:skewleft} & Figure~\ref{fig:skewright} \\
    AUCC ratio & Figure~\ref{fig:AUCCratioleft} & Figure~\ref{fig:AUCCratioright} \\
    CI skewness agreement & Figure~\ref{fig:kappaleft} & Figure~\ref{fig:kapparight} \\
    AUCC ratio skewness agreement & Figure~\ref{fig:kappa-auccratioleft} & Figure~\ref{fig:kappa-auccratioright} \\
    CI skewness correlation & Figure~\ref{fig:corleft} & Figure~\ref{fig:corright} \\
    %% AUCC ratio for Edgington & Figure~\ref{fig:CIauccrEdgingtonleft} & Figure~\ref{fig:CIauccrEdgington} \\
    %% Mean CI skewness for Edgington & Figure~\ref{fig:CIskewEdgingtonleft} & Figure~\ref{fig:CIskewEdgingtonright} \\
    \bottomrule
  \end{tabular}
\end{table}

\paragraph{Bias}
As discussed earlier, we considered two estimands in the skew normal conditions
-- the mean and the median of the true effect distribution. Random effects
meta-analysis and the Hartung-Knapp method were unbiased for the mean effect
(Figures~\ref{fig:biasmeanleft} and~\ref{fig:biasmeanright}) but biased for the
median effect (Figures~\ref{fig:biasmedleft} and~\ref{fig:biasmedright}), with
the amount of bias increasing with increasing $I^{2}$. Among the $p$-value
combination methods, Wilkinson's and Tippett's methods were substantially biased
for both the mean and median true effect, while Edgington's, Fisher's, and
Pearson's methods were also biased, but to a lesser extent. Interestingly,
Edgington's method appears to be positively biased for the mean effect and
negatively biased for the median effect (for the left-skew condition, and vice
versa for the right-skew condition), thus implicitly targeting an estimand
somewhere in between the mean and median of the skew normal distributions.
Furthermore, the bias of Edgington's method for the median effect was lower than
the bias of the random effects meta-analysis and Hartung-Knapp methods.

\paragraph{Coverage}
The random effects meta-analysis and the Hartung-Knapp method showed similar
patterns of coverage for the mean effect as under symmetric-study effects
conditions (Figures~\ref{fig:covmeanleft} and~\ref{fig:covmeanright}), but
different patterns for the median effect (Figures~\ref{fig:covmedleft}
and~\ref{fig:covmedright}). For the latter, their coverage did not approach the
nominal 95\% coverage as the number of studies increased, but actually worsened
after an initial increase. This makes sense since both methods are
targeting the mean rather than the median true effect, so their confidence
intervals become more concentrated around the mean effect with increasing number
of studies. Edgington's method showed comparable or better coverage for the mean
effect than random effects meta-analysis when $I^{2}$ was not too high, while it
generally showed better (but not nominal) coverage for the median true effect.
The remaining $p$-value combination methods seem to increase to too high
coverage for the mean effect with increasing number of studies, and show
coverage values all over the place for the median effect.

\paragraph{Summary}
Surprisingly, the performance of Edgington's method was worse when study effects
were simulated from a skewed distribution, possibly because the method targets
neither the mean nor the median of the distribution, but something in between.
In addition, our study showed that when the true study effects are simulated
from a skewed distribution and the estimand is the mean study effect, both
random effects meta-analysis and Hartung Knapp seem to be unbiased and show
similar coverage patterns as for symmetric study effect distributions, but when
the estimand of interest is the median study effect, both methods become biased
and their coverage becomes worse.

\begin{table}[!htb]
  \centering
  \caption{Simulation conditions for which convergence rate was not 100\% (i.e.,
    some methods did not produce a confidence interval in some repetitions).}
  \label{tab:convergence-skewed}

% latex table generated in R 4.4.2 by xtable 1.8-4 package
% Fri Feb 21 13:23:20 2025
\begin{tabular}{cccccc}
  \toprule
Convergence rate & Method & $I^2$ & $k$ & Distribution & Large studies \\ 
  \midrule
96.540\% & Pearson & 0.90 &  20 & Left-skewed &   2 \\ 
  96.874\% & Pearson & 0.90 &  50 & Left-skewed &   2 \\ 
  97.649\% & Pearson & 0.90 &  50 & Left-skewed &   1 \\ 
  98.482\% & Pearson & 0.90 &  20 & Left-skewed &   1 \\ 
  99.185\% & Pearson & 0.90 &   5 & Left-skewed &   2 \\ 
  99.215\% & Pearson & 0.90 &   5 & Left-skewed &   1 \\ 
  99.455\% & Pearson & 0.90 &  50 & Left-skewed &   0 \\ 
  99.650\% & Pearson & 0.90 &  10 & Left-skewed &   2 \\ 
  99.710\% & Wilkinson & 0.90 &  10 & Left-skewed &   2 \\ 
  99.725\% & Pearson & 0.90 &  10 & Left-skewed &   1 \\ 
  99.735\% & Wilkinson & 0.90 &  10 & Left-skewed &   1 \\ 
  99.820\% & Wilkinson & 0.90 &  20 & Left-skewed &   2 \\ 
  99.920\% & Wilkinson & 0.90 &  20 & Left-skewed &   1 \\ 
  99.970\% & Wilkinson & 0.90 &  50 & Left-skewed &   2 \\ 
  99.975\% & Wilkinson & 0.90 &  50 & Left-skewed &   1 \\ 
  99.980\% & Pearson & 0.90 &   5 & Right-skewed &   1 \\ 
  99.990\% & Pearson & 0.60 &  50 & Left-skewed &   2 \\ 
  99.990\% & Pearson & 0.90 &   5 & Right-skewed &   2 \\ 
  99.995\% & Pearson & 0.60 &  50 & Left-skewed &   1 \\ 
   \bottomrule
\end{tabular}

\end{table}

\begin{landscape}

%% additional symmetric results
%% -----------------------------------------------------------------------------

%% left-skewed results
%% -----------------------------------------------------------------------------

\begin{figure}[!htb]
\begin{knitrout}
\definecolor{shadecolor}{rgb}{0.969, 0.969, 0.969}\color{fgcolor}
\includegraphics[width=\maxwidth]{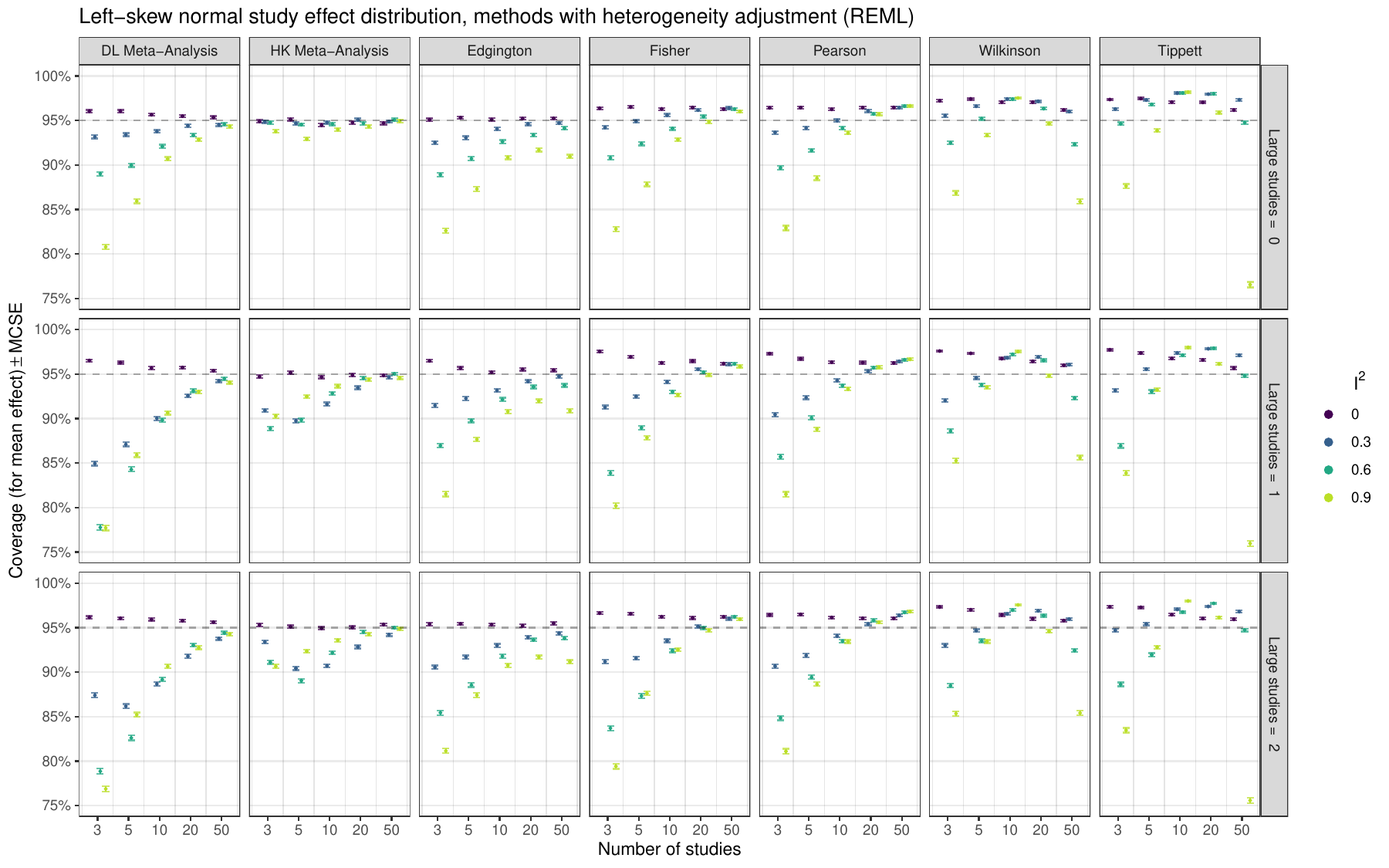} 
\end{knitrout}
\caption{Empirical coverage (for mean effect) of the 95\% confidence intervals
  based on 20'000 simulation repetitions.}
\label{fig:covmeanleft}
\end{figure}
\begin{figure}[!htb]
\begin{knitrout}
\definecolor{shadecolor}{rgb}{0.969, 0.969, 0.969}\color{fgcolor}
\includegraphics[width=\maxwidth]{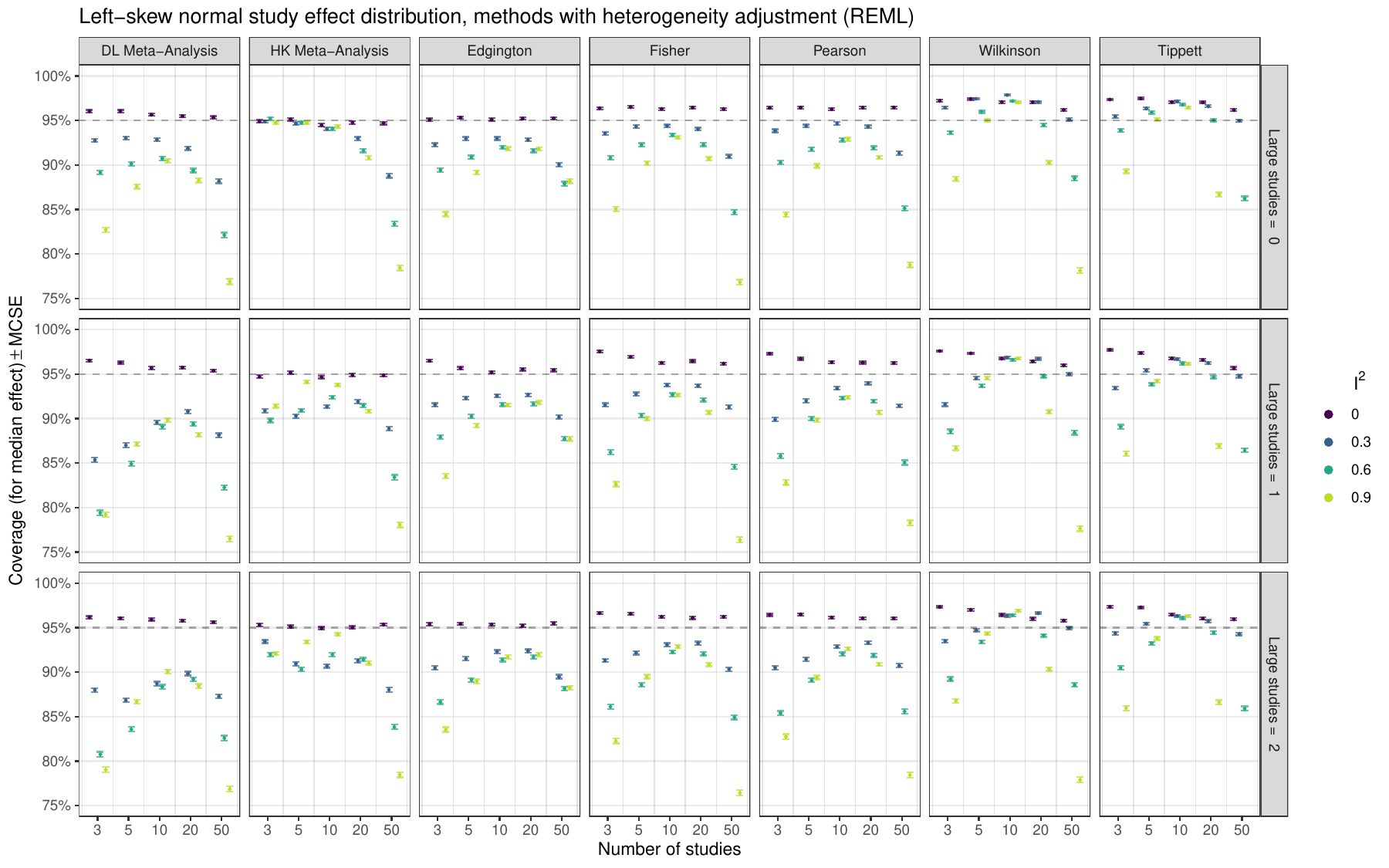} 
\end{knitrout}
\caption{Empirical coverage (for median effect) of the 95\% confidence intervals
  based on 20'000 simulation repetitions.}
\label{fig:covmedleft}
\end{figure}
\begin{figure}[!htb]
\begin{knitrout}
\definecolor{shadecolor}{rgb}{0.969, 0.969, 0.969}\color{fgcolor}
\includegraphics[width=\maxwidth]{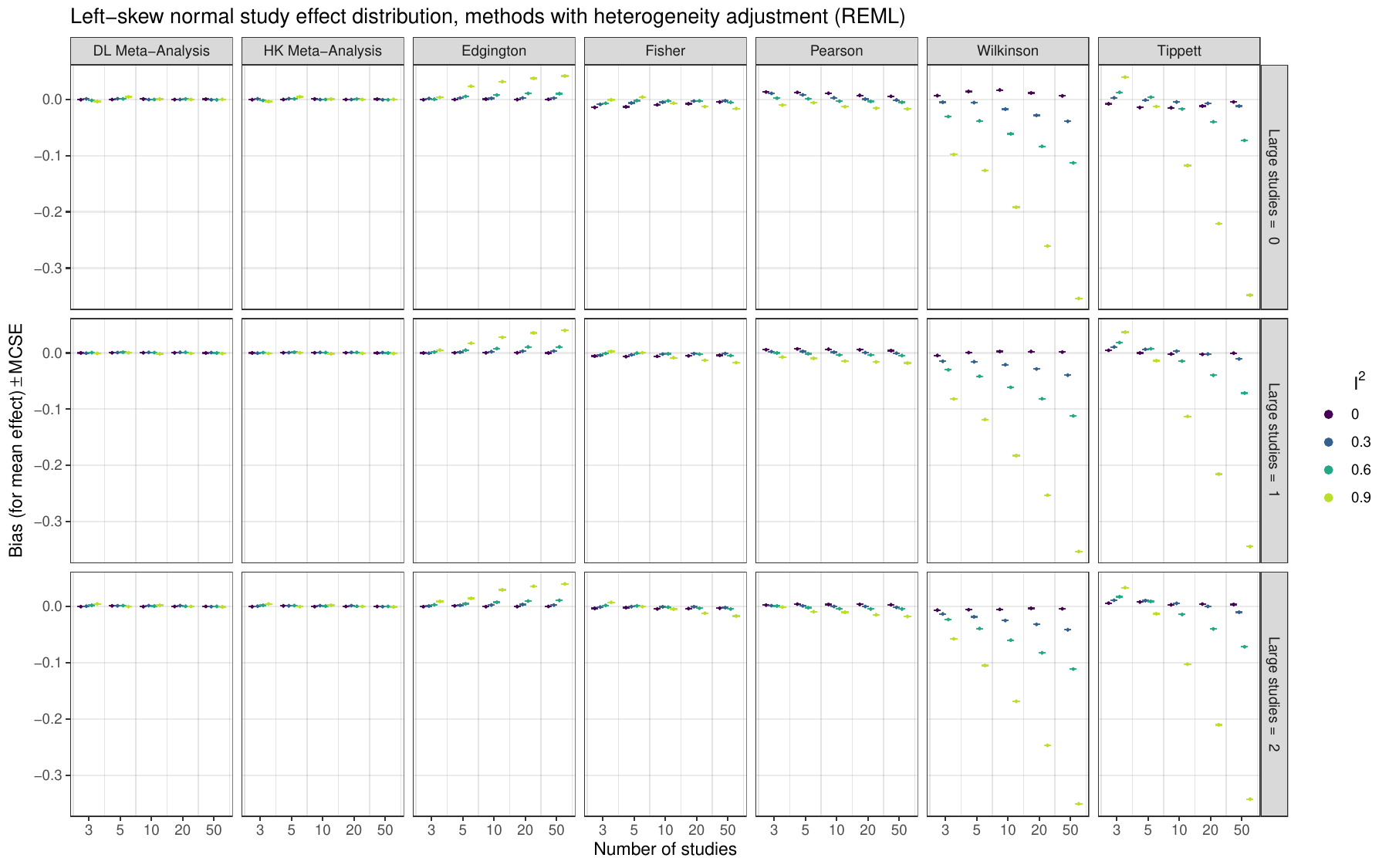} 
\end{knitrout}
\caption{Empirical bias of point estimates (for mean effect) based on
  20'000 simulation repetitions.}
\label{fig:biasmeanleft}
\end{figure}

\begin{figure}[!htb]
\begin{knitrout}
\definecolor{shadecolor}{rgb}{0.969, 0.969, 0.969}\color{fgcolor}
\includegraphics[width=\maxwidth]{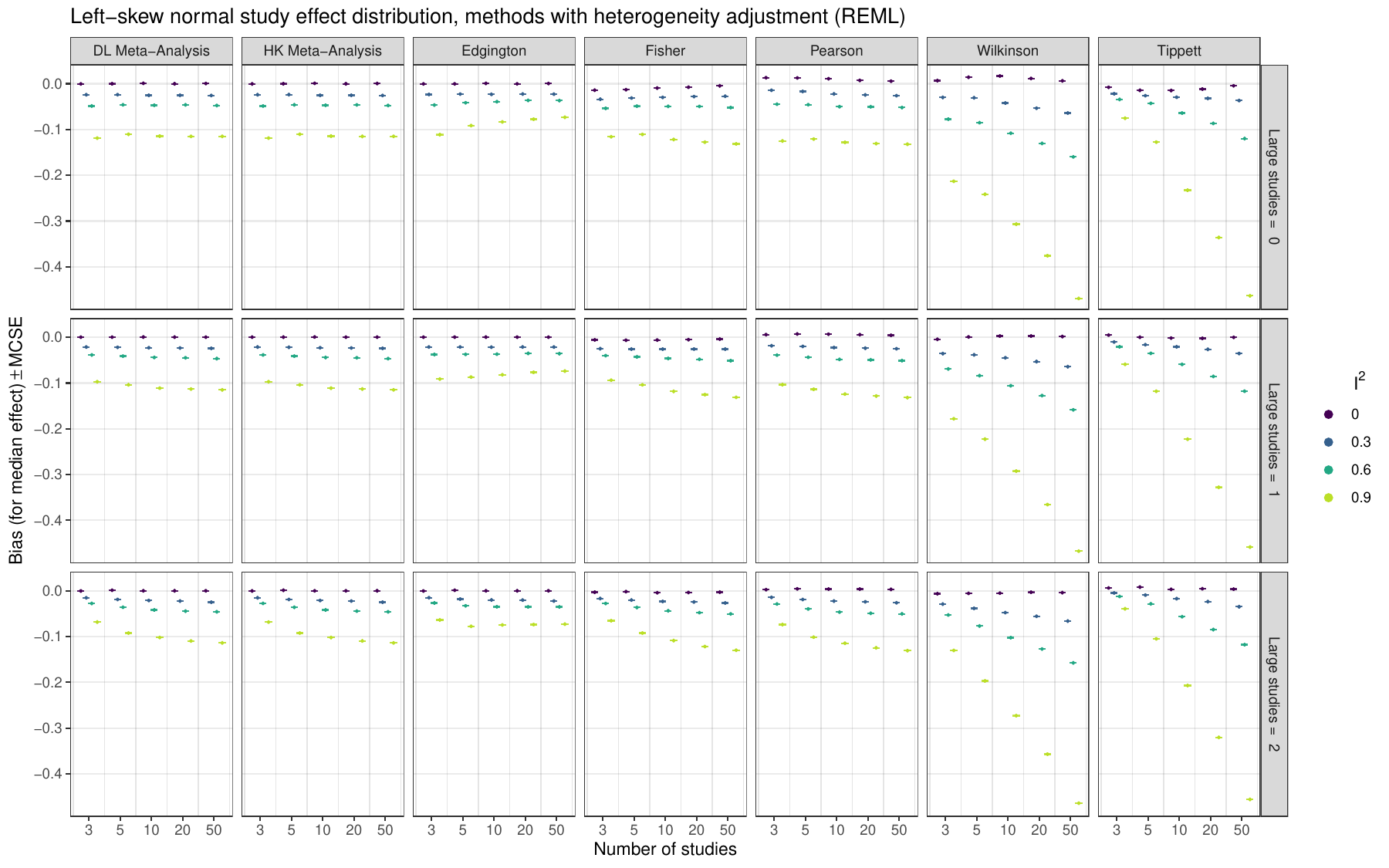} 
\end{knitrout}
\caption{Empirical bias of point estimates (for median effect) based on
  20'000 simulation repetitions.}
\label{fig:biasmedleft}
\end{figure}
\begin{figure}[!htb]
\begin{knitrout}
\definecolor{shadecolor}{rgb}{0.969, 0.969, 0.969}\color{fgcolor}
\includegraphics[width=\maxwidth]{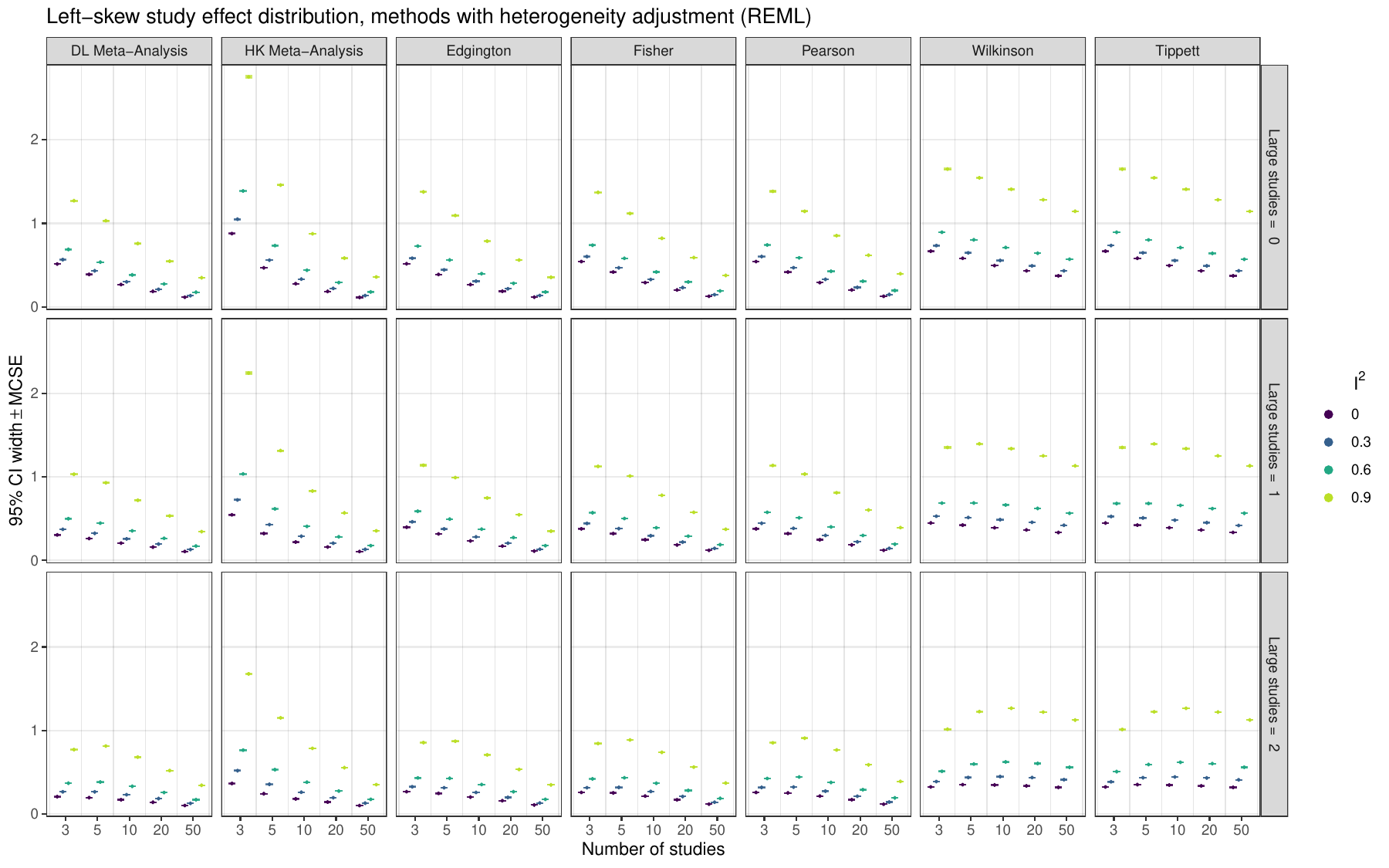} 
\end{knitrout}
\caption{Mean width of 95\% confidence intervals based on 20'000 simulation repetitions.}
\label{fig:widthleft}
\end{figure}
\begin{figure}[!htb]
\begin{knitrout}
\definecolor{shadecolor}{rgb}{0.969, 0.969, 0.969}\color{fgcolor}
\includegraphics[width=\maxwidth]{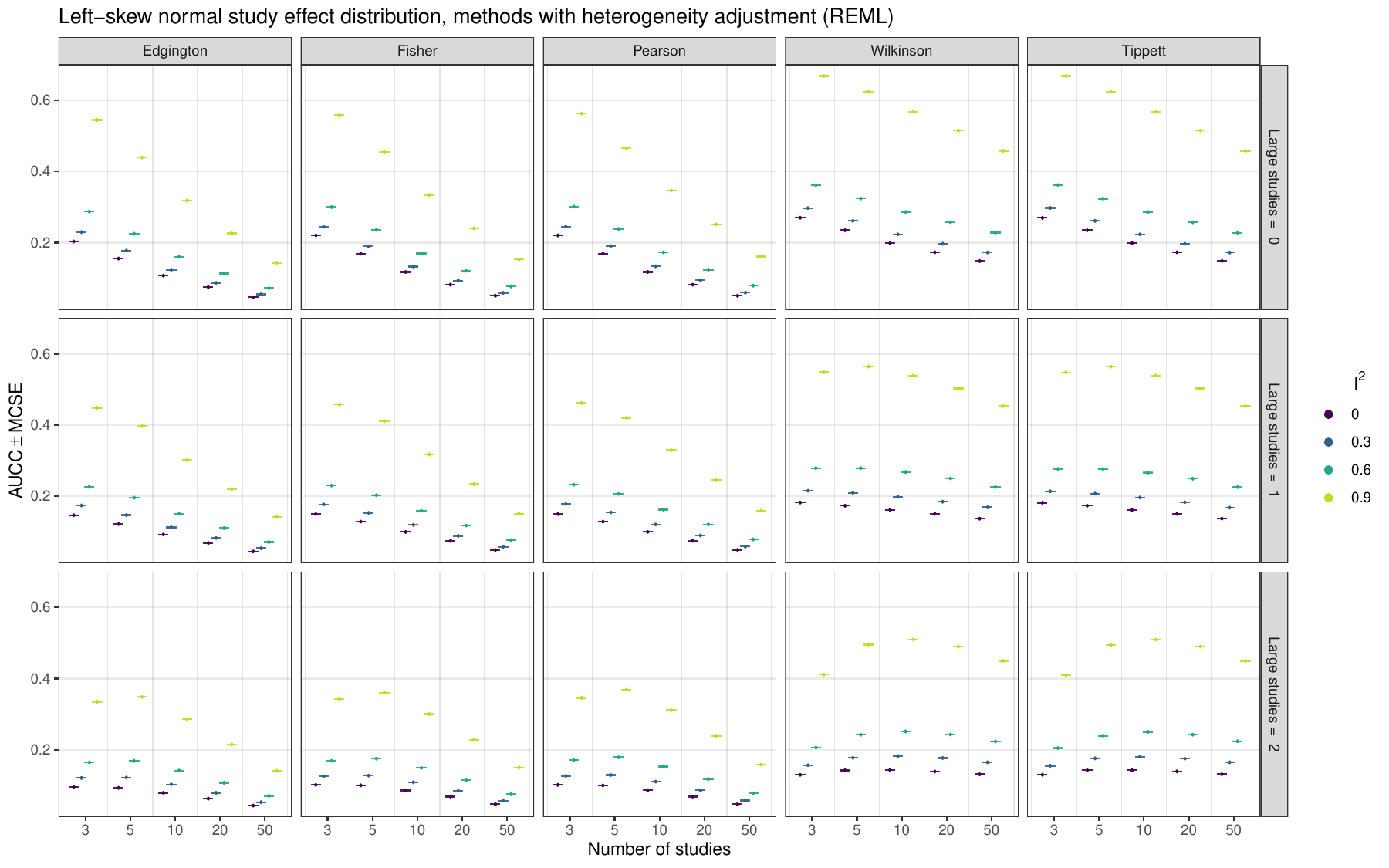} 
\end{knitrout}
\caption{Mean area under the confidence curve (AUCC) based on 20'000 simulation repetitions.}
\label{fig:AUCCleft}
\end{figure}

\begin{figure}[!htb]
\begin{knitrout}
\definecolor{shadecolor}{rgb}{0.969, 0.969, 0.969}\color{fgcolor}
\includegraphics[width=\maxwidth]{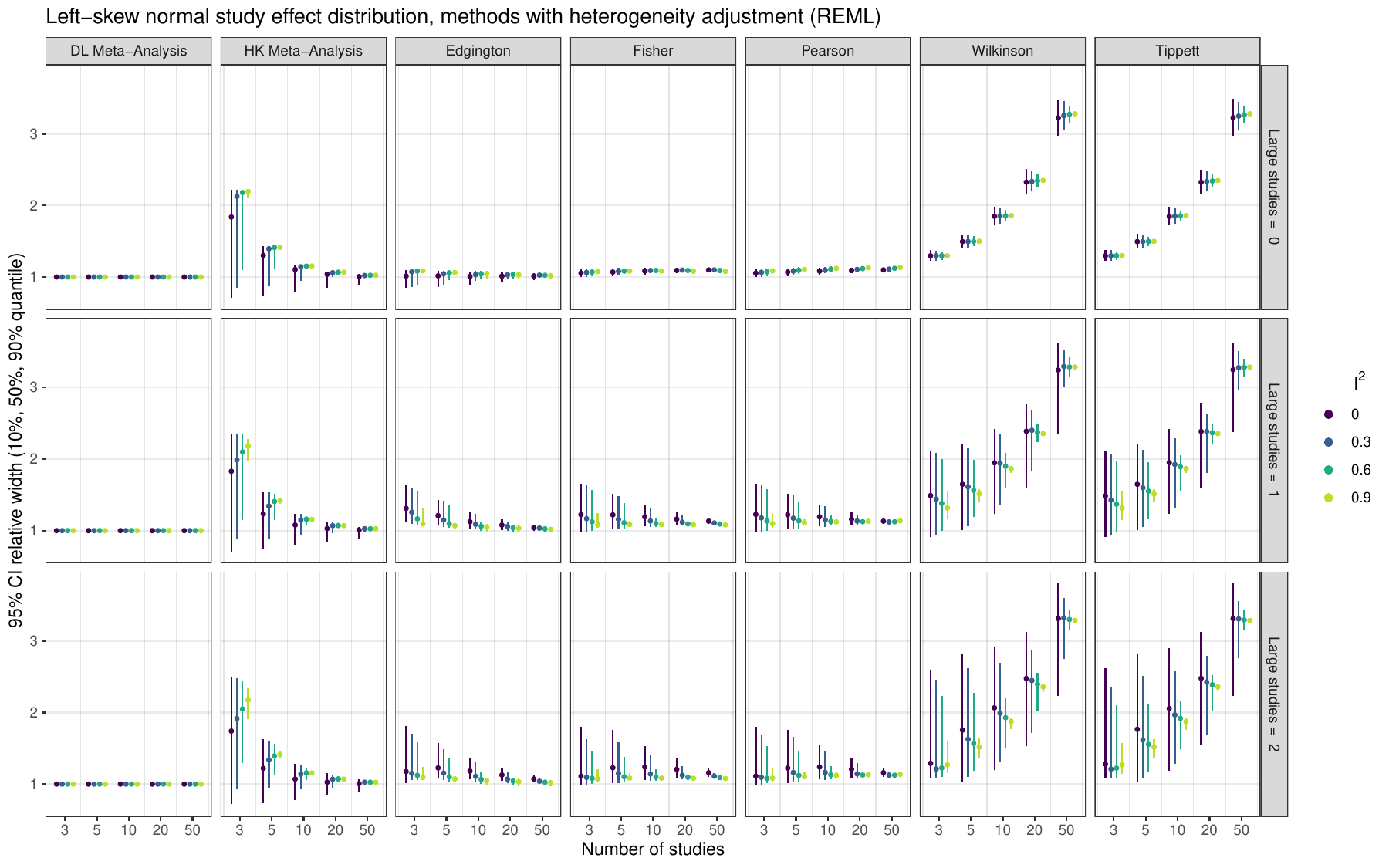} 
\end{knitrout}
\caption{Relative width of 95\% confidence intervals (relative to random effects
  meta-analysis) based on 20'000 simulation
  repetitions.}
\label{fig:relwidthleft}
\end{figure}

\begin{figure}[!htb]
\begin{knitrout}
\definecolor{shadecolor}{rgb}{0.969, 0.969, 0.969}\color{fgcolor}
\includegraphics[width=\maxwidth]{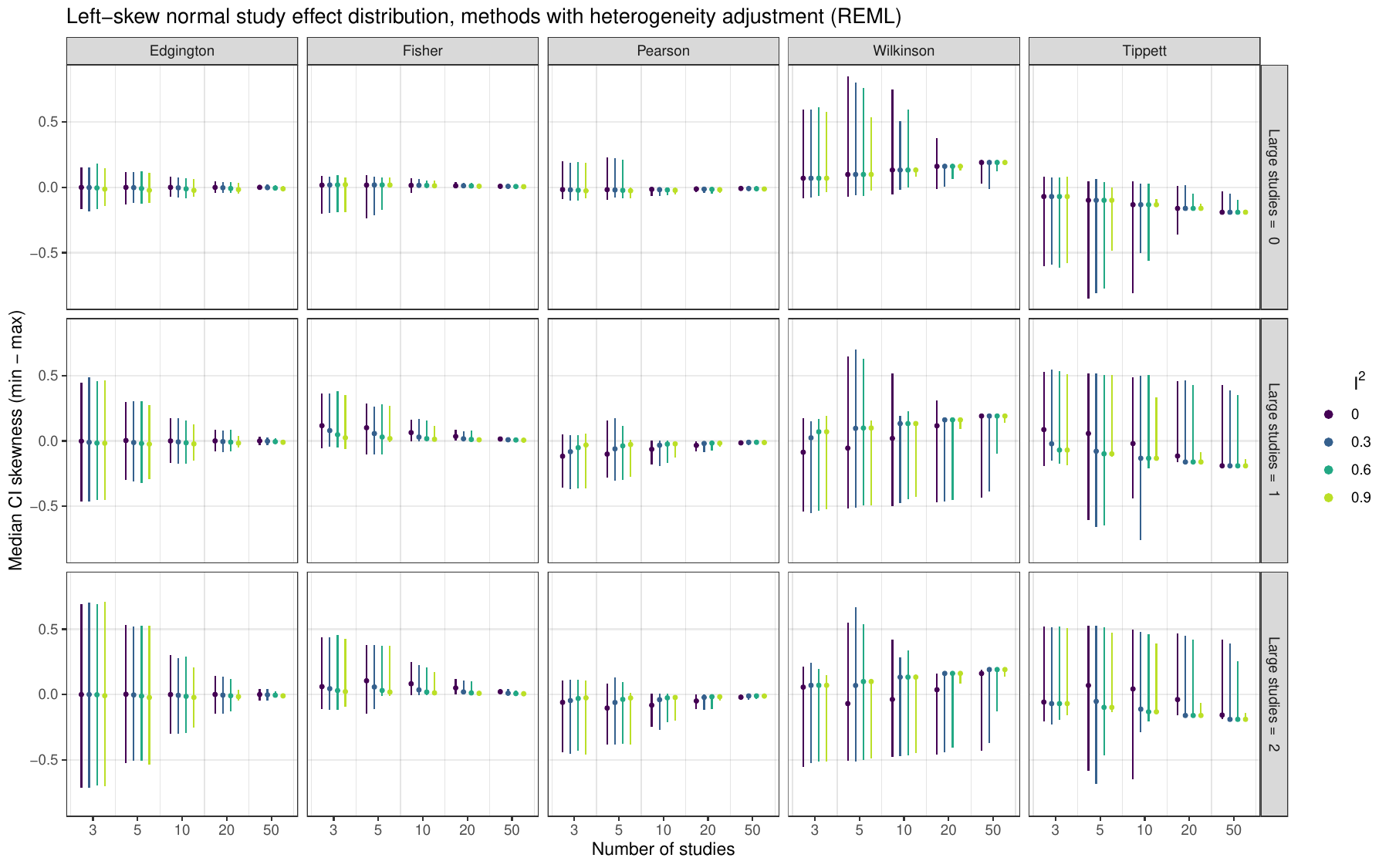} 
\end{knitrout}
\caption{Skewness of 95\% confidence intervals based on 20'000 simulation repetitions.}
\label{fig:skewleft}
\end{figure}
\begin{figure}[!htb]
\begin{knitrout}
\definecolor{shadecolor}{rgb}{0.969, 0.969, 0.969}\color{fgcolor}
\includegraphics[width=\maxwidth]{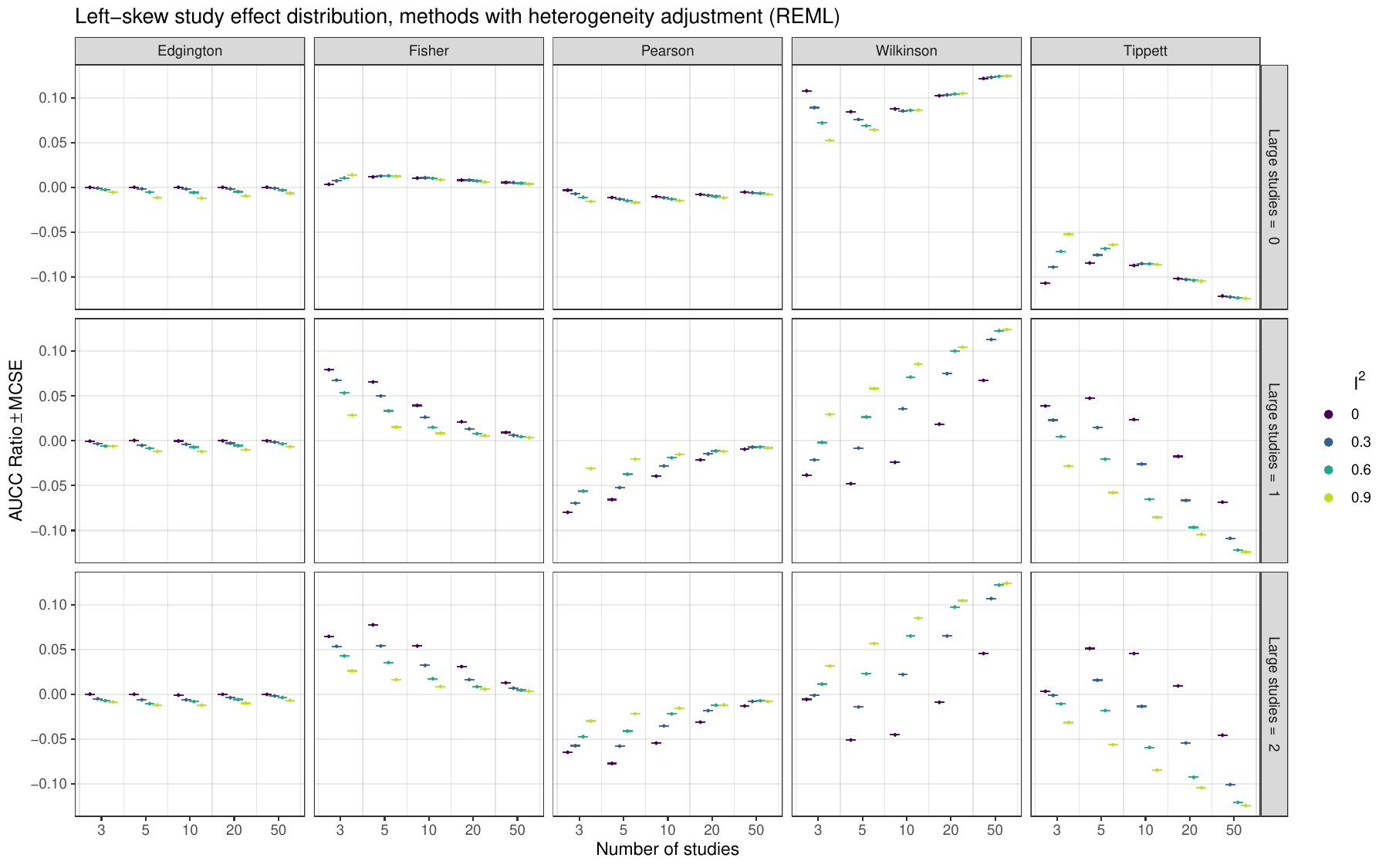} 
\end{knitrout}
\caption{Mean area under the confidence curve (AUCC) ratio based on 20'000 simulation repetitions.}
\label{fig:AUCCratioleft}
\end{figure}

\begin{figure}[!htb]
\begin{knitrout}
\definecolor{shadecolor}{rgb}{0.969, 0.969, 0.969}\color{fgcolor}
\includegraphics[width=\maxwidth]{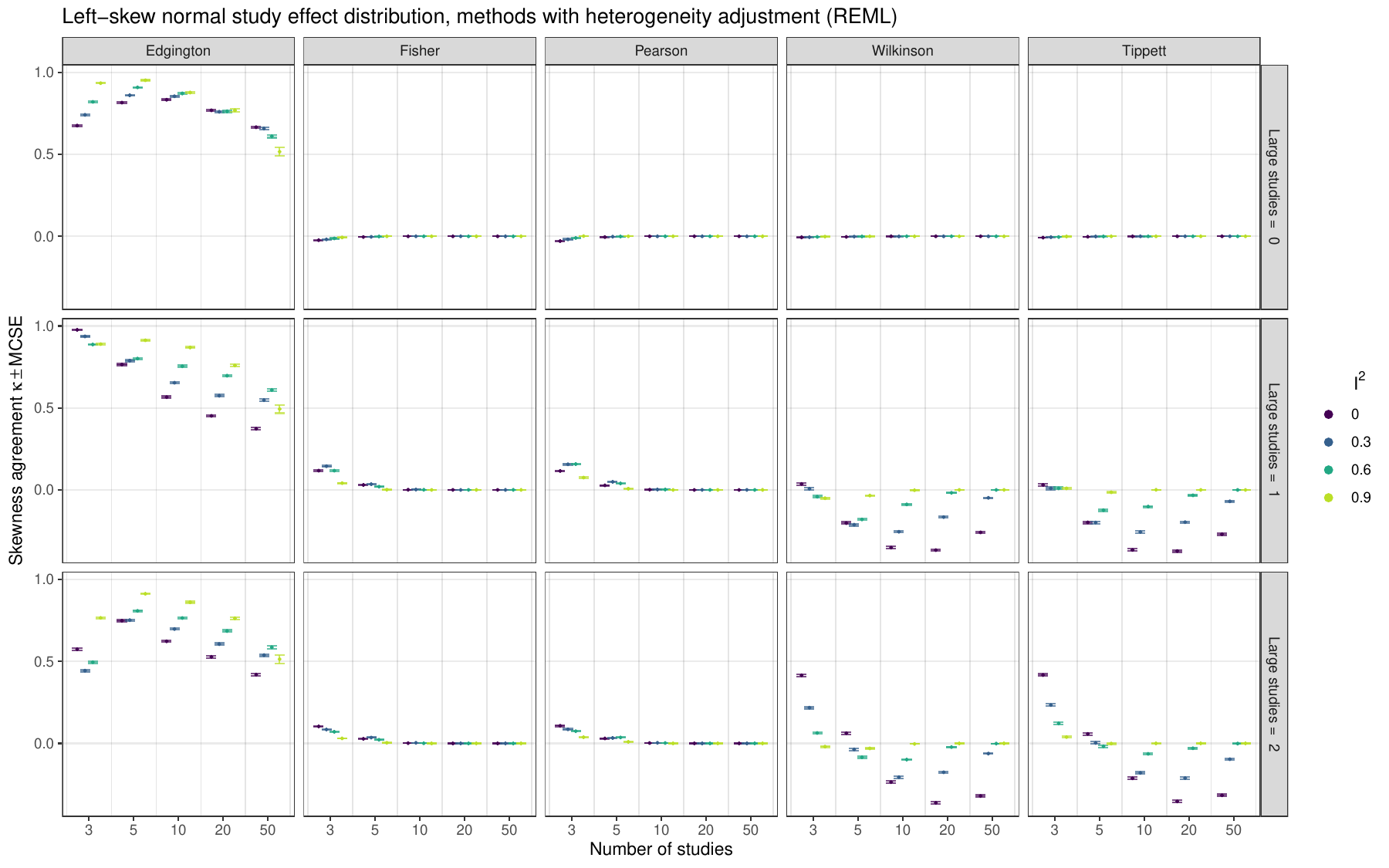} 
\end{knitrout}
\caption{Cohen's $\kappa$ sign agreement between 95\% confidence interval
  skewness and data skewness based on 20'000
  simulation repetitions.}
\label{fig:kappaleft}
\end{figure}

\begin{figure}[!htb]
\begin{knitrout}
\definecolor{shadecolor}{rgb}{0.969, 0.969, 0.969}\color{fgcolor}
\includegraphics[width=\maxwidth]{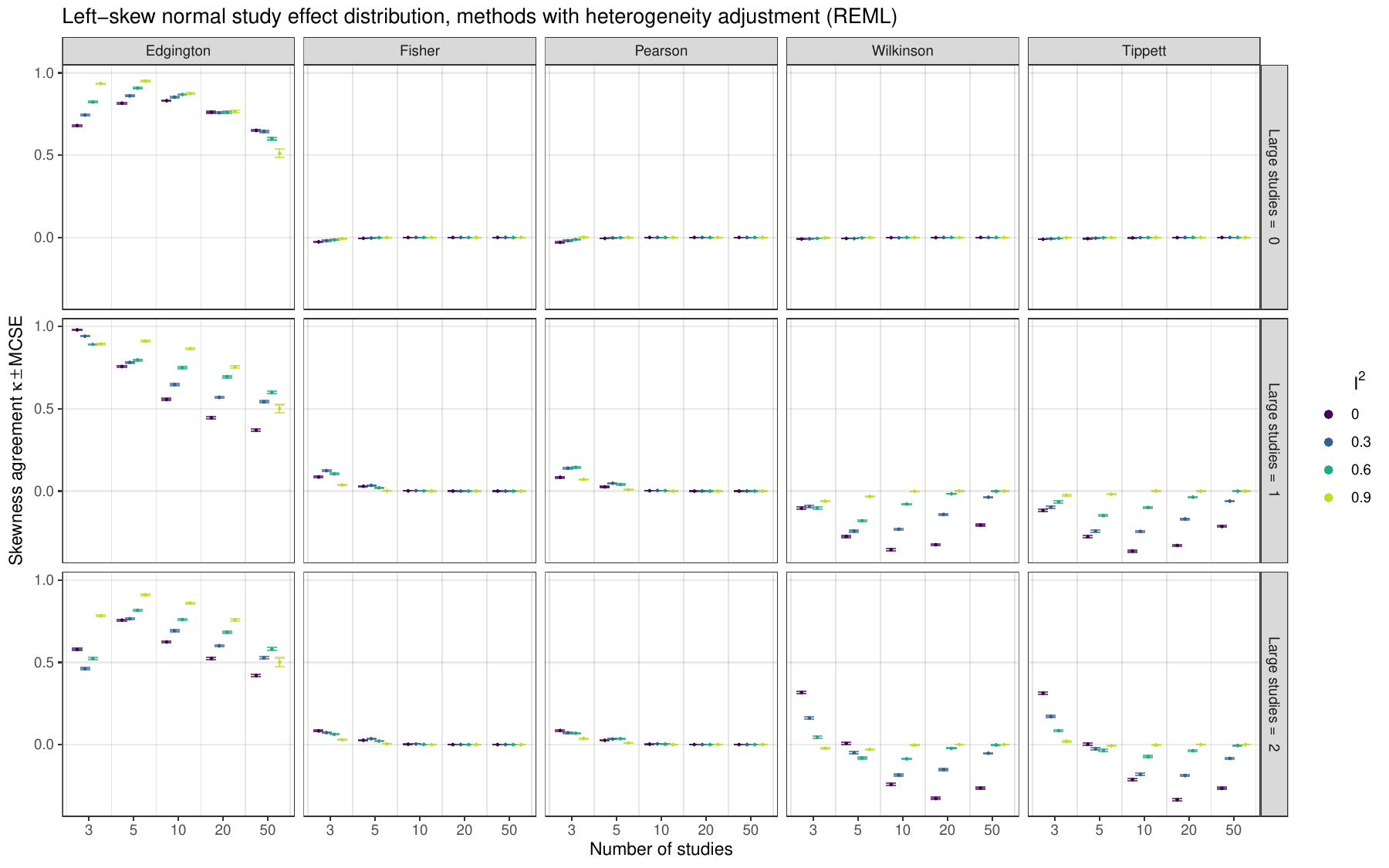} 
\end{knitrout}
\caption{Cohen's $\kappa$ sign agreement between AUCC ratio skewness and data
  skewness based on 20'000 simulation
  repetitions.}
\label{fig:kappa-auccratioleft}
\end{figure}
\begin{figure}[!htb]
\begin{knitrout}
\definecolor{shadecolor}{rgb}{0.969, 0.969, 0.969}\color{fgcolor}
\includegraphics[width=\maxwidth]{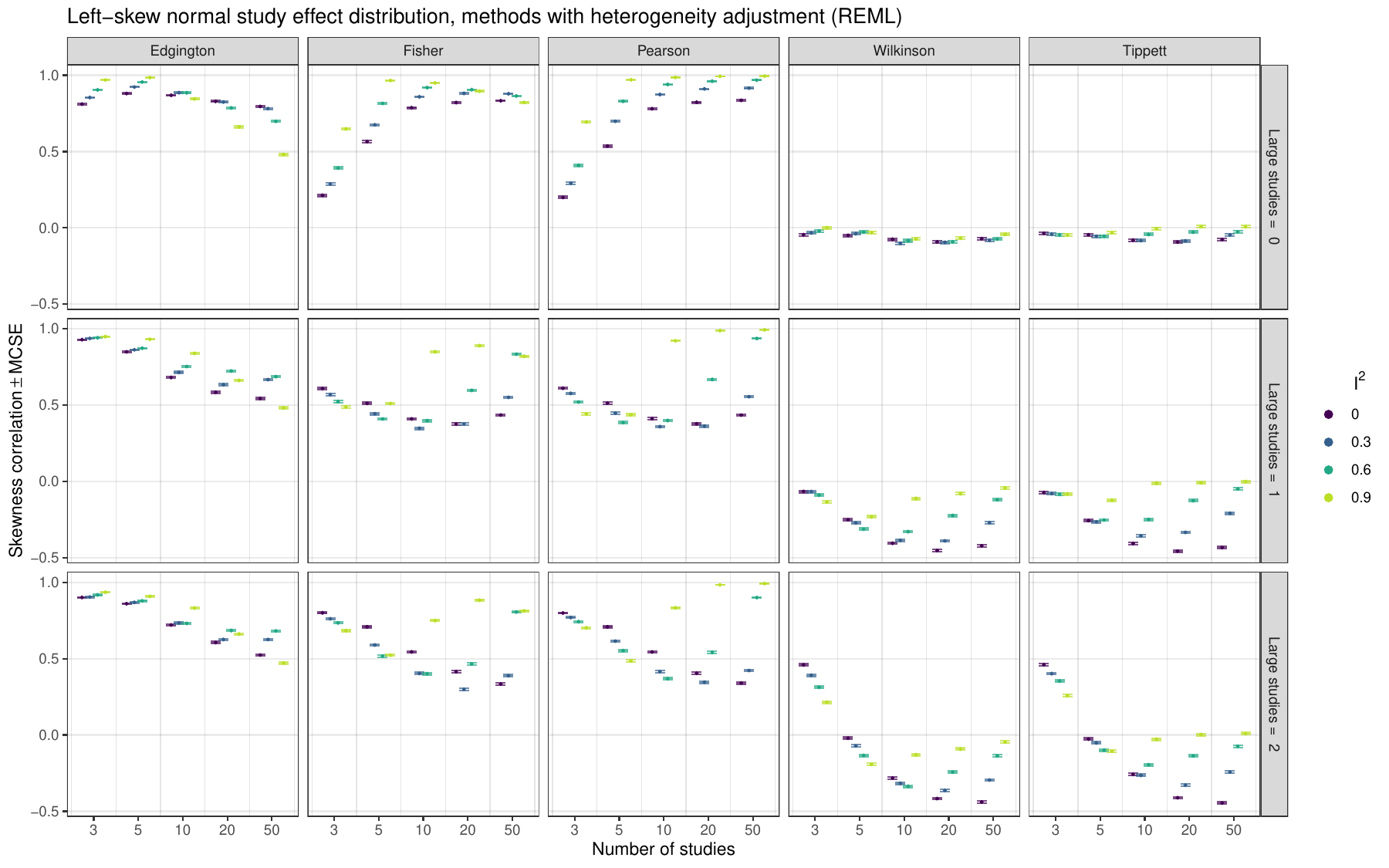} 
\end{knitrout}
\caption{Correlation between 95\% confidence interval skewness and data skewness
  based on 20'000 simulation repetitions.}
\label{fig:corleft}
\end{figure}

%% right-skewed results
%% -----------------------------------------------------------------------------

\begin{figure}[!htb]
\begin{knitrout}
\definecolor{shadecolor}{rgb}{0.969, 0.969, 0.969}\color{fgcolor}
\includegraphics[width=\maxwidth]{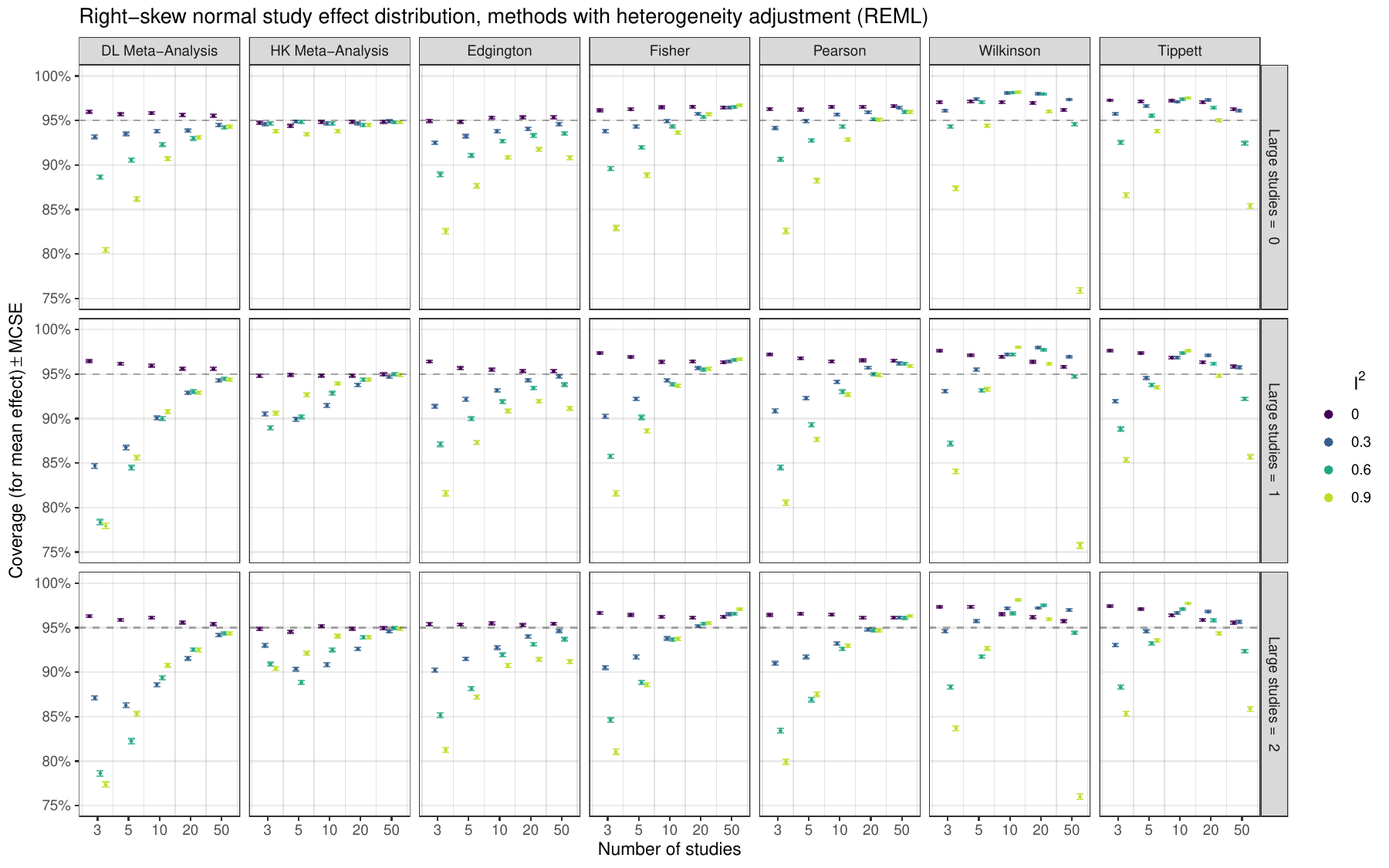} 
\end{knitrout}
\caption{Empirical coverage (for mean effect) of the 95\% confidence intervals
  based on 20'000 simulation repetitions.}
\label{fig:covmeanright}
\end{figure}
\begin{figure}[!htb]
\begin{knitrout}
\definecolor{shadecolor}{rgb}{0.969, 0.969, 0.969}\color{fgcolor}
\includegraphics[width=\maxwidth]{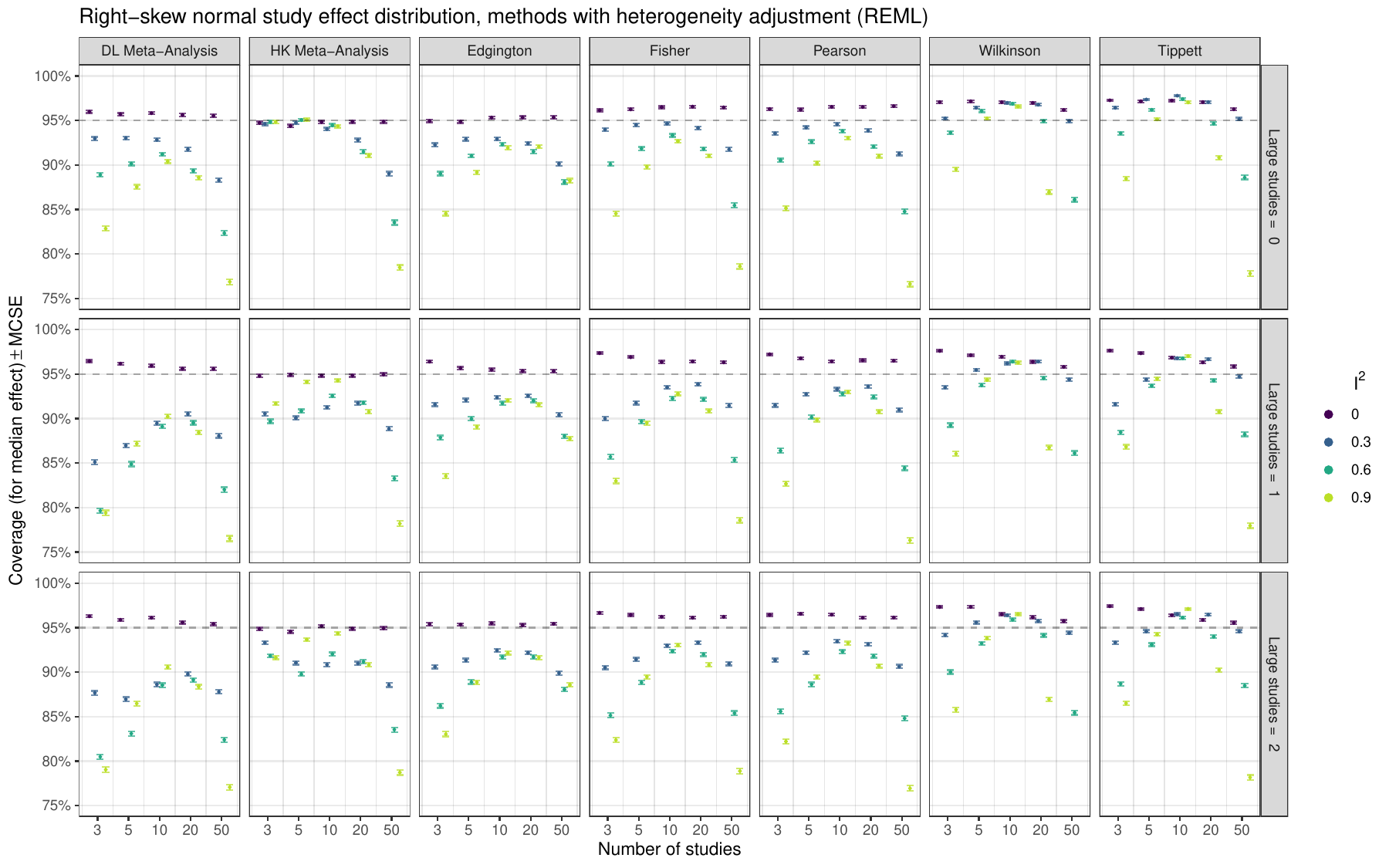} 
\end{knitrout}
\caption{Empirical coverage (for median effect) of the 95\% confidence intervals
  based on 20'000 simulation repetitions.}
\label{fig:covmedright}
\end{figure}
\begin{figure}[!htb]
\begin{knitrout}
\definecolor{shadecolor}{rgb}{0.969, 0.969, 0.969}\color{fgcolor}
\includegraphics[width=\maxwidth]{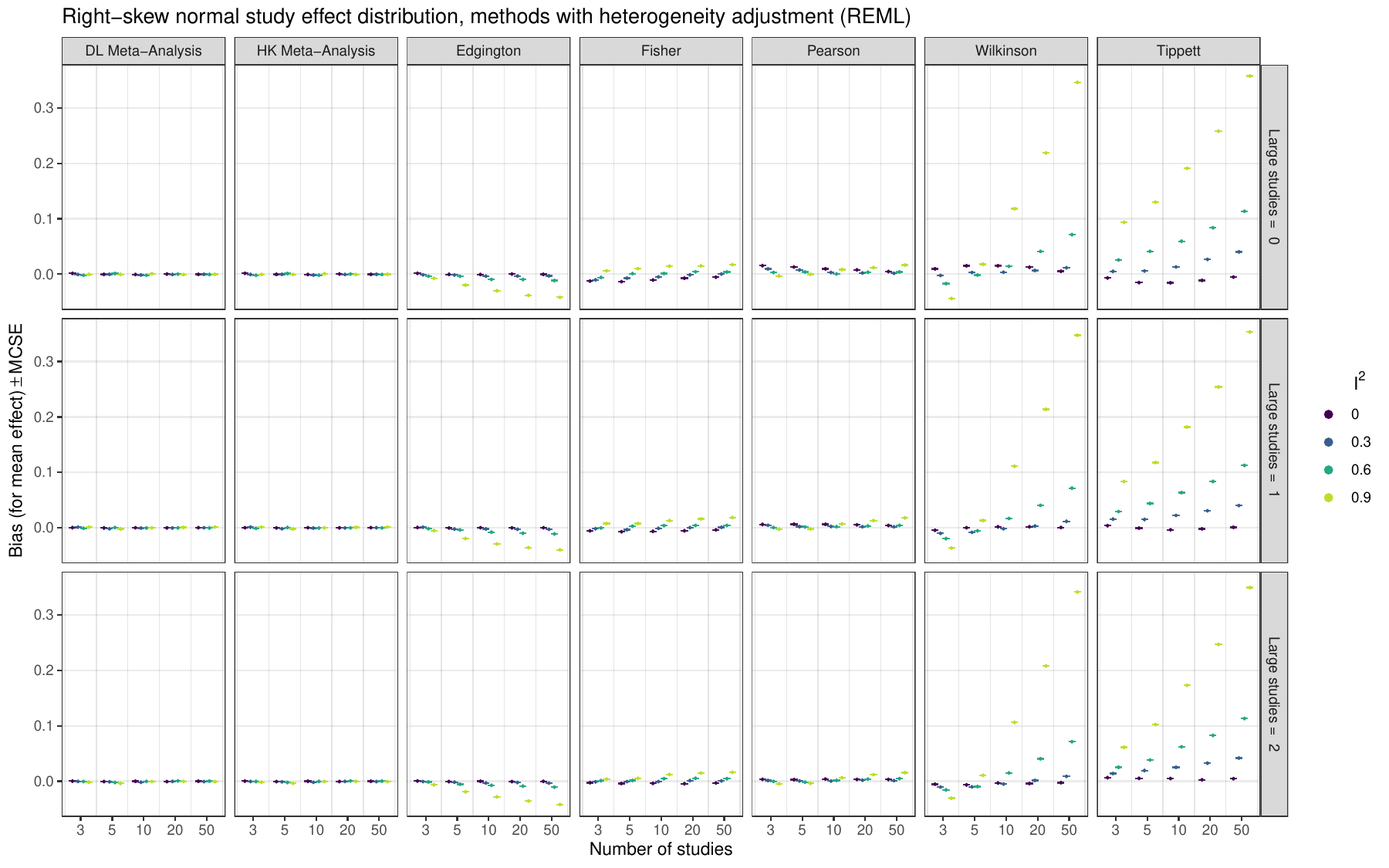} 
\end{knitrout}
\caption{Empirical bias of point estimates (for mean effect) based on
  20'000 simulation repetitions.}
\label{fig:biasmeanright}
\end{figure}

\begin{figure}[!htb]
\begin{knitrout}
\definecolor{shadecolor}{rgb}{0.969, 0.969, 0.969}\color{fgcolor}
\includegraphics[width=\maxwidth]{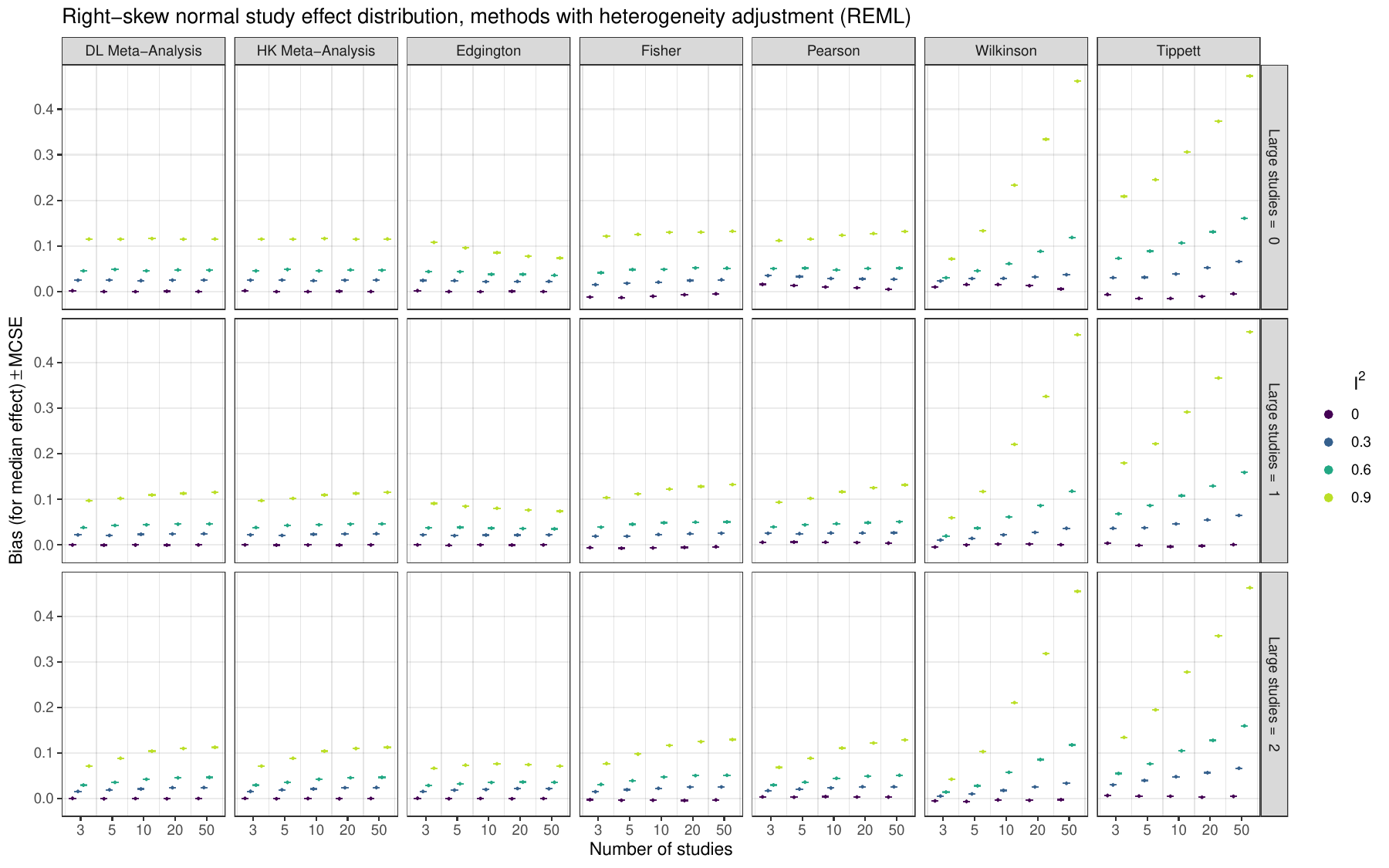} 
\end{knitrout}
\caption{Empirical bias of point estimates (for median effect) based on
  20'000 simulation repetitions.}
\label{fig:biasmedright}
\end{figure}
\begin{figure}[!htb]
\begin{knitrout}
\definecolor{shadecolor}{rgb}{0.969, 0.969, 0.969}\color{fgcolor}
\includegraphics[width=\maxwidth]{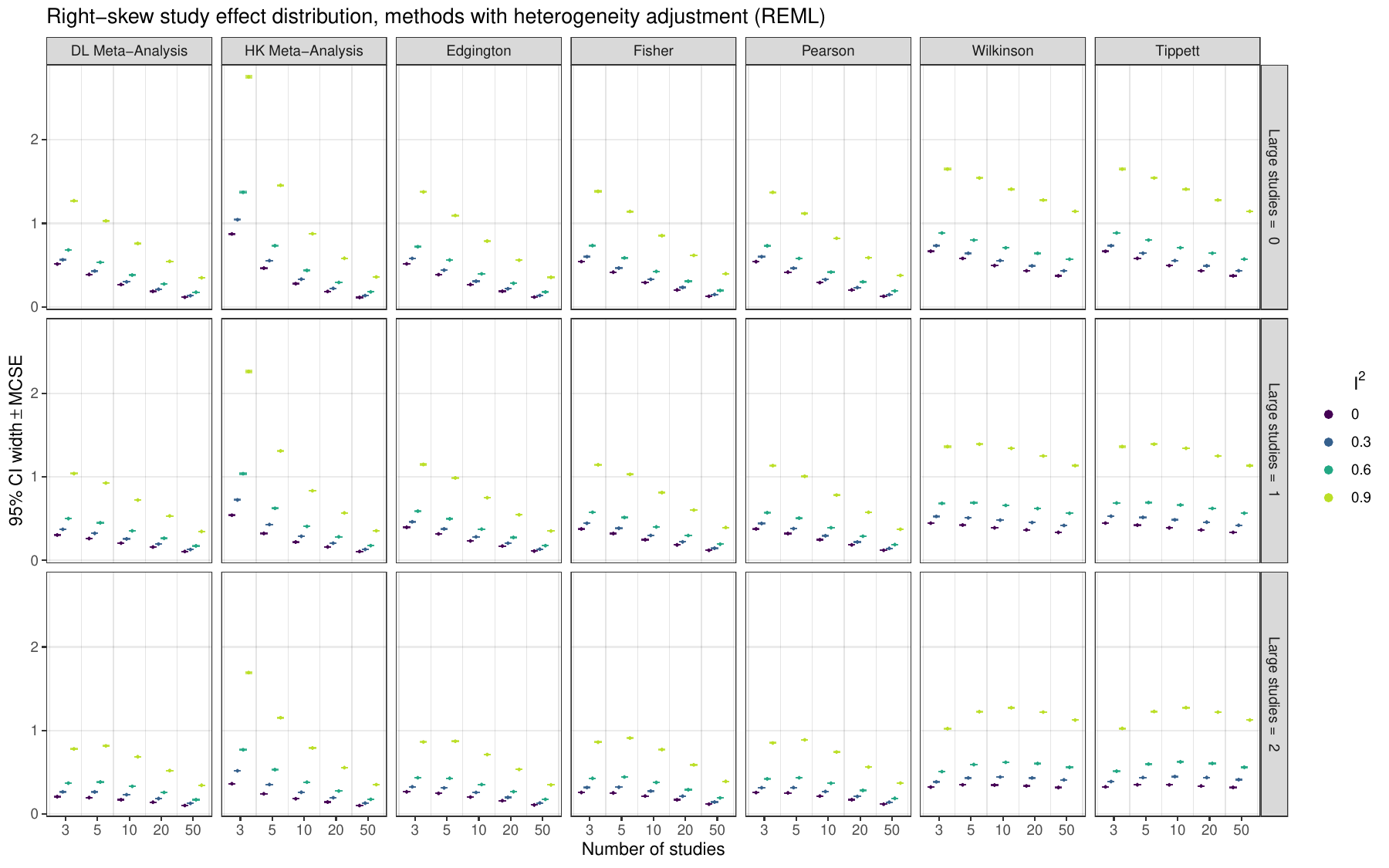} 
\end{knitrout}
\caption{Mean width of 95\% confidence intervals based on 20'000 simulation repetitions.}
\label{fig:widthright}
\end{figure}
\begin{figure}[!htb]
\begin{knitrout}
\definecolor{shadecolor}{rgb}{0.969, 0.969, 0.969}\color{fgcolor}
\includegraphics[width=\maxwidth]{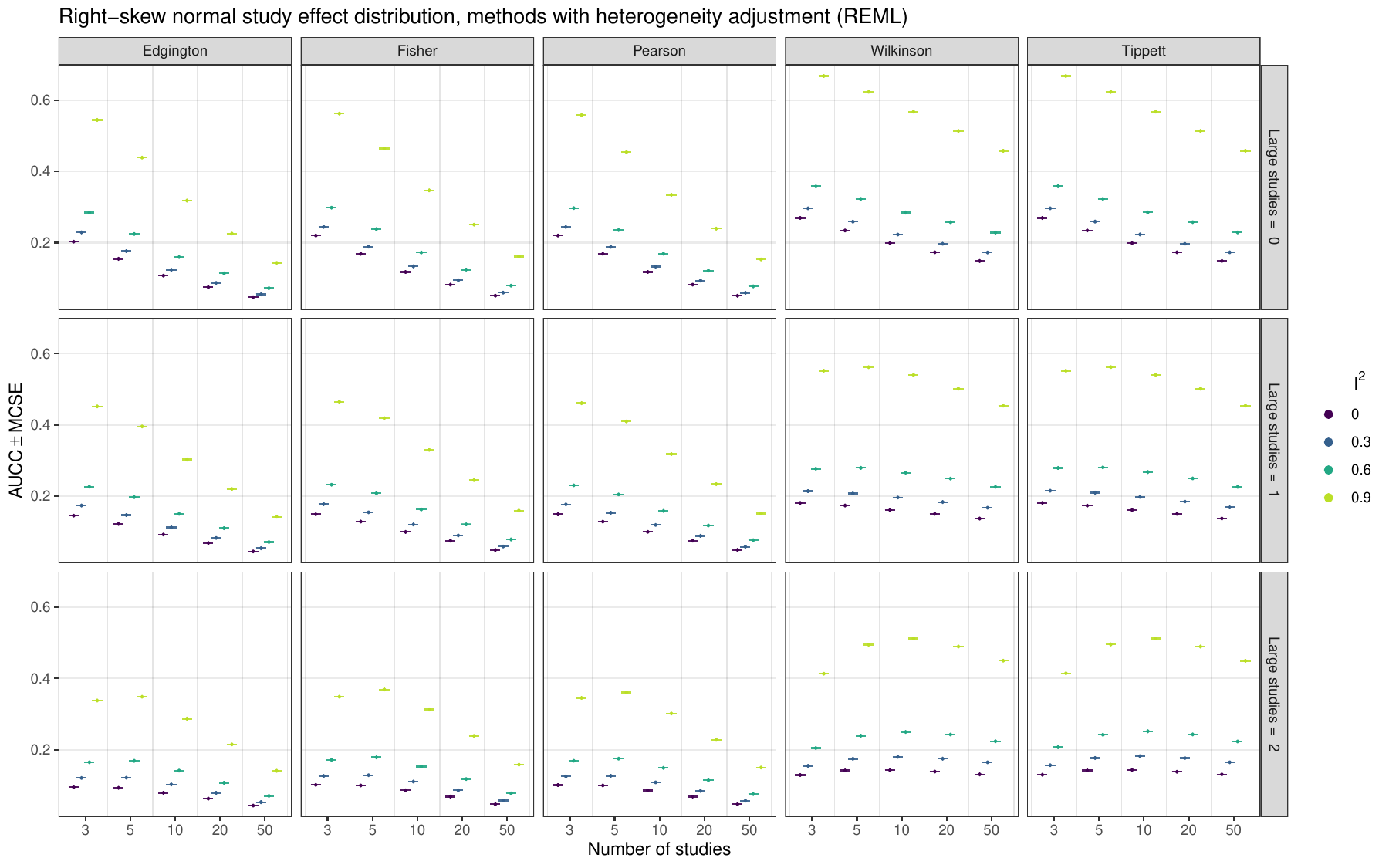} 
\end{knitrout}
\caption{Mean area under the confidence curve (AUCC) based on 20'000 simulation repetitions.}
\label{fig:AUCCright}
\end{figure}
\begin{figure}[!htb]
\begin{knitrout}
\definecolor{shadecolor}{rgb}{0.969, 0.969, 0.969}\color{fgcolor}
\includegraphics[width=\maxwidth]{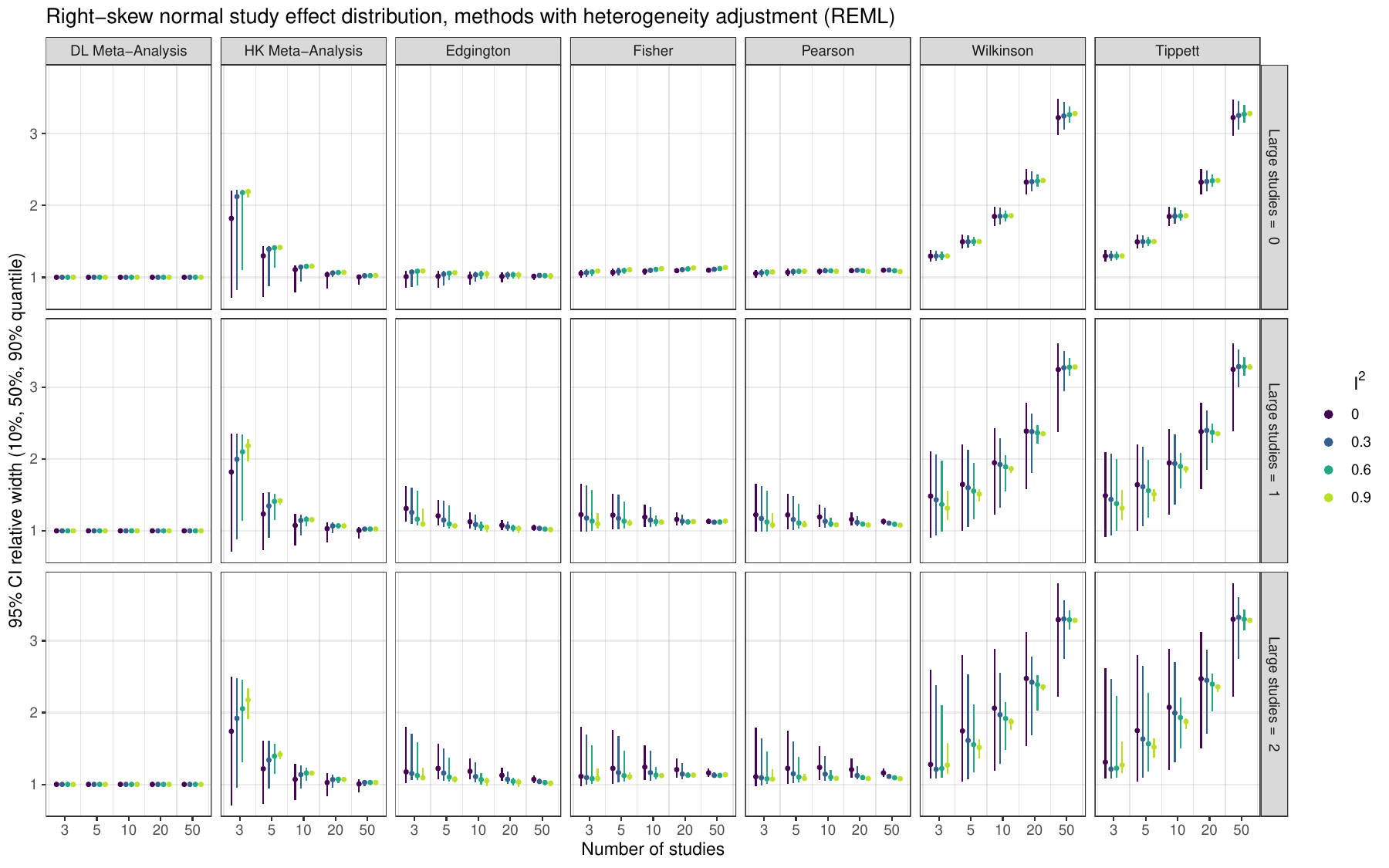} 
\end{knitrout}
\caption{Relative width of 95\% confidence intervals (relative to random effects
  meta-analysis) based on 20'000 simulation
  repetitions.}
\label{fig:relwidthright}
\end{figure}

\begin{figure}[!htb]
\begin{knitrout}
\definecolor{shadecolor}{rgb}{0.969, 0.969, 0.969}\color{fgcolor}
\includegraphics[width=\maxwidth]{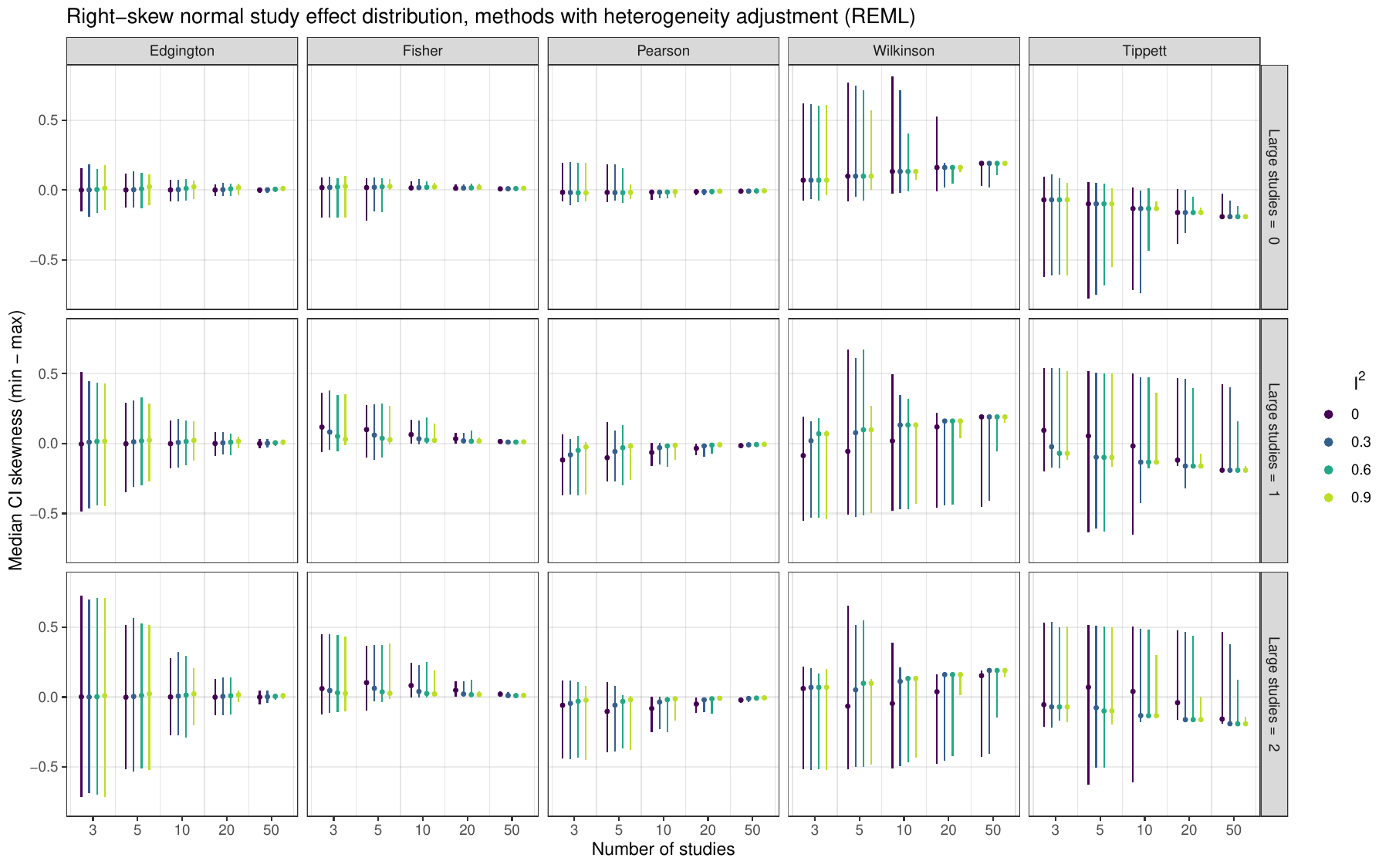} 
\end{knitrout}
\caption{Skewness of 95\% confidence intervals based on 20'000 simulation repetitions.}
\label{fig:skewright}
\end{figure}
\begin{figure}[!htb]
\begin{knitrout}
\definecolor{shadecolor}{rgb}{0.969, 0.969, 0.969}\color{fgcolor}
\includegraphics[width=\maxwidth]{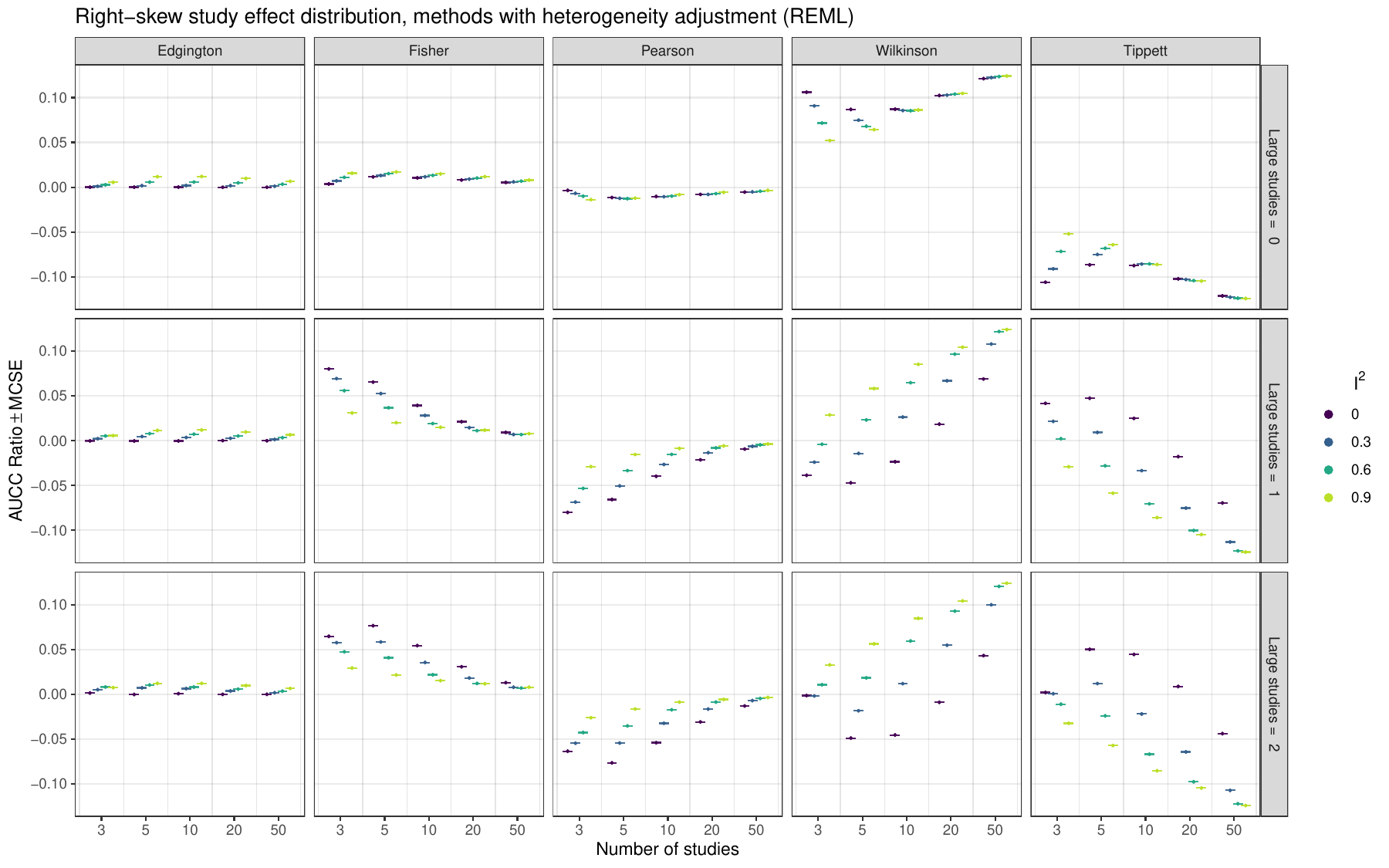} 
\end{knitrout}
\caption{Mean area under the confidence curve (AUCC) ratio based on 20'000 simulation repetitions.}
\label{fig:AUCCratioright}
\end{figure}

\begin{figure}[!htb]
\begin{knitrout}
\definecolor{shadecolor}{rgb}{0.969, 0.969, 0.969}\color{fgcolor}
\includegraphics[width=\maxwidth]{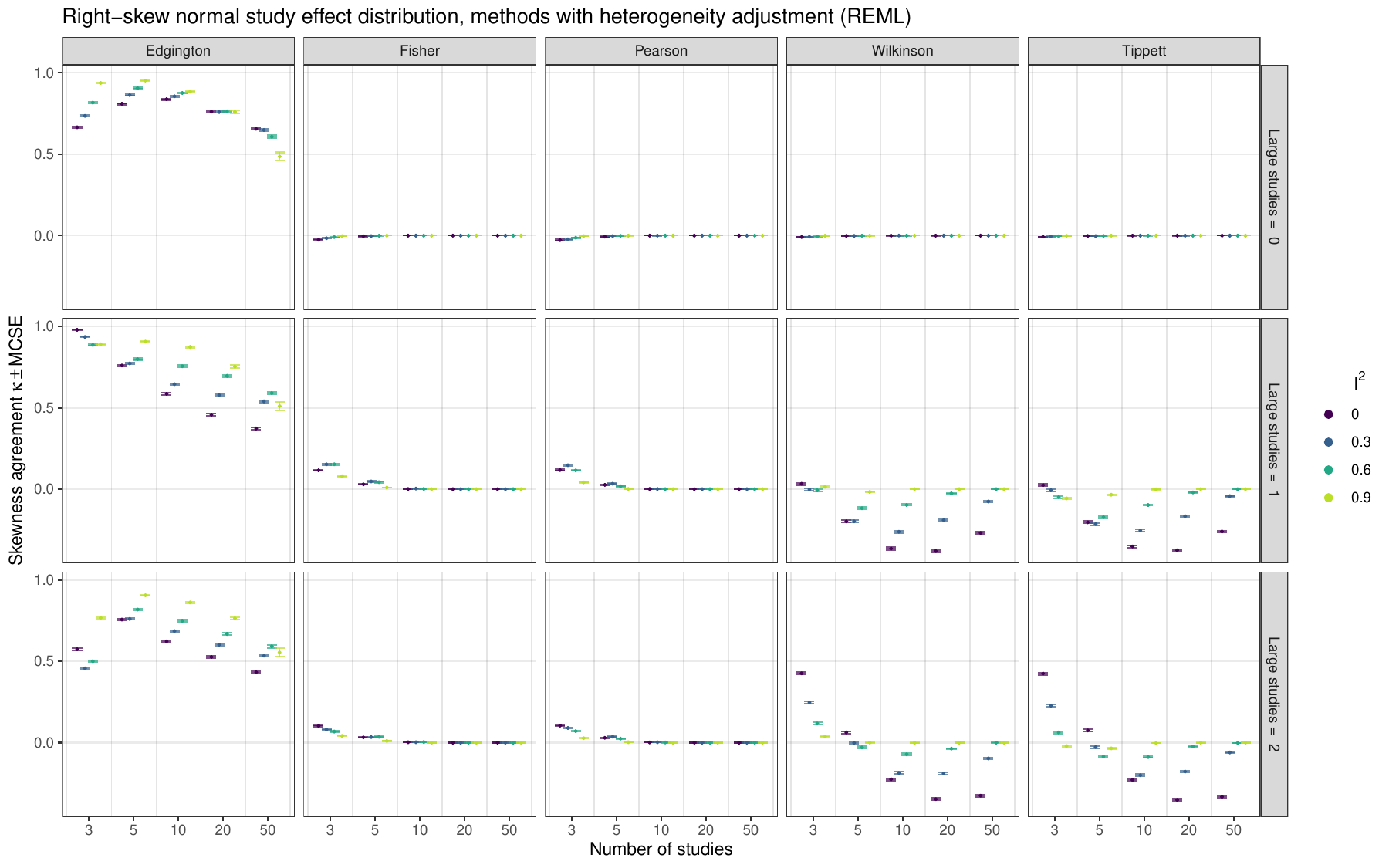} 
\end{knitrout}
\caption{Cohen's $\kappa$ sign agreement between 95\% confidence interval
  skewness and data skewness based on 20'000
  simulation repetitions.}
\label{fig:kapparight}
\end{figure}

\begin{figure}[!htb]
\begin{knitrout}
\definecolor{shadecolor}{rgb}{0.969, 0.969, 0.969}\color{fgcolor}
\includegraphics[width=\maxwidth]{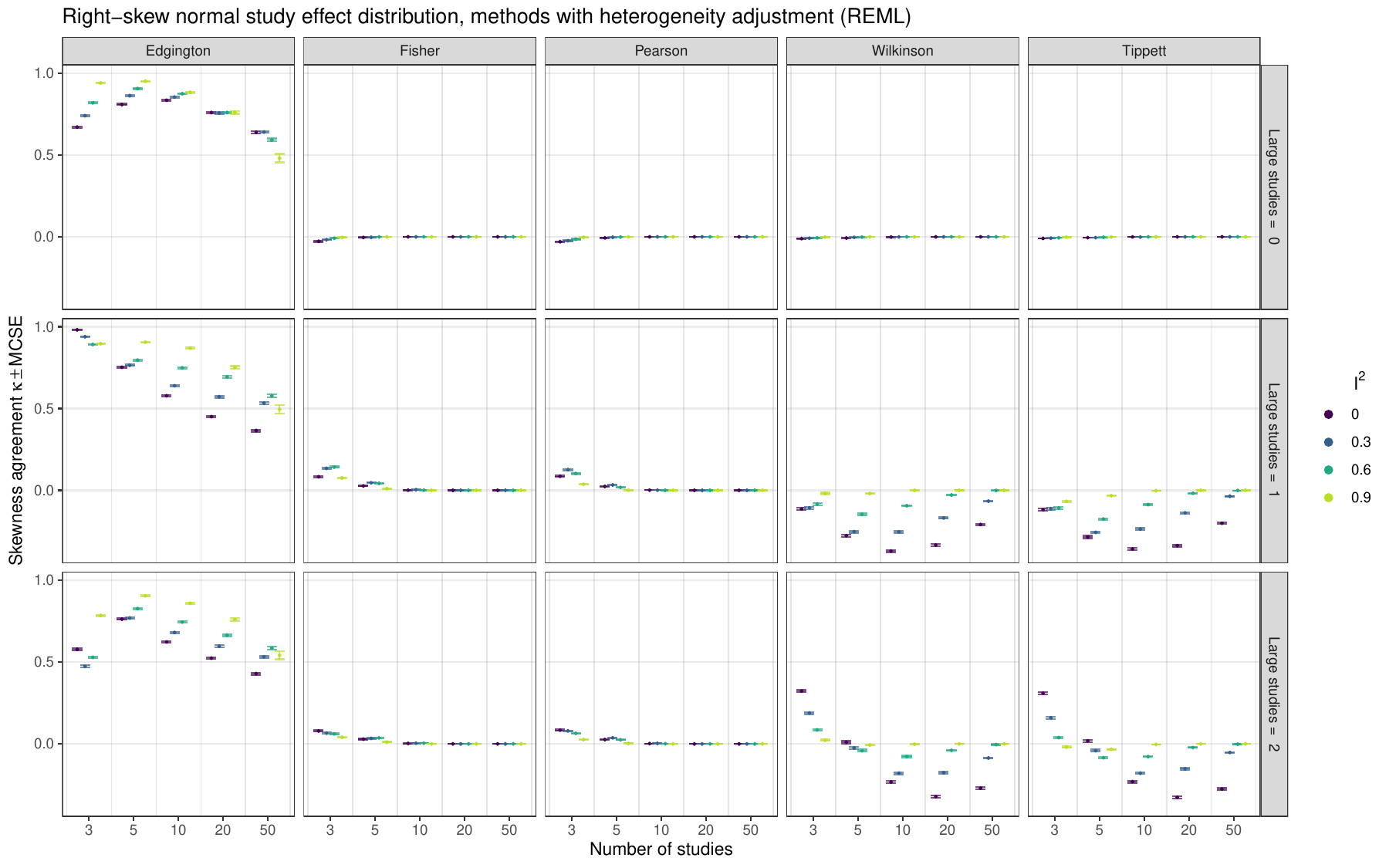} 
\end{knitrout}
\caption{Cohen's $\kappa$ sign agreement between AUCC ratio skewness and data
  skewness based on 20'000 simulation
  repetitions.}
\label{fig:kappa-auccratioright}
\end{figure}

\begin{figure}[!htb]
\begin{knitrout}
\definecolor{shadecolor}{rgb}{0.969, 0.969, 0.969}\color{fgcolor}
\includegraphics[width=\maxwidth]{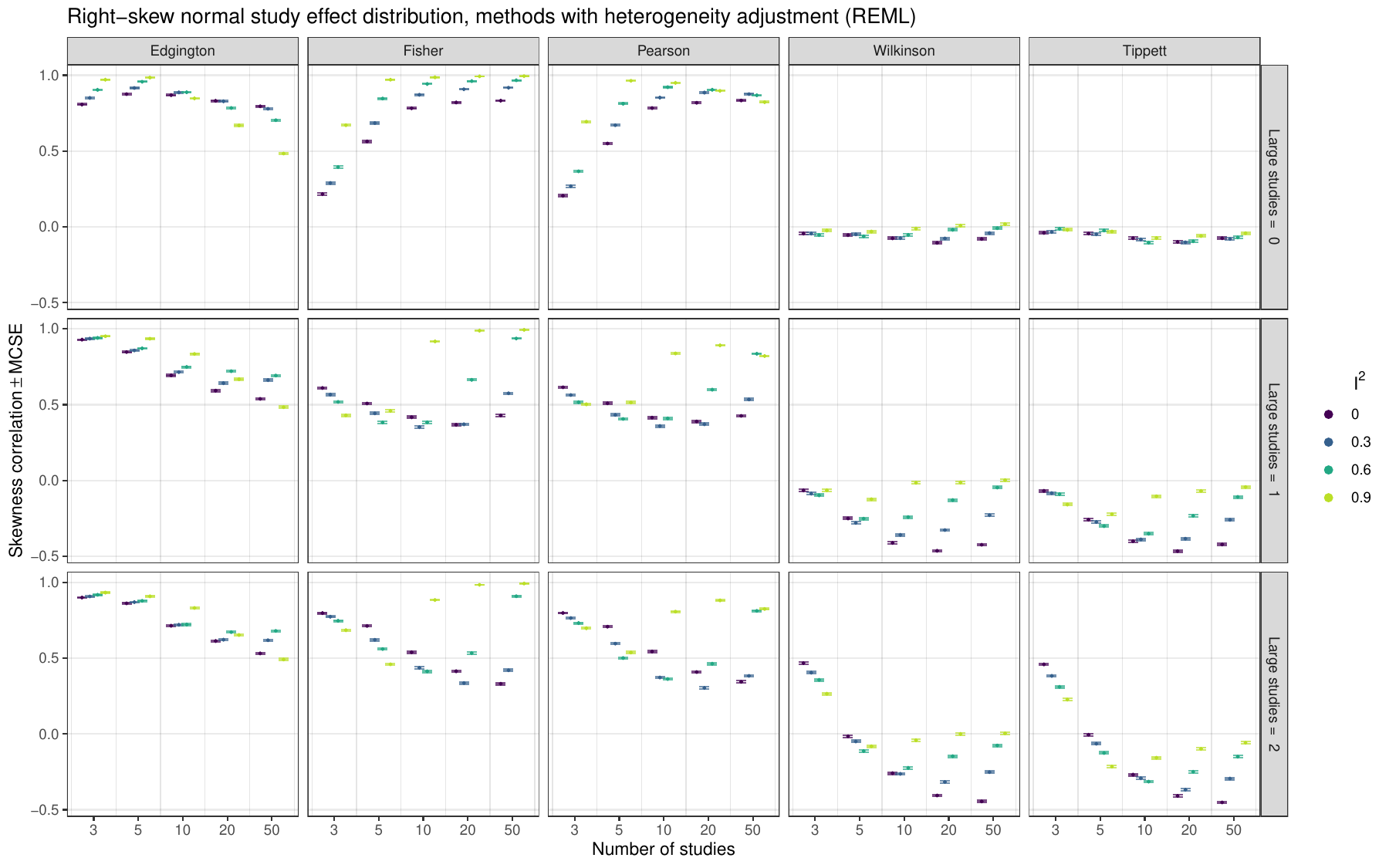} 
\end{knitrout}
\caption{Correlation between 95\% confidence interval skewness and data skewness
  based on 20'000 simulation repetitions.}
\label{fig:corright}
\end{figure}
\end{landscape}

\end{appendix}

\end{document}